\documentclass[traditabstract]{aa}
\usepackage{graphicx}
\usepackage{enumerate}
\usepackage{epsfig}
\usepackage{color}
\usepackage{txfonts}
\usepackage{multirow}
\usepackage{float}
\usepackage[export]{adjustbox}

\begin{document}
\title{The disc origin of the Milky Way bulge: \\ On the necessity of the thick disc}
\titlerunning{Why a thick disc is necessary for the Milky Way bulge}

\author{P.~Di Matteo\inst{1},  F.~Fragkoudi\inst{2}, S.~Khoperskov\inst{3}, B.~Ciambur\inst{4},  M.~Haywood\inst{1}, F.~Combes\inst{4,5}, A.~G$\rm \acute{o}$mez\inst{1}}

\authorrunning{P. Di Matteo et al.}

\institute{GEPI, Observatoire de Paris, PSL Research University, CNRS,
Place Jules Janssen, 92190 Meudon, France\\
\email{paola.dimatteo@obspm.fr}
\and Max-Planck-Institut f$\rm \ddot u$r Astrophysik, Karl-Schwarzschild-Str 1, D-85748 Garching bei M$\rm \ddot u$nchen, Germany
\and Max Planck Institute for Extraterrestrial Physics, 85741 Garching, Germany
\and Observatoire de Paris, LERMA, CNRS, PSL Research University, UPMC, Sorbonne Univ., F-75014, Paris, France
\and College de France, 11 Place Marcelin Berthelot, 75005, Paris, France
}

\date{Accepted, Received}

\abstract{In the Milky Way bulge, metal-rich stars form a strong bar and are more peanut-shaped than metal-poor stars. It has been recently claimed that this behavior is driven by the initial (i.e. before bar formation) in-plane radial velocity dispersion of these populations, rather than by their initial vertical random motions. This has led to the suggestion that a thick disc is not necessary to explain the characteristics of the Milky Way bulge.   We rediscuss this issue by analyzing two dissipationless N-body simulations of boxy/peanut-shaped bulges formed from composite stellar discs, made of kinematically cold and hot stellar populations. These two models represent two extreme cases,  namely one where all three components of the disc have fixed vertical velocity dispersion and different in-plane radial  dispersion,  and another where they all have fixed radial dispersion and different vertical random motions (thickness), and are intended to quantify the drivers of the main features observed in composite boxy/peanut-shaped bulges and their origin. We quantify  the mapping  into a boxy/peanut-shaped bulge of disc populations in these two cases, and we 
conclude that initial vertical random motions are as important as in-plane random motions in determining the relative contribution of cold and hot disc populations with height above the plane, the metallicity and age trends. Previous statements emphasizing the dominant role of in-plane motions in determining these trends are not confirmed. However,
 significant differences exist in the morphology and strength of the resulting boxy/peanut-shaped bulges. In particular, the model where disc populations have initially only different in-plane random motions, but similar thickness, results into a boxy/peanut bulge where all populations have a similar peanut shape, independently on their initial kinematics -- or metallicity. This is at odds with the trends observed in the Milky Way bulge.  We discuss the reasons behind these differences, and also predict the signatures that these two extreme initial conditions would leave on the vertical age and metallicity gradients of disc stars, outside the bulge region.   As a consequence of this analysis,  we conclude that -- given our current knowledge of the Milky Way bulge and of the properties of its main stellar components  --   a metal-poor, kinematically (radial and \emph{vertical}) hot component, that is a thick disc, is necessary in the Milky Way before bar formation, supporting the scenario traced in  previous works. Boxy/peanut shape bulges and their surrounding regions are fossil records of the conditions present at early times in disc galaxies, and by dissecting their stellar components by  chemical compositions, and/or age, it is potentially possible to reconstruct their early state. }


\maketitle

\section{Introduction}\label{intro}

The Milky Way (hereafter MW)  bulge is a boxy/peanut-shaped bulge \citep{okuda77, maihara78, weiland94, dwek95, binney97, babusiaux05, lopez05, rattenbury07, cao13, wegg13, portail15b, portail17a,  ciambur17, portail17b}, made up, for the vast majority, of stars originated in its disc. Several works indeed now agree in limiting the contribution of a classical bulge to a small percentage of its total mass \citep{shen10, kunder12, dimatteo14, kunder16, debattista16, gomez18}. Also the contribution of the stellar halo, whose density is expected to peak in the inner few kiloparsecs of the Galaxy, seems to be marginal, not exceeding few percent of the bulge total mass \citep{ness13spop}. While there is growing consensus on these general results, the details remain more controversial. One of the points still debated concerns the origin of the metal-poor ($\rm -1~dex \le [Fe/H] \le 0~dex$), $\alpha$-enhanced population observed in the bulge. Compared to the metal-rich population ($\rm [Fe/H] > 0$), metal-poor stars appear kinematically hotter at latitudes $\rm b \lesssim -4~deg$, and colder closer to the midplane \citep{ness13kin, rojas14, zoccali17}, not strongly peanut-shaped \citep[][but with an evident difference between stars with metallicities below and above $\sim -0.5$~dex, see \citet{ness13spop}]{rojas14}, and dominate the outer bulge region. Their proportion, relative to the total population, decreases with decreasing height from the Galactic plane until at least latitudes $\rm b \simeq -5~deg$ \citep{ness13spop}, and then seems to show an inversion, becoming again dominant at very low latitudes \citep{zoccali17}.\\ In \citet{dimatteo14, dimatteo15} we have proposed  that the metal-poor population observed in the bulge is the Galactic thick disc, mapped into the boxy/peanut-shaped (hereafter B/P)   bulge differently from the metal-rich stars, because of its hotter kinematics. In particular, we have suggested  \citep[see also][ for this same interpretation]{ness13spop} that stars with metallicity between $-1\le \rm [Fe/H] \le -0.5$~dex can be associated to the old Galactic thick disc  \citep[ages greater than about 10 Gyr, metallicities below -0.5~dex, see][]{haywood13} --  
 while bulge stars with metallicities in the range $-0.5 \le \rm [Fe/H] \le 0$ would correspond to stars of the young Galactic thick disc \citep[ages between 8 and 10 Gyr, metallicities between $-0.5$~dex and solar, see][]{haywood13}. The reasons behind this suggestion have been reviewed in \citet{dimatteo16}, and we refer the reader to that paper for a discussion on the disc-bulge connection in this picture.  In particular, in \citet{dimatteo16} and \citet{fragkoudi17}  we have shown that a composite disc galaxy made initially of a kinematically cold, intermediate and hot disc -- mimicking respectively the thin disc, the young and the old thick disc of the Galaxy -- can reproduce the morphological and populations trends observed in the MW bulge. Namely: the colder the population, the stronger the bar, and -- at all heights where a B/P morphology is observed -- the hotter the population, the weaker the B/P structure. 
Also the chemical patterns and the metallicity distribution function of bulge stars can be understood simply as a consequence of the differential mapping of the composite Galactic (thin+thick) disc into the B/P bulge, modulo their  scale-lengths and heights \citep[see][]{fragkoudi17b, haywood18, fragkoudi18}. In this context,  the agreement of the metallicity map that this scenario predicts  with that of the MW bulge, as  derived from APOGEE data, is remarkable  \citep{fragkoudi18}.\\
Our findings, based on models tailored to match the MW bulge and inner disc populations, agree with the general result that in a composite disc, populations with different kinematics are mapped differently into a bar and B/P bulge --  a process referred to as \emph{kinematic fractionation} by  \citet{debattista16}. This result, firstly shown by \citet{bekki11}, has been since then indeed confirmed and extended in a number of other works \citep{dimatteo16, fragkoudi17, athanassoula17, buck18}. However, in  \citet{debattista16} it is suggested that the main driver of this different response to the bar perturbation is  the radial, in-plane random motion of stars, rather then their vertical ones. In their view,  a composite disc made of stellar populations with the same initial thickness but different in-plane random motions constitutes a  sufficient condition to reproduce the trends observed in the MW bulge, and  they conclude that, as a consequence, a thick disc --  that is a disc both radially \emph{and} vertically hot -- is not necessary to explain the MW bulge main characteristics.  \\

In this paper we re-investigate this issue, about the driver of the trends observed in the MW bulge. First, we aim at understanding whether the in-plane, radial random motions of disc stars constitute the main drivers of the trends observed in the MW bulge, as suggested, and the role played by vertical motions.
Second, by comparing some of the current properties of the MW bulge with the models predictions, we aim at finding signatures and suggesting criteria to constrain the kinematic conditions present in early MW disc.
  
 For this study, we analyse two dissipationless N-body simulations of composite disc galaxies. These two models represent two extreme cases, namely one where all three components of the disc have fixed vertical velocity dispersion and different planar/radial (in-plane) dispersion, as in \citet{debattista16}, and another where they all have fixed radial dispersion and different vertical random motions (thickness). Their analysis  reveals  that disc populations with different initial vertical random motions and same in-plane random motions are subject,  at first order, to a similar mapping into a B/P-shaped bulge, as that experienced by stellar populations with equal initial thickness and different in-plane motions. More specifically, both initial conditions lead to composite B/P bulges, where formerly disc stars with initially the highest velocity dispersions redistribute in a thicker B/P structure and weaker bar, than disc stars which have initially a cold kinematics. As a consequence of this similar mapping, in both models, the B/P bulges show vertical metallicity/age gradients, and a pinching of the metallicity maps along the bulge minor axis, as also observed \citet{gonzalez17}.
Despite these same trends, we note, however, some significant differences in the morphology of the resulting B/P structure in the two cases. In particular, when the disc is made of stellar populations with the same initial thickness, which are differentiated only in their in-plane motions, as in \citet{debattista16}, the peanut structure appears at the same height above the plane, for all populations, independently on their kinematics, and its strength and shape is remarkably similar.  This is at odds with what is observed in the MW bulge. We discuss the reasons behind this different behaviour  and how the coupling of radial and vertical motions acts in the two cases.

These two models \emph{constitute a first step}  to explore to what extent in-plane and vertical random motions reshape the properties of B/P bulges and their surrounding discs, but none of them evolves into properties compatible with those of MW disc. \emph{As we discuss,}  it is indeed necessary that disc populations both radially and vertically warm -- such as those currently found in the MW thick disc -- were  present in the Galaxy at the beginning of its secular evolution, in order to reproduce some of the trends observed in the Galaxy.

The paper is organized as follows : in Sect.~\ref{models} we describe the two models analyzed in this paper, their properties, and the adopted numerical methods; in Sect.~\ref{results} we present the results, by discussing first some main general trends (morphology, metallicity, ages) of  their B/P bulges and their surrounding discs (Sect.~\ref{bp}), then the strength of the B/P shape and its dependence on the initial disc kinematics (Sect.~\ref{strength})   and, finally, in Sect.~\ref{conclusions}, we outline our conclusions.

\section{Models}\label{models}

\begin{figure*}
\begin{center}
\includegraphics[clip=true, trim = 4mm 0mm 15mm 0mm, width=0.3\linewidth]{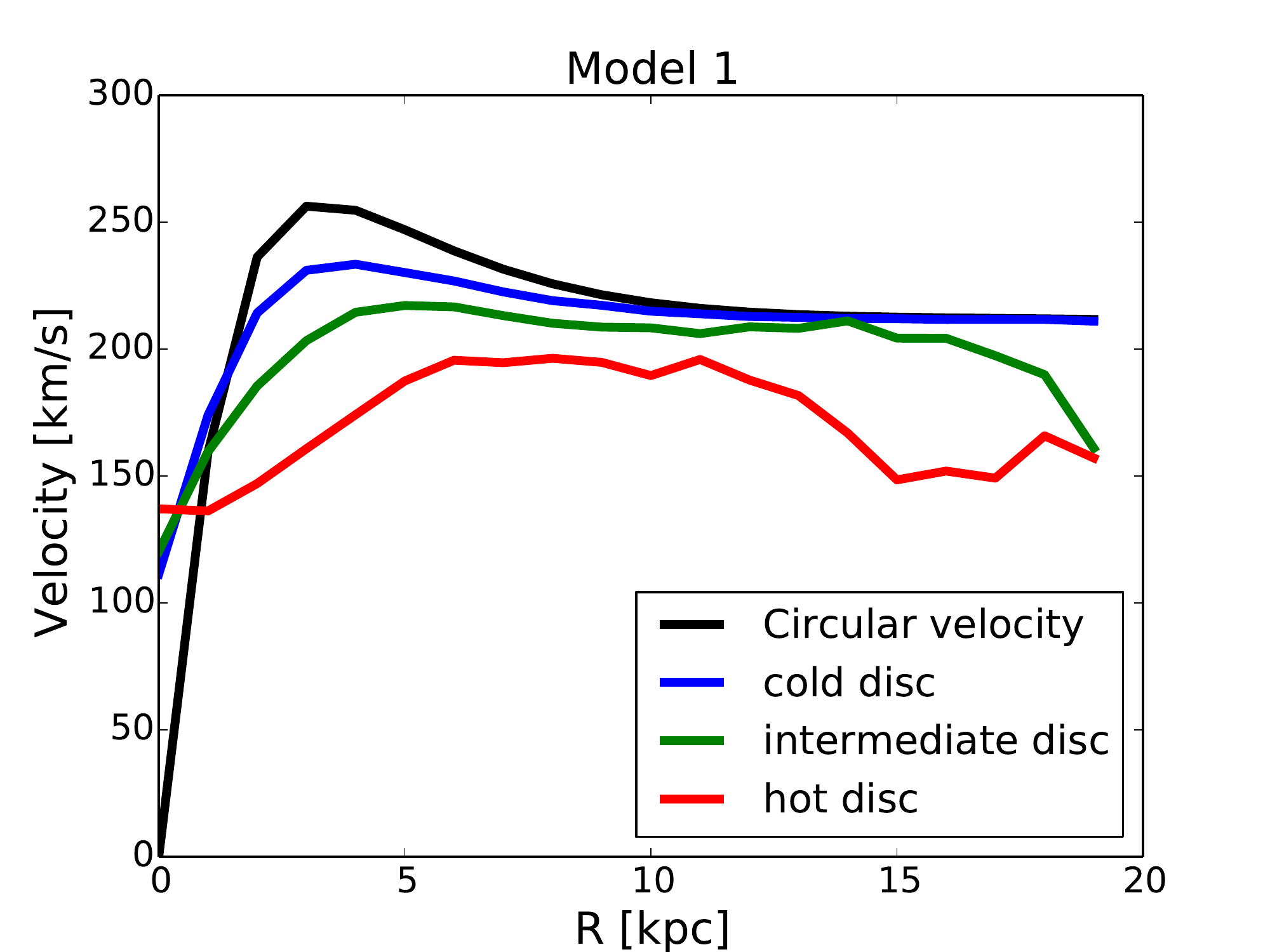}
\includegraphics[clip=true, trim = 4mm 0mm 15mm 0mm, width=0.3\linewidth]{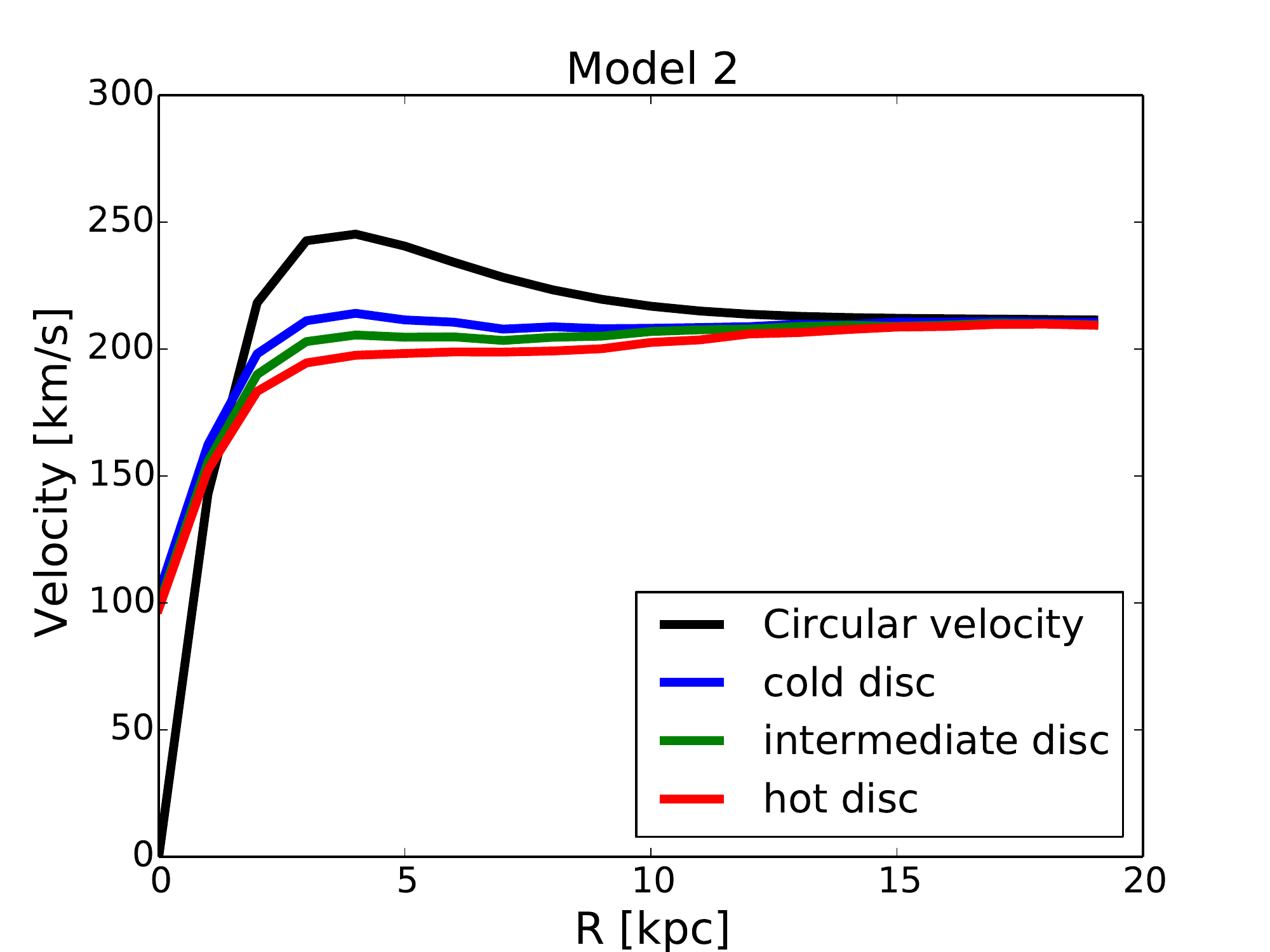}

\includegraphics[clip=true, trim = 4mm 0mm 15mm 0mm, width=0.3\linewidth]{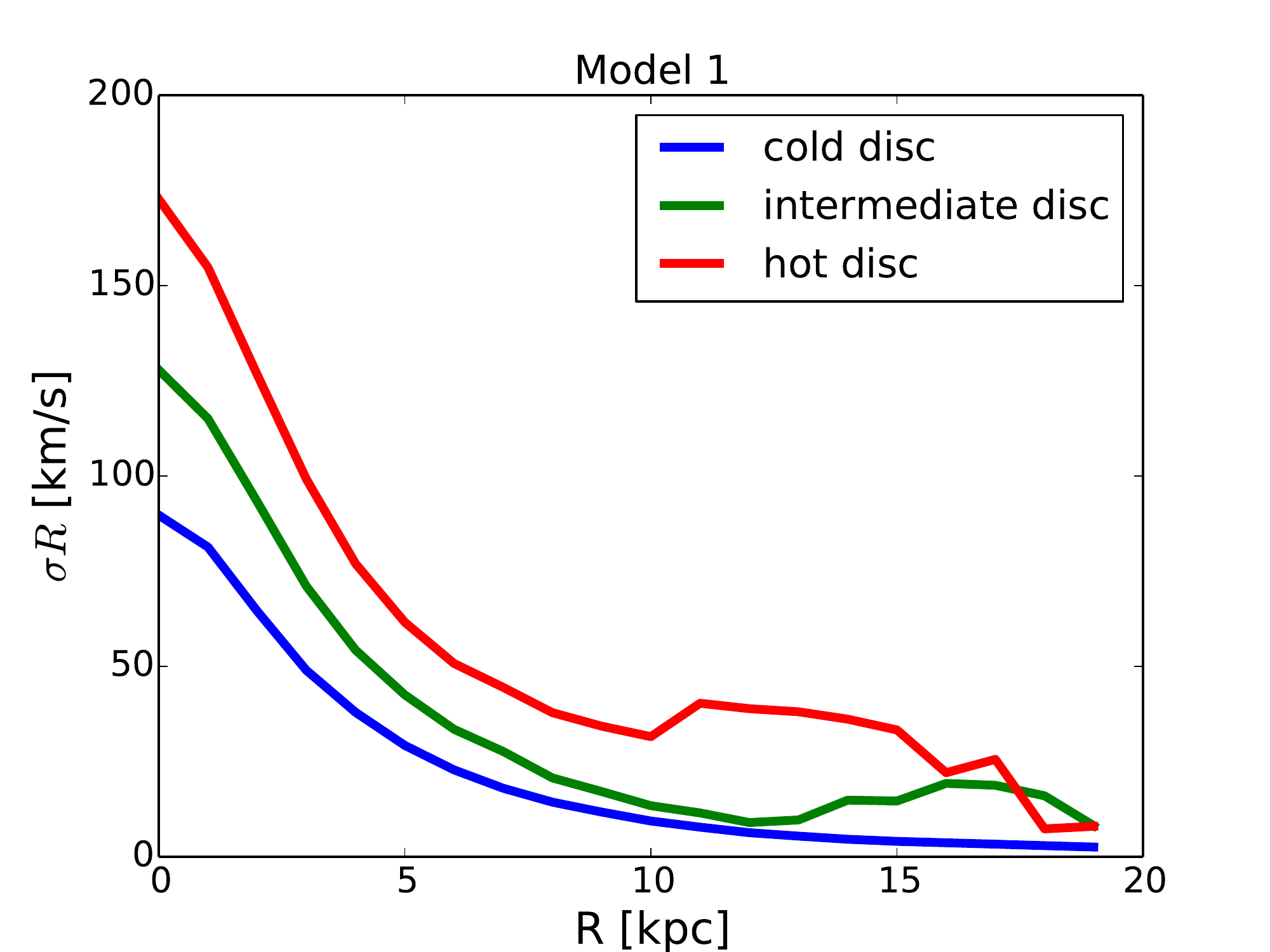}
\includegraphics[clip=true, trim = 4mm 0mm 15mm 0mm, width=0.3\linewidth]{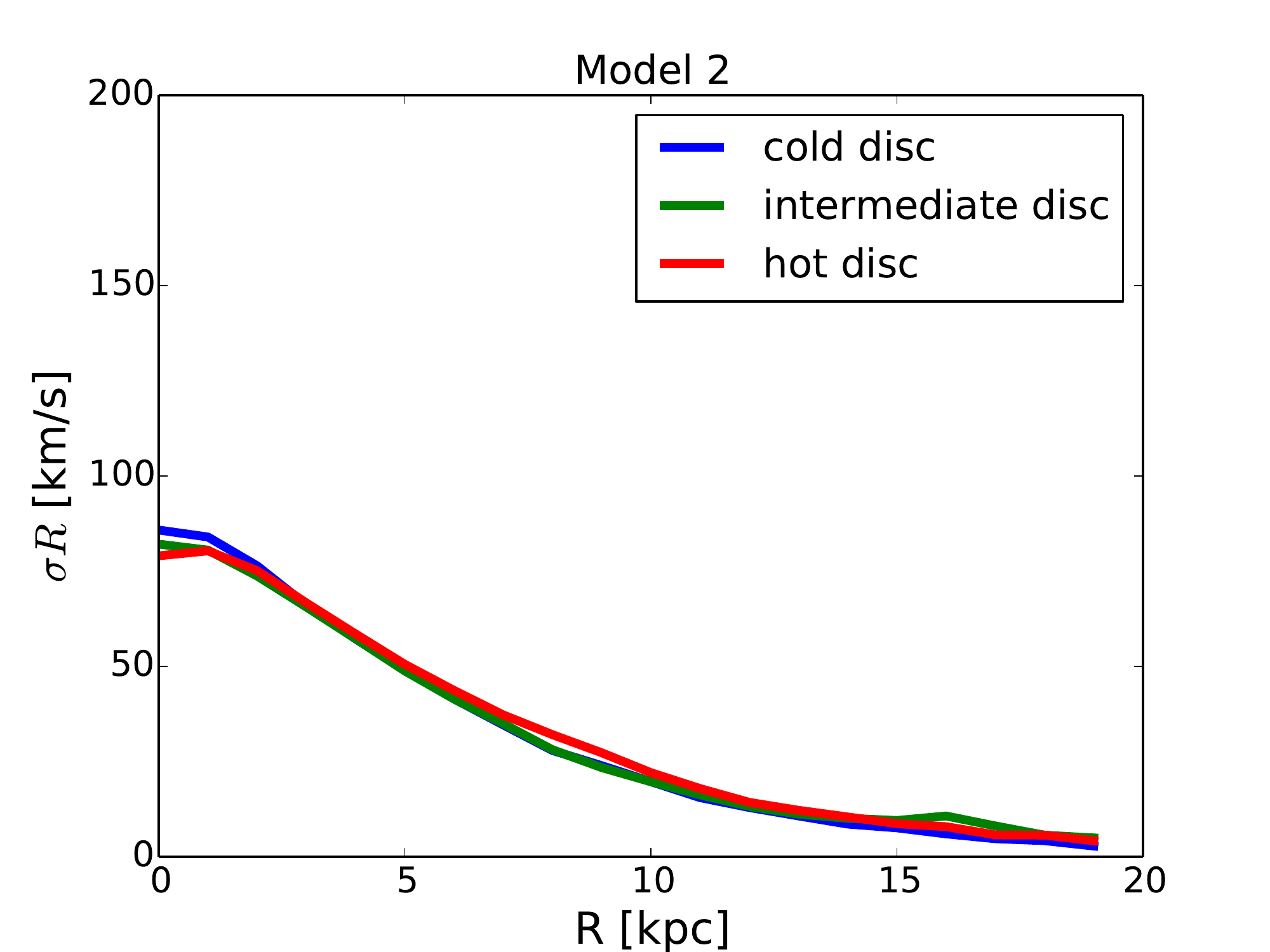}

\includegraphics[clip=true, trim = 4mm 0mm 15mm 0mm, width=0.3\linewidth]{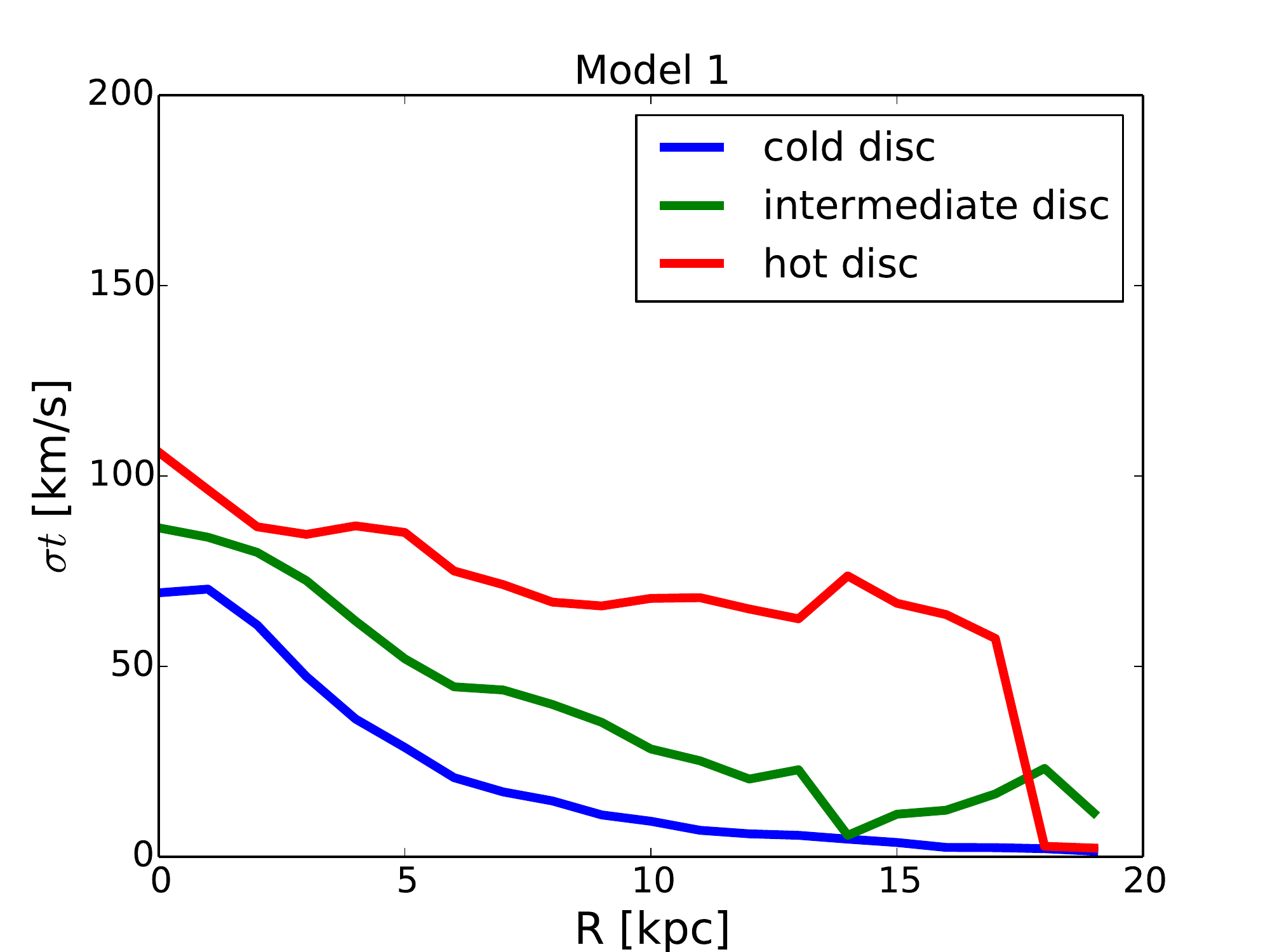}
\includegraphics[clip=true, trim = 4mm 0mm 15mm 0mm, width=0.3\linewidth]{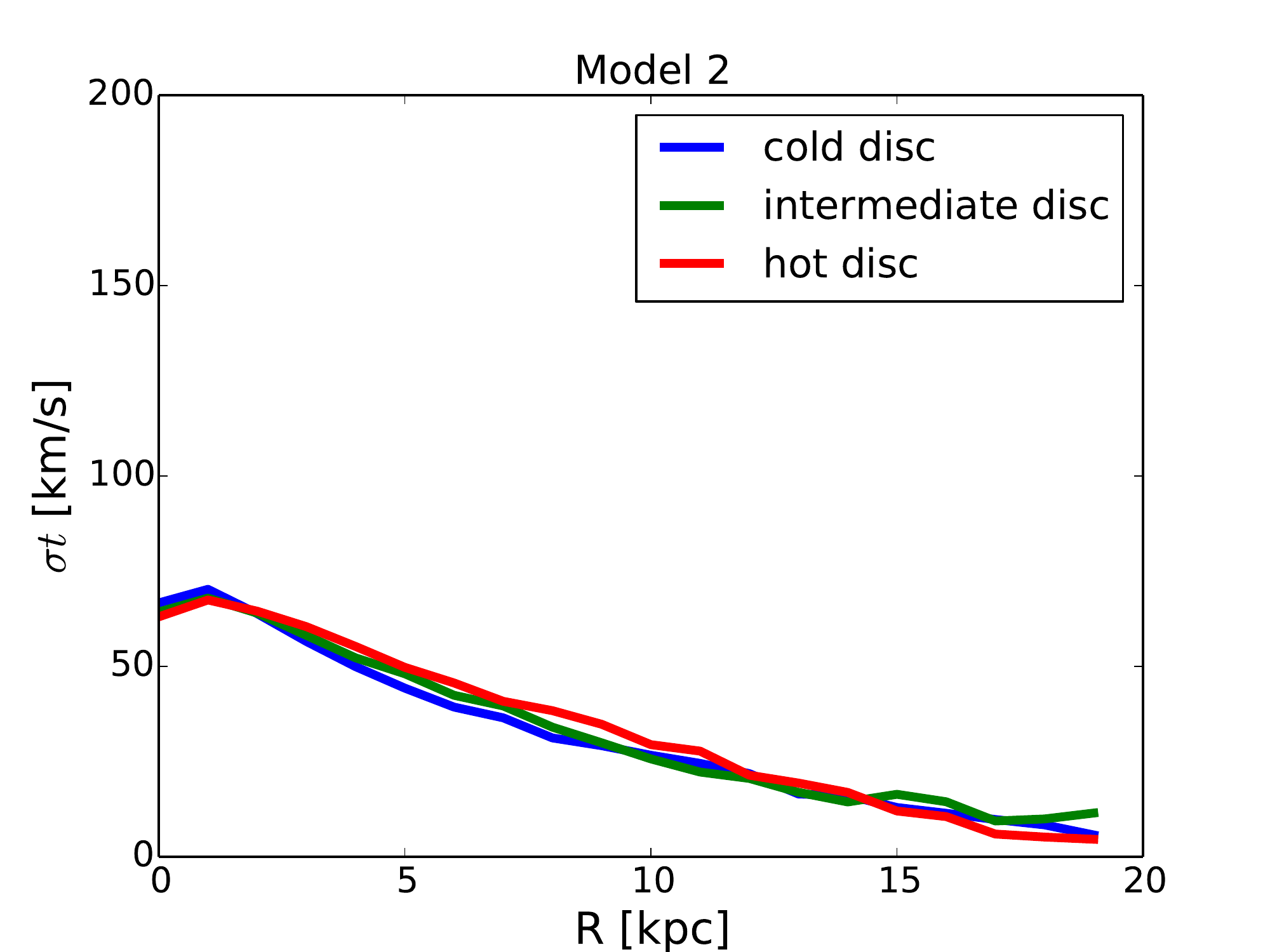}

\includegraphics[clip=true, trim = 4mm 0mm 15mm 0mm, width=0.3\linewidth]{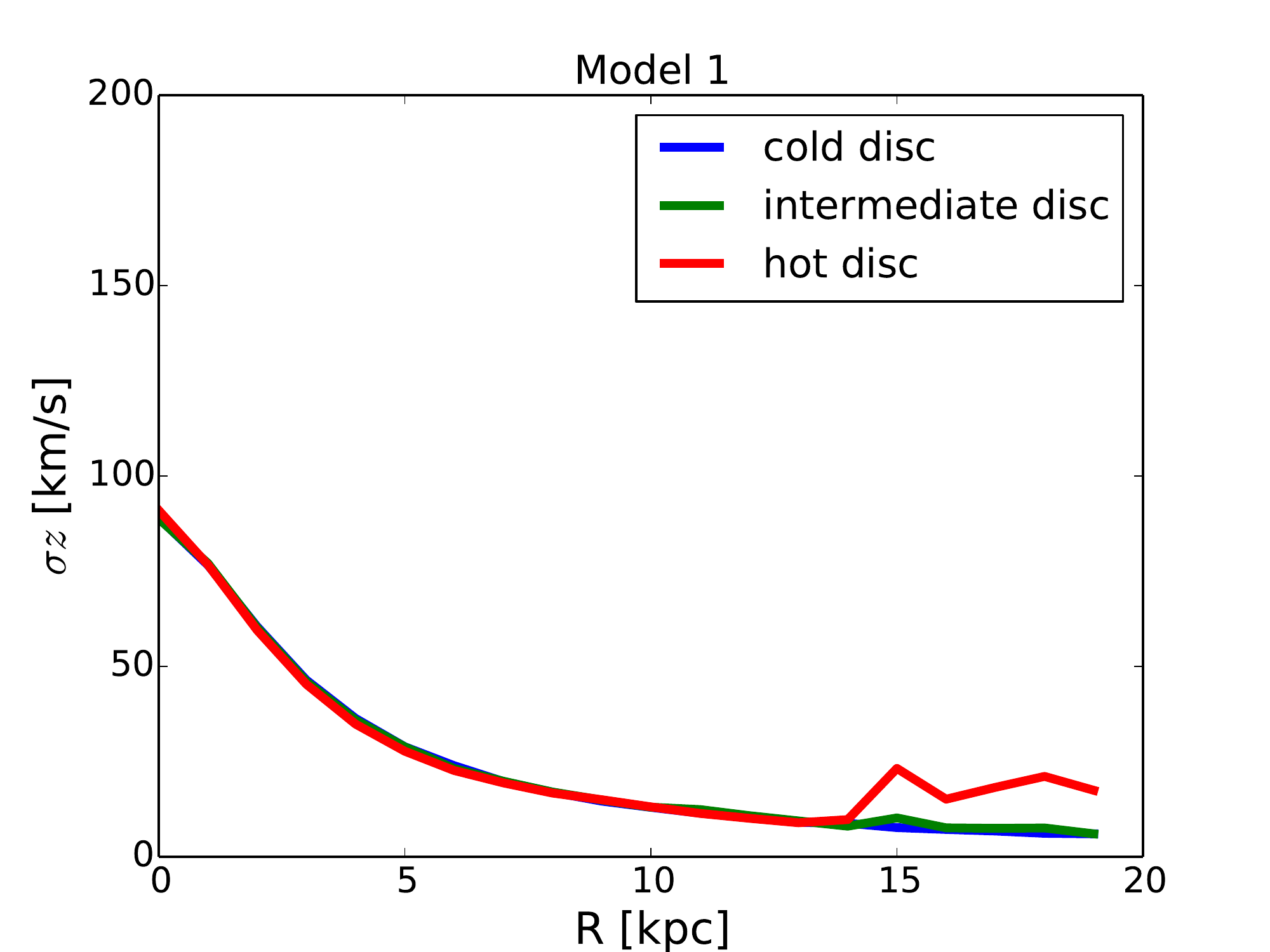}
\includegraphics[clip=true, trim = 4mm 0mm 15mm 0mm, width=0.3\linewidth]{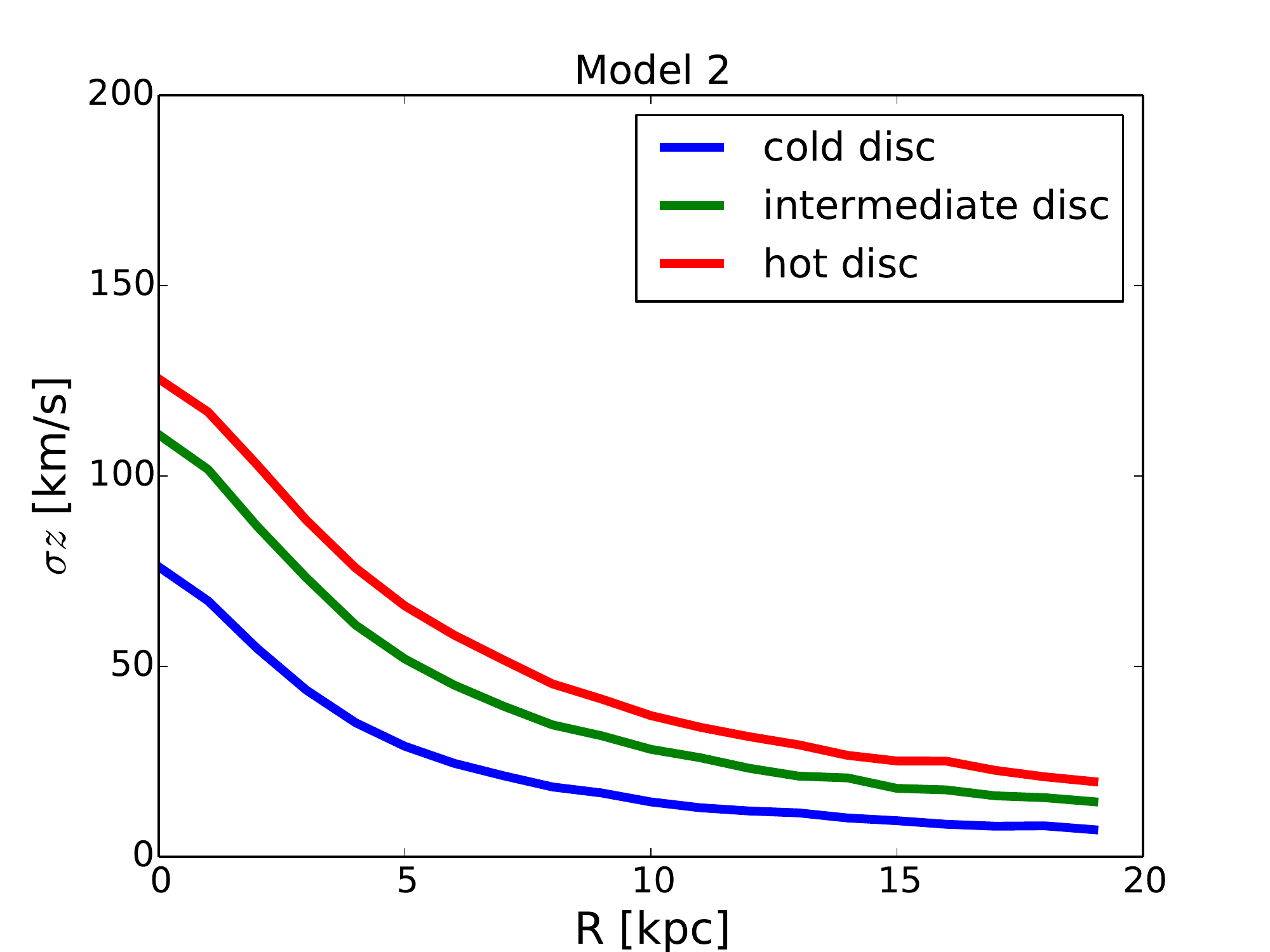}
\end{center}
\caption{\emph{Initial conditions of our composite stellar discs. From top to the bottom, } \emph{First row:} Circular velocity (\emph{black curve}) of the model, together with the rotation curve of the cold (\emph{blue}), intermediate (\emph{green}) and hot (\emph{red}) discs; \emph{Second row:} In-plane radial velocity dispersion of the cold (\emph{blue}), intermediate (\emph{green}) and hot (\emph{red}) discs,  as a function of the distance $R$ from the galaxy centre;  \emph{Third row: } In-plane tangential velocity dispersion of the cold (\emph{blue}), intermediate (\emph{green}) and hot (\emph{red}) discs,  as a function of the distance $R$ from the galaxy centre; \emph{Bottom row: } Vertical velocity dispersion of the cold (\emph{blue}), intermediate (\emph{green}) and hot (\emph{red}) discs,  as a function of the distance $R$ from the galaxy centre. Model 1 is shown on the left column, Model 2 on the right column, as indicated. }
\label{IC}
\end{figure*}

\begin{table}\label{galparamtable}
\begin{center}
\begin{tabular}{lcccc}
\hline
\hline
& \multicolumn{4}{c}{Model 1}\\
\hline
               & $M$ & $a$ & $h$ & $N$ \\ 
\hline         
\\
Cold disc & 18.3 & 2. & 0.25 & 5M\\
Intermediate disc & 11.2 & 2. & 0.25 & 3M\\
Hot disc & 8.1 & 2. & 0.25 & 2M\\
Dark halo & 160. & 0. & 21. & 5M\\
\\
&  \multicolumn{4}{c}{Model 2}\\
\hline
               & $M$ & $a$ & $h$ & $N$ \\ 
\hline         
\\
Cold disc & 18.3 & 2. & 0.25 & 5M\\
Intermediate disc & 11.2 & 2. & 0.6 & 3M\\
Hot disc & 8.1 & 2. & 0.9 & 2M\\
Dark halo & 160. & 0. & 21. & 5M\\
\vspace{0.05cm}\\
\hline

\end{tabular}
\caption{Masses, characteristic scale lengths and heights and number of particles, for the different components in Model 1 and Model 2. All masses are in units of 2.3$\times10^9M_\odot$, distances in kpc.}\label{galparamtable}
\end{center}
\end{table}

\begin{figure*}
\begin{center}
\includegraphics[clip=true, trim = 0mm 0mm 0mm 0mm, width=0.9\linewidth]{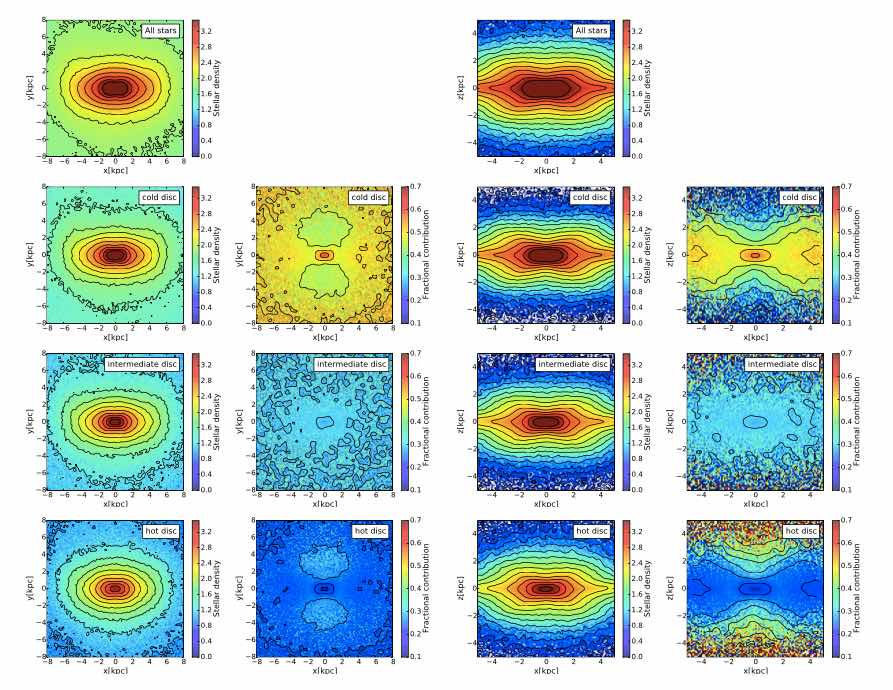}
\end{center}
\caption{\emph{Model 1, morphology of the bar-B/P region at $t=5$~Gyr. First and third column, from top to bottom:} face-on (\emph{col 1}) and edge-on (\emph{col 3}) absolute stellar densities of all stars (\emph{top panel}), and stars initially in the cold (\emph{second panel}), intermediate (\emph{third panel}) and hot (\emph{bottom panel}) discs.  \emph{Second and fourth column, from top to bottom:} Fraction of cold (\emph{second panel}), intermediate (\emph{third  panel}), and hot disc (\emph{bottom panel}) stars in the B/P bulge, seen face-on (\emph{col 2}) and edge-on (\emph{col 4}). The fraction is defined as the ratio of cold, intermediate or hot disc stars to the total  (i.e. cold+intermediate+hot) disc stars in the B/P bulge. In the edge-on maps, only stars in the bar ($|x|\le 5.5$~kpc and $|y|\le 3$~kpc) have been selected.  } 
\label{ModI_xyzmaps}
\end{figure*}

\begin{figure}
\includegraphics[clip=true, trim = 5mm 0mm 5mm 0mm, width=0.48\linewidth]{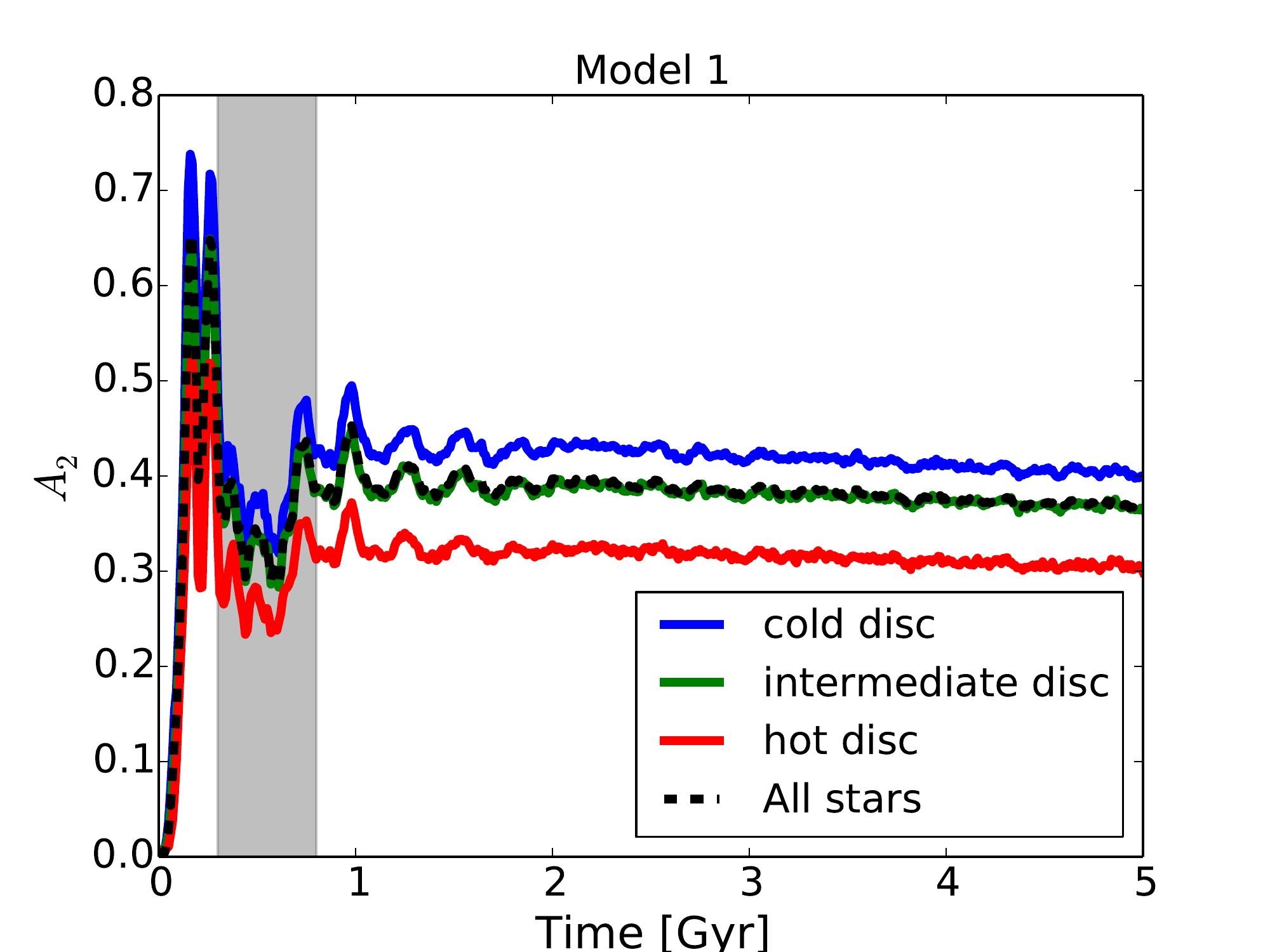}
\includegraphics[clip=true, trim = 5mm 0mm 5mm 0mm, width=0.48\linewidth]{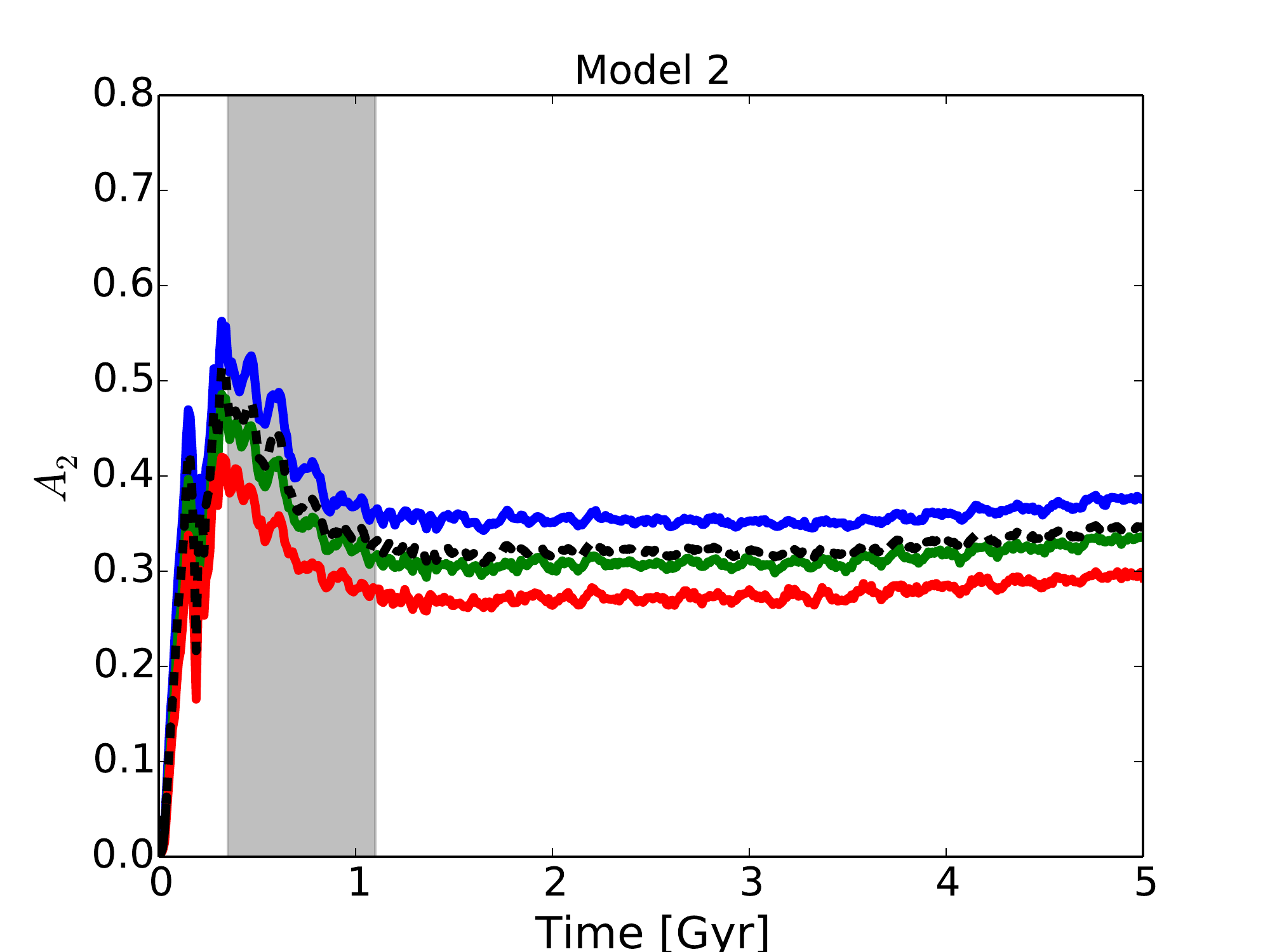}\\
\includegraphics[clip=true, trim = 5mm 0mm 5mm 0mm, width=0.48\linewidth]{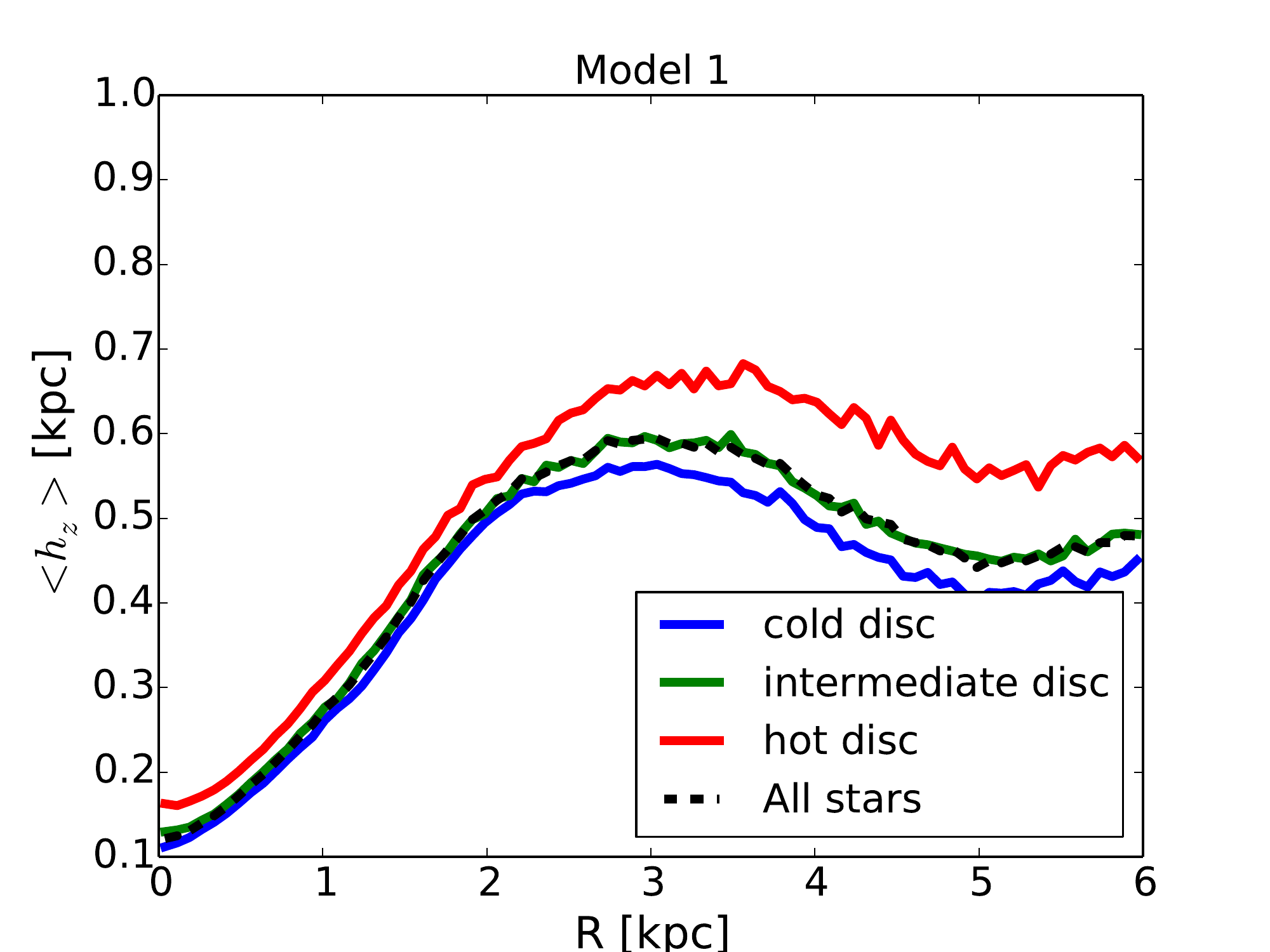}
\includegraphics[clip=true, trim = 5mm 0mm 5mm 0mm, width=0.48\linewidth]{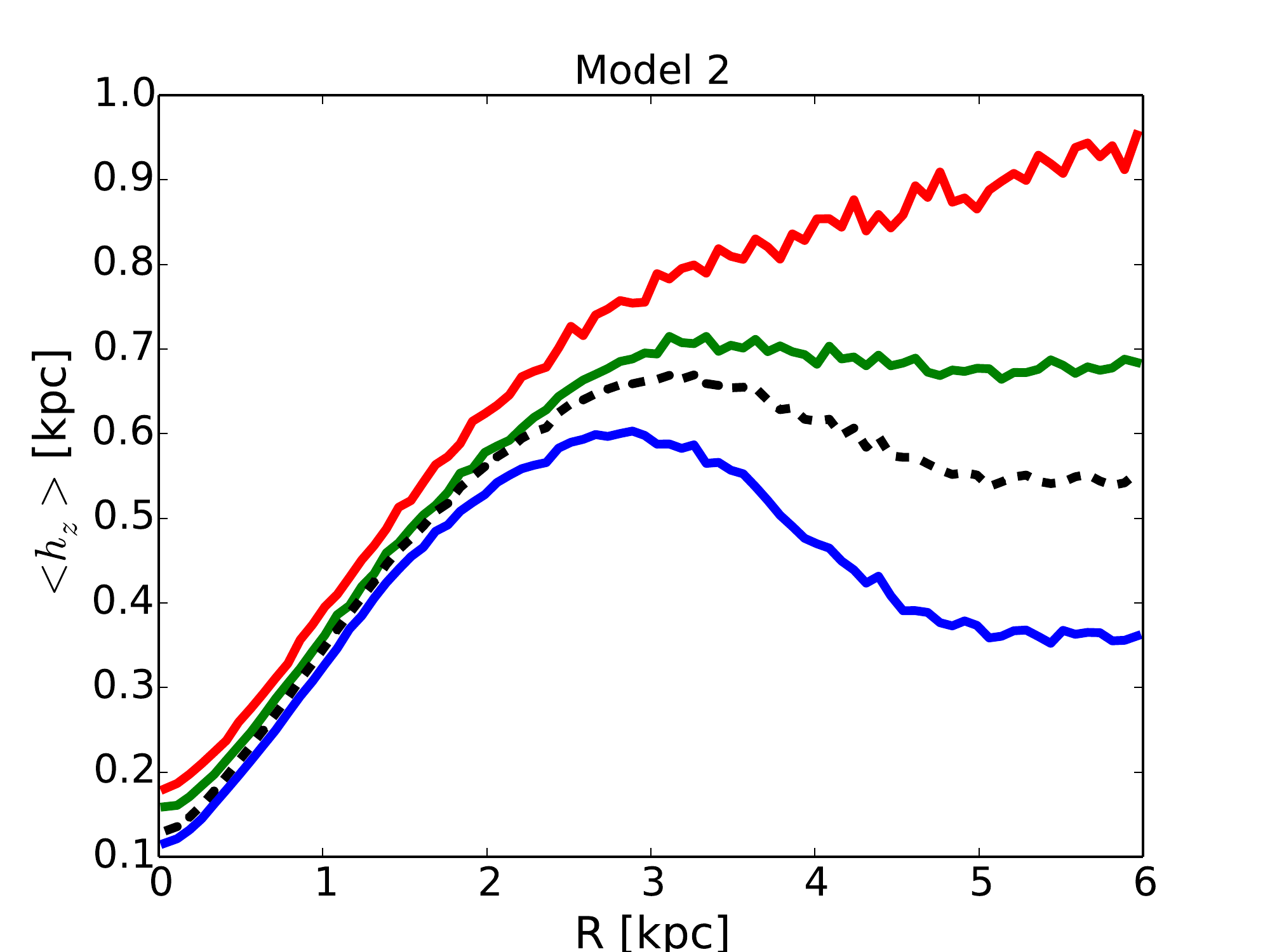}
\caption{\emph{First row: } Bar strength $A_2$  as a function of time, for all stars (dashed black curve), stars initially in the cold (blue curve), intermediate (green curve), and hot (red curve) discs. \emph{Second row: }Height profiles in the B/P region of all stars, and  stars initially in the cold, intermediate, and hot discs. Models 1 and 2 are shown, respectively, on the left and right columns. In the panels on the first row, the grey area indicates the time of disc bending. At the end of this phase of instability, a B/P-shaped bulge is formed in both models, and its morphology does not change significantly after that time. }\label{Mod_bar}
\end{figure}

\begin{figure*}
\begin{center}
\includegraphics[clip=true, trim = 0mm 0mm 0mm 0mm, width=0.9\linewidth]{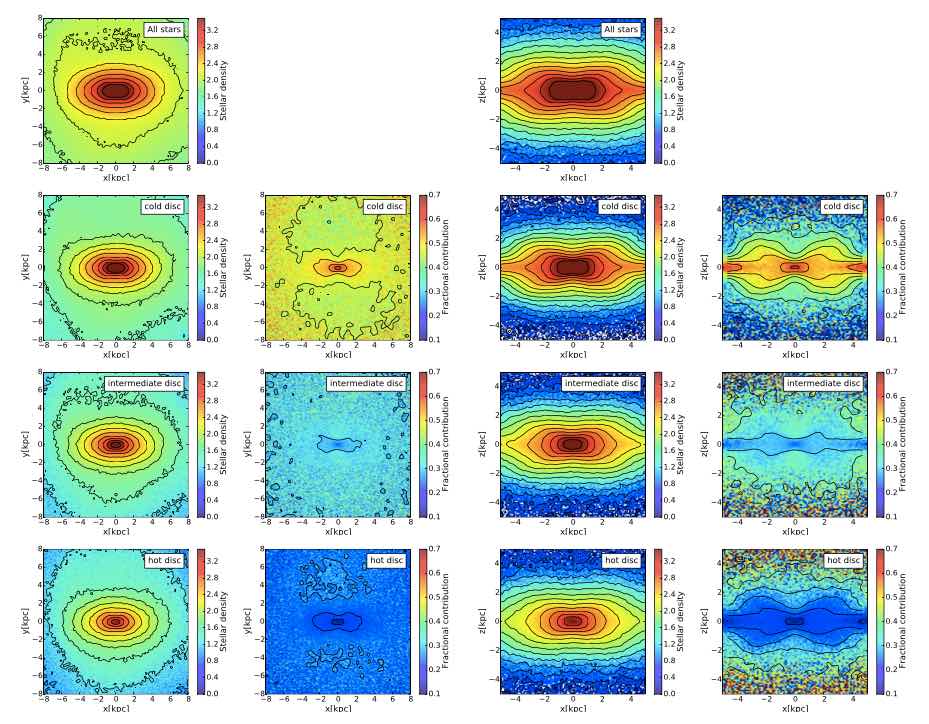}
\end{center}
\caption{Same as Fig.~\ref{ModI_xyzmaps}, but for Model 2.}\label{ModII_xyzmaps}
\end{figure*}

In this paper, we analyze two dissipationless simulations of galaxies made of the superposition of discs of different initial  kinematics. In Model 1, the galaxy is made of three stellar disc components, of equal initial thickness (and hence initial vertical velocity dispersion), but different initial in-plane motions (i.e. different radial and tangential velocity dispersions). In Model 2, the galaxy is also made of three stellar discs, but with equal initial in-plane motions and different initial thickness. These two models do not aim at representing a realistic model for the MW at the time of the formation of the bar, but are intended to constitute two significantly different initial conditions, needed to explore the drivers of the main characteristics observed in the MW bulge. In particular Model 1 has a similar set-up to the models analyzed in \citet{debattista16} (i.e. equal thickness, different in-plane motions), thus allowing a comparison with this work. \\In each model, the three discs are represented by Miyamoto-Nagai density profiles \citep{miyamoto75}, with characteristic heights  and scale lengths as given in Table~\ref{galparamtable}. In each of the two models, the cold, intermediate and hot discs have masses which represent, respectively, 50\%, 30\% and 20\% of the total disc mass. In this context, the choice of the mass of each of these discs is arbitrary, and the conclusions of this work do not depend on it. However, we use these percentanges because they are reminiscent of those of the thin, young, and old thick discs respectively (for a derivation of the Galactic mass growth and relation with the thin and thick discs we refer the reader to Snaith et al 2014, 2015). To allow all comparisons with the models that will be presented in forthcoming papers, we have thus decided to keep the same mass repartition between the three discs as that used in those papers. \\
These composite discs are then embedded in a dark matter halo, modelled as a Plummer sphere \citep{plummer11}, and whose parameters are reported in Table~\ref{galparamtable}. The resulting rotation curves are shown in Fig.~\ref{IC}. Face-on and edge-on maps of stars in the bulge region at the initial time are given in Appendix~\ref{initparam-app}, where we also discuss further details about the adopted parameters and their impact on the final bulge morphology. \\
To generate initial conditions, we employed the iterative method described in \citet{rodionov09}. This method allows to generate initial conditions at equilibrium with the required density profiles and/or kinematics constraint, avoiding the relaxation processes -- and departure from initial conditions -- often observed in N-body models of disc galaxies. It is particularly useful for building composite discs, each with a specific kinematic and/or density profile, allowing a full control on their initial state. For the models presented here, the initial velocity dispersion profiles of the cold, intermediate and hot populations are reported in Fig.~\ref{IC}. In Model 1 the initial thickness of the three discs is the same, and thus the corresponding vertical velocity dispersion profile; the in-plane motions (radial and tangential) are however different. On the contrary, in Model 2, the vertical velocity dispersions are different, while the in-plane motions are initially the same. We investigate in the following how these two models evolve secularly, and how these (initially different) composite discs are mapped into the B/P bulge. We refer the reader to Appendix~\ref{rodionov} for a discussion on the velocity dispersion structure that
  is imposed for the initial discs and on the final convergence of the method. \\
Both simulations have been run with a recently developed parallel MPI Tree-code which takes into account the adaptive spatial decomposition
of particle space between nodes. The multi-node Treecode is based on the 256-bit AVX instructions which significantly speed up the floating point vector operations and sorting algorithms (Khoperskov et al. in prep).  Fifteen million particles have been employed for each model, 10 millions in the disc components, and 5 millions in the dark matter halo.  A  time step $\Delta t=2\times 10^5 $~yr has been adopted, together with a gravitational smoothing length of 50 pc. Both  simulations have been run over a time scale of 5~Gyr. In less than 1~Gyr from the  beginning of the simulations, a bar and a B/P bulge forms in both models. We  then analyze the final configuration, at $t=5$~Gyr, to quantify the properties of the B/P bulge and surrounding disc. 

\begin{figure*}
\begin{center}
\includegraphics[clip=true, trim = 0mm 0mm 0mm 0mm, width=0.5\linewidth]{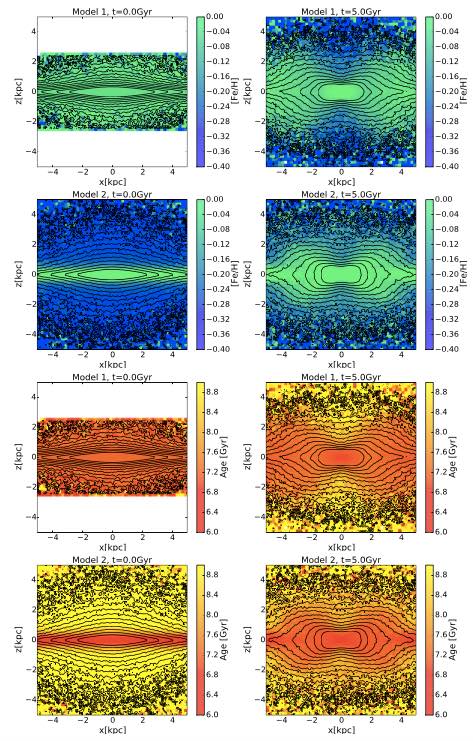}
\end{center}
\caption{Edge-on metallicity and age maps for stars in Model 1 (\emph{first and third row}) and for Model 2 (\emph{second and fourth row}), at time $t=0$ (\emph{first column}) and at time $t=5$~Gyr (\emph{second column}). Only stars in the bar ($|x|\le 5.5$~kpc and $|y|\le 3$~kpc) have been selected. In all these maps, when present the bar is oriented side-on. Isodensity contours are shown in black. }
\label{stellarpop}
\end{figure*}

\begin{figure}
\begin{center}
\includegraphics[clip=true, trim = 0mm 0mm 0mm 0mm, width=0.9\linewidth]{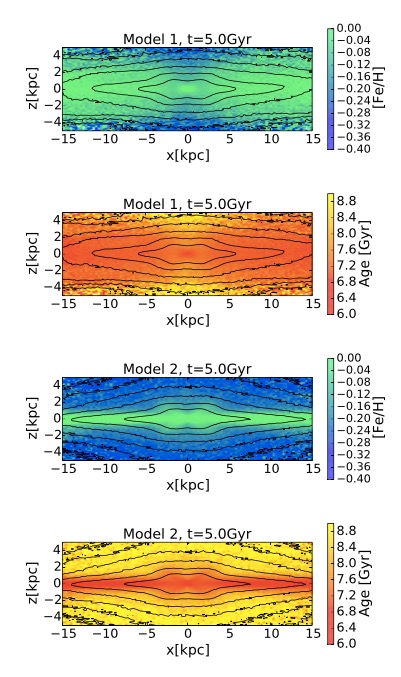}
\end{center}
\caption{Edge-on metallicity and age maps at the final time of the simulation for  Model 1 (respectively, first and second row), and Model 2  (third and fourth row). Both the B/P region and the surrounding disc are shown in the maps. Isodensity contours are shown in black.}
\label{metageBIG}
\end{figure}

\section{Results}\label{results}
\subsection{Multiple populations with different in-plane or vertical kinematics:  relative contribution to the B/P bulge,  metallicity and age trends}\label{bp}

The face-on and edge-on projection of stars in the bulge region for Model 1 are shown in Fig.~\ref{ModI_xyzmaps}. We recall that in this model, the simulated galaxy is made of three discs, with different in-plane kinematics (the radial velocity dispersion increases moving from the cold disc to the hot disc, see Fig.~\ref{IC}), but same vertical velocity dispersions. This set-up is thus similar to those analyzed in \citet{debattista16}, and indeed the morphology of the bulge in Fig.~\ref{ModI_xyzmaps} agrees qualitatively with the idealized model presented in their paper\footnote{Note however that the two models remain different in the details, thus illustrating the possible variety of models and morphologies present in each class.}. \\Our analysis for this model confirms the findings of \citet{debattista16} \citep[see also][for earlier works on the strength of bars in single component discs, and their dependence on the initial disc kinematics]{combes90, athanassoula13}. Stars with different in-plane kinematics are mapped differently in the B/P bulge. When viewed face-on, the colder the initial disc, the stronger the bar. This can be appreciated by looking at the isodensity contours that are more elongated for the initially cold disc, but is also quantified by means of the $m=2$ coefficient, $A_2$, of the Fourier analysis of the face-on density maps  (Fig.~\ref{Mod_bar}, left column) which shows that the colder the disc is initially, the higher the corresponding  $A_2$ value.  In the edge-on projection, stars with initially the coldest kinematics show the thinnest distribution, as we have verified by estimating  $<h_z>$, the median of the absolute value of the $z$-component of their positions, as a function of the distance $R$ from the galaxy centre (Fig.~\ref{ModI_xyzmaps}). Because of this different response to the bar perturbation, stars belonging initially to discs with different in-plane kinematics contribute differentially to the B/P bulge, both in its face-on and edge-on projection. In agreement with \citet{debattista16}, the colder the disc, the higher its relative contribution to the B/P structure is. Stars in the initially kinematically hottest disc dominate at larger heights above the disc. As a consequence their relative contribution along the B/P minor axis increases with height above the plane (see Fig.~\ref{ModI_xyzmaps}, fourth column. This trend -- namely the  larger thickness of the initially in-plane hot disc with respect to the colder ones -- comes from the fact that in the inner regions of the simulated disc the azimuthal and vertical frequencies, $\Omega$ and $\nu$, respectively, are similar (see Appendix~\ref{freq-app}), and thus in-plane and vertical motions are not decoupled, as usually assumed in the case of very thin, or infinitely thin discs \citep[see][]{binney87}. As a result, a disc which has initially only a  hot in-plane kinematics will rapidly relax into a disc also vertically hot, because of this coupling.  \\

The analysis of Model 2, interestingly reveals that all the trends observed for Model 1 are satisfied also in this case (Fig.~\ref{ModII_xyzmaps}). When viewed face-on, the lower the initial vertical velocity dispersion of the disc, the stronger is the bar -- as measured by means of the $A_2$ coefficient, as a function of time (Fig.~\ref{Mod_bar}, right column). When viewed edge-on, the lower the initial vertical velocity dispersion of the disc, the thinner the B/P structure. Note however the difference in the thickness of cold and hot populations in the two cases: in Model 2, outside the B/P region, the hot population has a higher thickness than that measured for the cold population, while in Model 1 the difference is milder. For example,  at $R=6$~kpc the hot population has a thickness about three times higher than that measured for the cold population, while in Model 1 the hot population is only 1.3 times thicker than the cold one, at the same radius.
Finally, because the three discs in Model 2 have, by construction, different initial thickness, their relative contribution to the B/P bulge depends on their initial kinematics: in particular, the weight of stars with initially  the hottest kinematics increases with the height above the plane. \\

The similarity in the trends of cold and hot populations with height above the plane, found for the two models, has also some important implications in the mean metallicity and age of stars across the B/P bulge, as we investigate in the following. 
To this aim, we assign to stars in each disc a metallicity and age, as follows: the cold, intermediate and hot discs have gaussian metallicity distributions with means respectively at $\rm [Fe/H]=0.3,~-0.2~and~-0.6$~dex, and dispersions $\rm \sigma_{[Fe/H]}=0.1, 0.3~and~0.3$; for the ages, we assume for the cold, intermediate and hot discs uniform distributions in the intervals $\rm [0., 8.],~[8., 10.]~and~[10., 13.]$~Gyr, respectively. The metallicity mean values and dispersions are similar to those found by \citet{ness13spop} for components A, B and C in the Galactic bulge. Since, as recalled in the introduction, in our scenario we associate these three bulge components\footnote{By "components" here we mean simply stars associated to different metallicity intervals, as defined by \citet{ness13spop} for  A, B and C. } to  the Galactic disc(s)  -- respectively the thin, young thick and old thick discs, which, as robustly established, show a decreasing metallicity with increasing (in-plane and vertical) velocity dispersions -- the choice to assign these metallicity values to our three discs is natural. We emphasize however that other choices would have been possible, and that the trends discussed in the following are not impacted by the specific metallicity values employed, as far as the three modeled discs have different chemistry and ages (i.e.  a trend between metallicity/age and velocity dispersion must exist). Since age distributions in the Galactic bulge -- as well as in most of the Galactic disc - are still missing, we decide to assign them in the simplest way, simply relating our modeled discs to the thin, young and old thick discs, respectively, and making use of the age estimates of (thin and thick) disc stars at solar vicinity, as given by \citet{haywood13}. 
With these choices of metallicities and ages, half of the disc stellar mass has metallicities below solar and ages below 8~Gyr, in agreement with the results presented in \citet{snaith14, haywood16a} for the inner Galactic disc (inside 6--7~kpc from the Galactic centre).  \\
Before the formation of the bar and of the B/P structure, the edge-on metallicity and age maps of Model 1 and Model 2  are shown in Fig.~\ref{stellarpop}. Because in Model 1 all discs have initially the same thickness, no trend is visible when the modeled galaxy is seen edge-on, and the vertical gradient along the bar major axis is null. For Model 2, however, because the discs have different initial vertical velocity dispersions, a vertical gradient is in place \emph{ab initio} both in the metallicity and age maps, with mean metallicities(/ages) decreasing(/increasing) with height above the plane. As an example, for Model 2 along the bulge minor axis the initial metallicity gradient is -0.20~dex/kpc, and the age gradient is 1.72 Gyr/kpc\footnote{Both gradients have been estimated comparing the values of metallicity (/age) at $z=0$ and $z=2$~kpc from the galaxy plane.}. We note however that in Model 2 the metallicity (/age) gradient is mostly due the presence of very metal-rich (/young) stars close to the galaxy midplane (at heights below 500~pc), while for larger heights the gradient is nearly flat also in this case. 
Once the B/P bulge is formed, the metallicity and age maps of Models 1 and 2 appear rather different from their initial state. In both models, X-shaped metallicity and ages maps are clearly visible \citep[as found also in the models by][]{athanassoula17, debattista16}, with metal-rich (young stars) associated to the initially cold disc mostly redistributed in the peanut configuration. Along the bulge minor axis, as one moves vertically from the plane, the metallicity and age maps become rapidly dominated by the metal-poor, old intermediate and hot discs, giving rise to a vertical negative (/positive) metallicity (/age) gradient. It is striking to note that not only are the global metallicity and age trends similar in the two models  (Fig.~\ref{stellarpop}), but the absolute values and strengths of the gradients as well. Except for the outermost bulge region outside the B/P lobes, that we will discuss in the following section, at all $x \lesssim |4|$~kpc the vertical metallicity and age profiles appear very similar, with comparable slopes. As an example, along the bulge minor axis, between $z=0$  and $z=2$~kpc from the galaxy midplane, the metallicity gradient  is equal to $-0.089$~dex/kpc in Model 1 and to $-0.091$~dex/kpc in Model 2. The age gradients are respectively $0.721$~Gyr/kpc for Model 1 and $0.782$~Gyr/kpc for Model 2. The age and metallicity maps generated in these two models are thus remarkably similar, despite the significantly different initial conditions adopted, and this is valid for all choices of the initial metallicity and age distributions in the disc populations, since here metallicity and age are simply tags of the stellar particles, that trace their spatial distribution. \\
As a final remark, we emphasize that both models produces B/P bulges whose metallicity/age maps  appear more pinched/peanut-shaped than the stellar density distribution itself (cf. metallicity/age maps with isodensity contours in Fig.~\ref{stellarpop}). This feature has been observed in external bulges \citep{gonzalez17}, and reported also in N-body simulations \citep{debattista16, aumer17}, and here we show that it is not necessarily the  signature of a pre-existing disc with stellar populations differentiated in their in-plane random motions, as previously suggested. 
 
From this analysis we thus conclude that \emph{initial vertical random motions are as important as in-plane random motions in determining the overall bulge population, metallicity and age trends}, and that previous statements emphasizing the dominant role of in-plane motions in determining these trends are not confirmed by our analysis. 

Finally, we would like to shortly comment on the metallicity/age trends that these two models generate outside the B/P bulge, in the surrounding disc. \\
In Fig.~\ref{metageBIG}, we show the large-scale -- i.e. over the whole extent of the galaxy -- edge-on metallicity and age  maps, for both models.  In Model 1, as shown in Appendix~\ref{discs-app},  during the whole evolution, outside the bulge region, the populations are never differentiated in their vertical velocity dispersion, and thus scale heights. This implies that the vertical  metallicity and age profiles of the disc do not exhibit any vertical gradient, as result of combining different populations. The consequence of the absence of any vertical differentiation in kinematics present in Model 1 is that  the mean metallicity and age of the disc do not vary with $z$: this feature is already evident at the borders of the B/P bulge, and persists all over the disc. So, while in the B/P bulge, the two models predict similar trends of metallicity/age with height above the plane, significant differences are expected outside the B/P region, and in the surrounding disc. In particular, the absence of any dependence of metallicity or age with height above the plane -- as observed in Model~1 -- is in sharp contrast to what we know about the MW disc : both at the solar vicinity and on a several--kpc scale metallicity and ages vary with height above the plane, and with the vertical velocity dispersion of the corresponding populations \citep{stromberg46, spitzer51, nordstrom04, seabroke07, holmberg07, holmberg09, bovy12c, haywood13, sharma14, bovy16, martig16, ness16, mackereth17}. To generate vertical metallicity/age gradients in the disc and in the region outside the lobes of the B/P bulge, it is necessary that populations of different metallicities/ages are differentiated in their vertical velocity dispersion \emph{ab initio}, before being trapped in the bar instability, otherwise only by secular evolution processes this relation cannot be put in place afterwards.

\begin{figure}
\includegraphics[clip=true, trim = 0mm 0mm 0mm 0mm, width=0.48\linewidth]{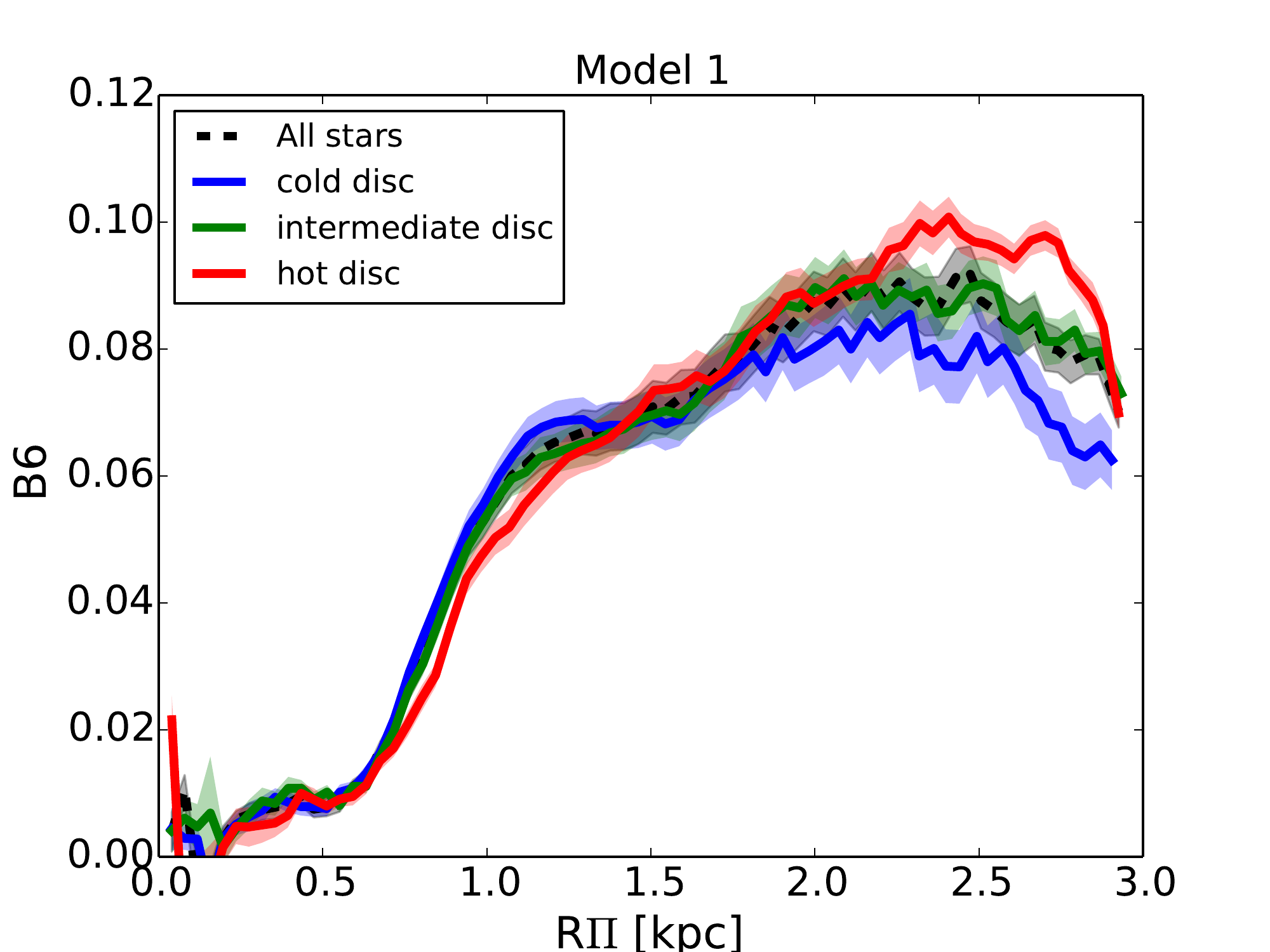}
\includegraphics[clip=true, trim = 0mm 0mm 0mm 0mm, width=0.48\linewidth]{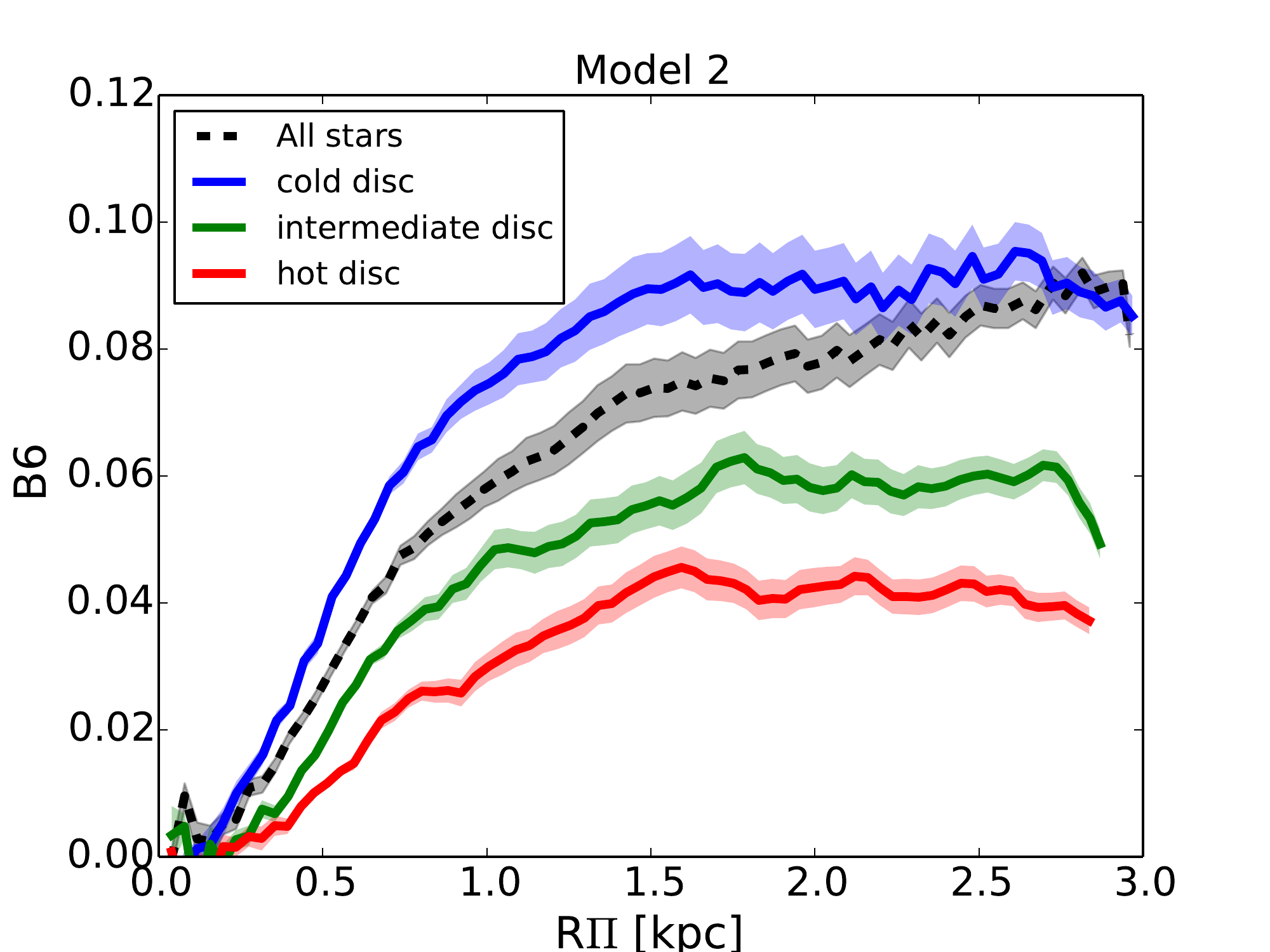}\\
\includegraphics[clip=true, trim = 0mm 0mm 0mm 0mm, width=0.48\linewidth]{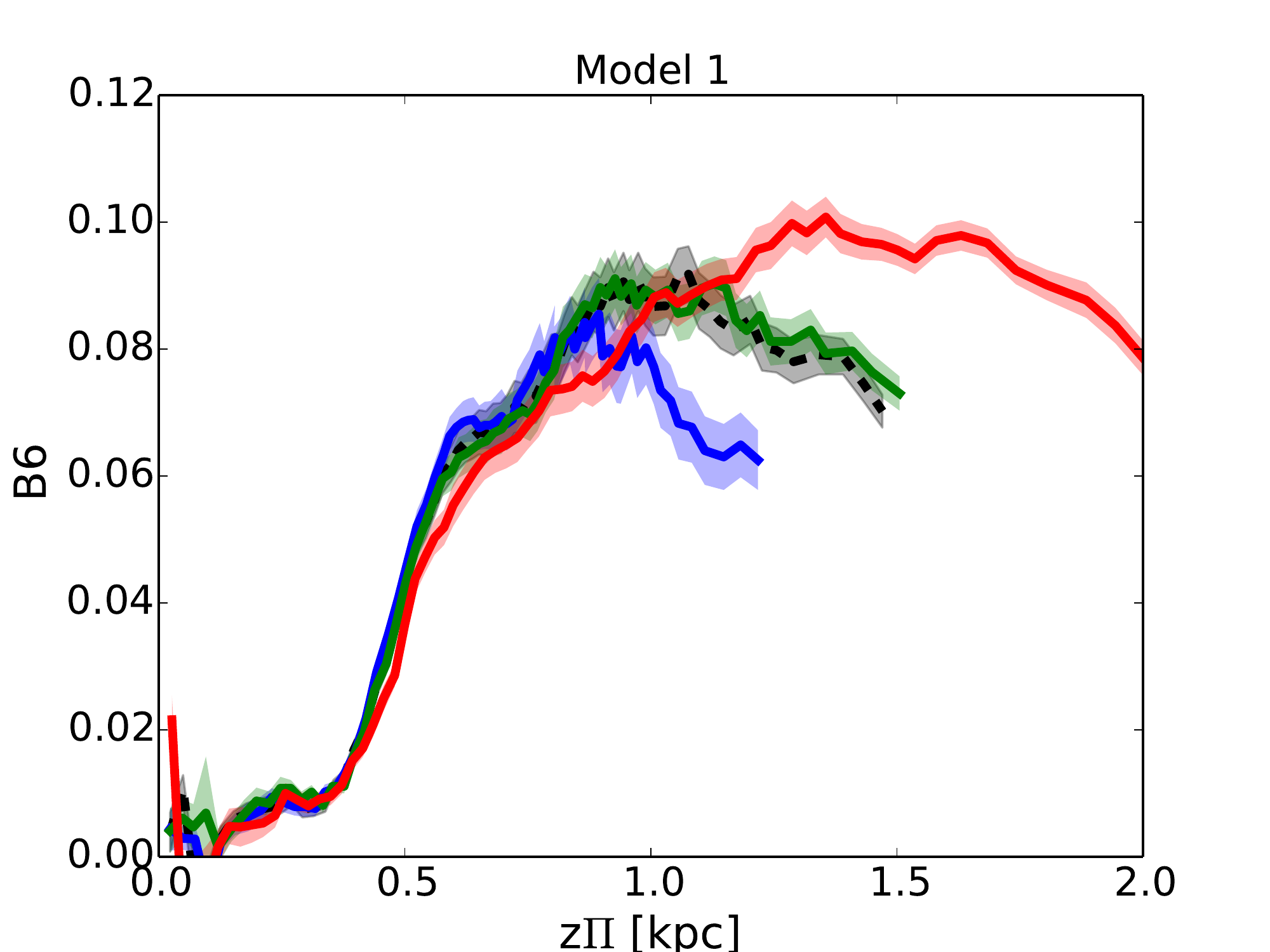}
\includegraphics[clip=true, trim = 0mm 0mm 0mm 0mm, width=0.48\linewidth]{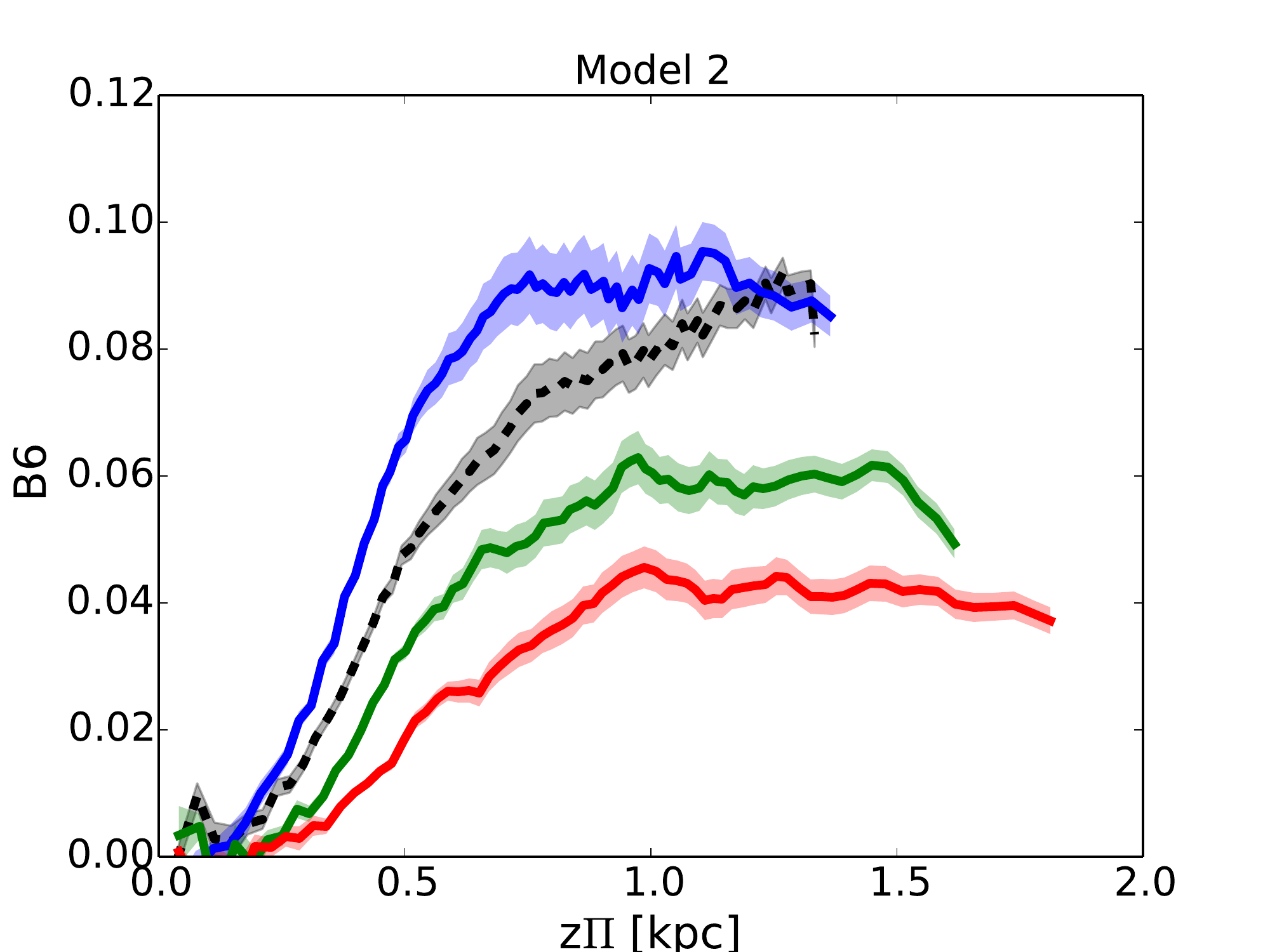}
\caption{\emph{First row:} B/P strength, quantified by means of the $B_6$ parameter, as a function of the projected radial length of the peanut, $R_\Pi$, for Model 1 (left panel) and Model  2 (right panel). \emph{Second row:} B/P strength as a function of the projected height of the peanut, $z_\Pi$, for Model 1 (left panel) and Model  2 (right panel). In all plots, the strength of the peanut is shown at $t=5$~Gyr for all stars (black curves), and stars initially belonging to the cold  (blues curves), intermediate  (green curves) and hot  (red curves) discs. In all panels, error bars are represented by shaded areas and indicate the intensity rms scatter along each isophote after the subtraction of the best-fitting pure ellipse with harmonic terms up to, and including, the $6$th order.  }\label{Bsix}
\end{figure}

\begin{figure}
\begin{center}
\includegraphics[clip=true, trim = 5mm 0mm 20mm 0mm, width=0.6\linewidth]{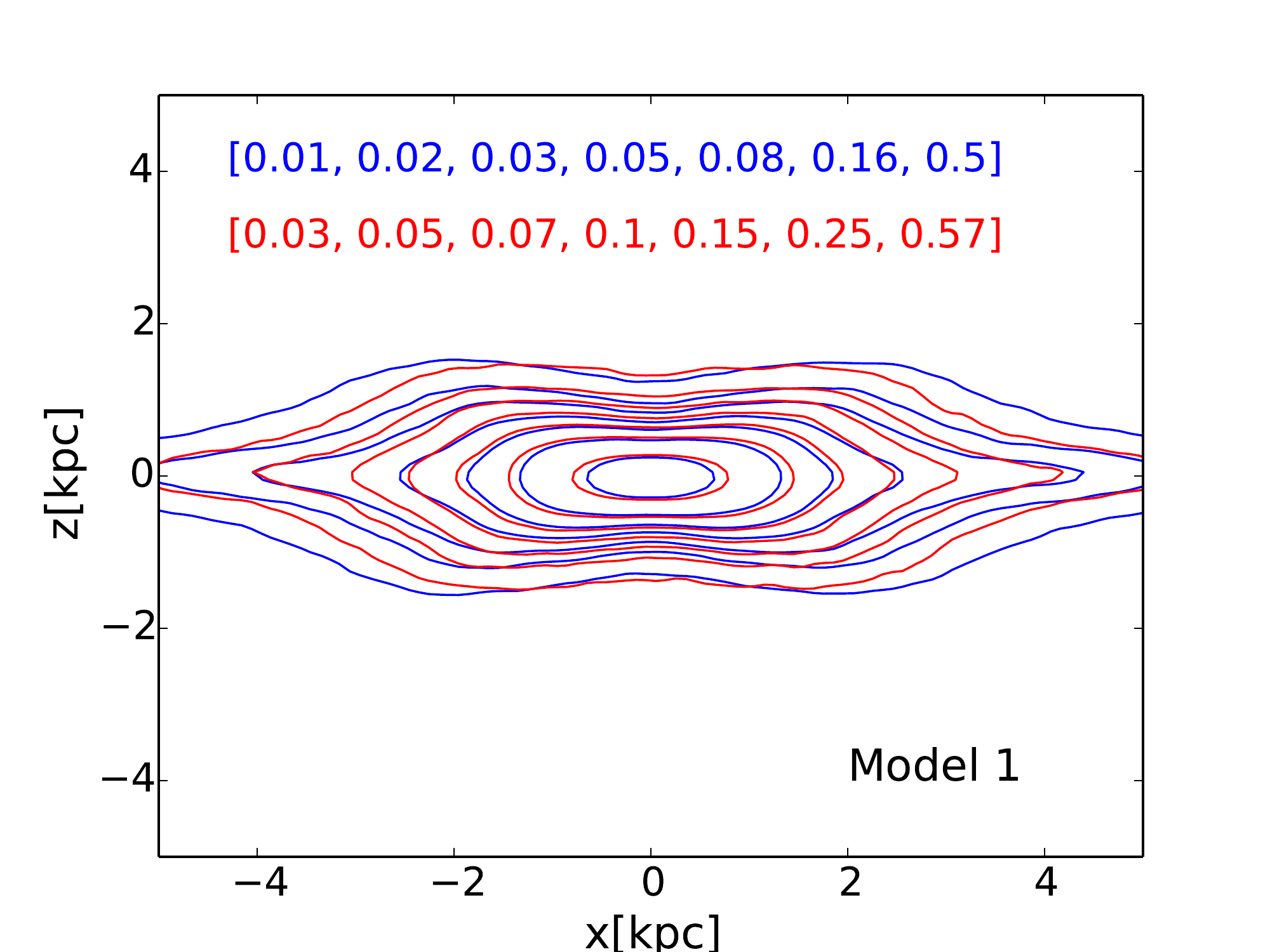}
\includegraphics[clip=true, trim = 5mm 0mm 20mm 0mm, width=0.6\linewidth]{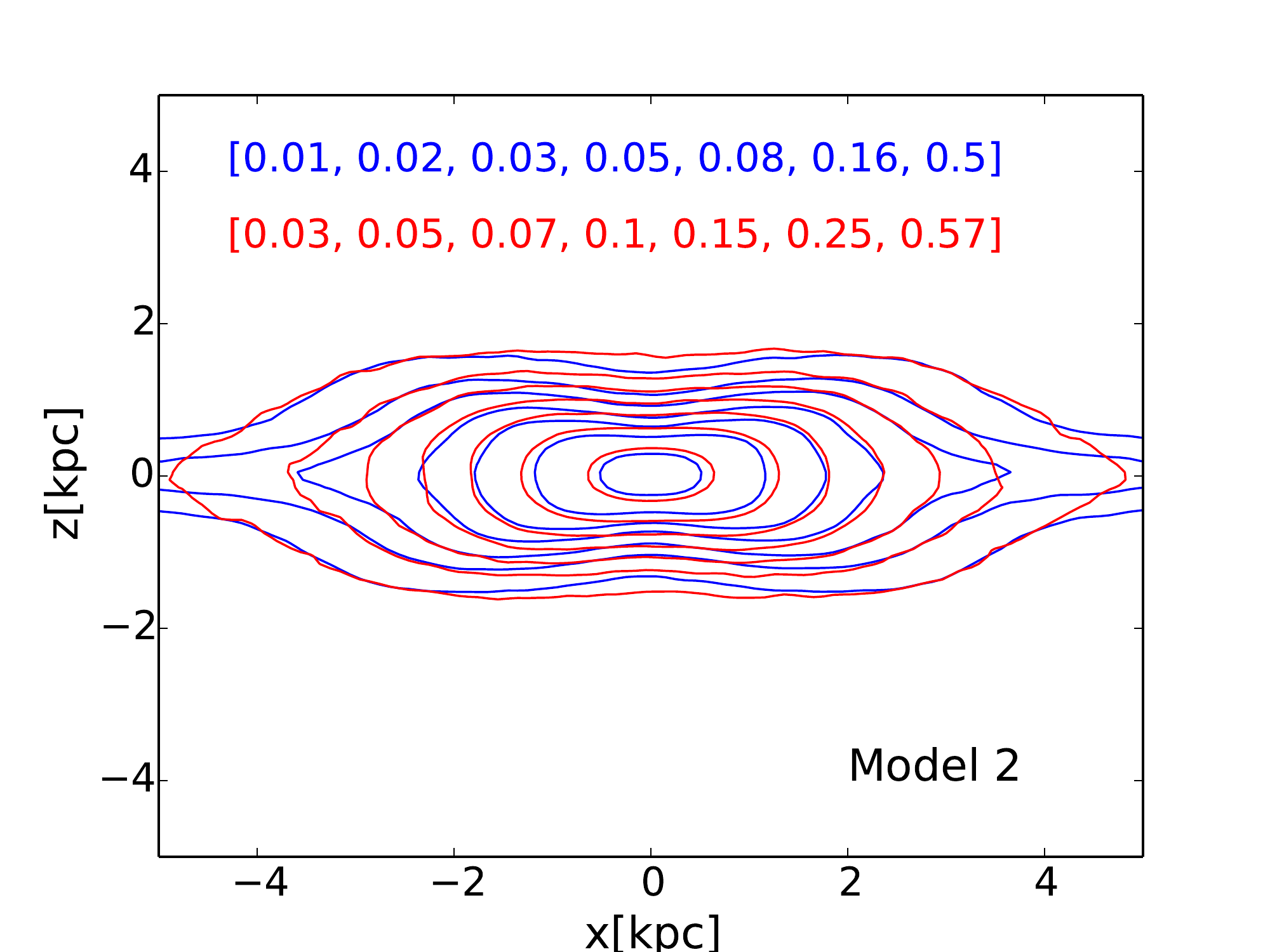}
\caption{Isodensity contours of the B/P bulge in Model~1 (\emph{top panel}) and Model~2 (\emph{bottom panel}), for the initially kinematically cold  (\emph{blue contours})  and hot  (\emph{red contours}) disc populations, at $t=5$~Gyr.  The contours of the intermediates disc are not shown,  but they are bracketed by those of the cold and hot disc populations. The numbers on the top left of each panel indicate the values of the isodensity contours, for cold and hot populations, normalized to their central density. Only stellar particles in the bar region ($|x| \le 5.5$~kpc and $|y| \le 3$~kpc) have been selected for these plots. }\label{normiso}
\end{center}
\end{figure}

\begin{figure*}
\includegraphics[clip=true, trim = 5mm 0mm 5mm 0mm, width=0.5\linewidth]{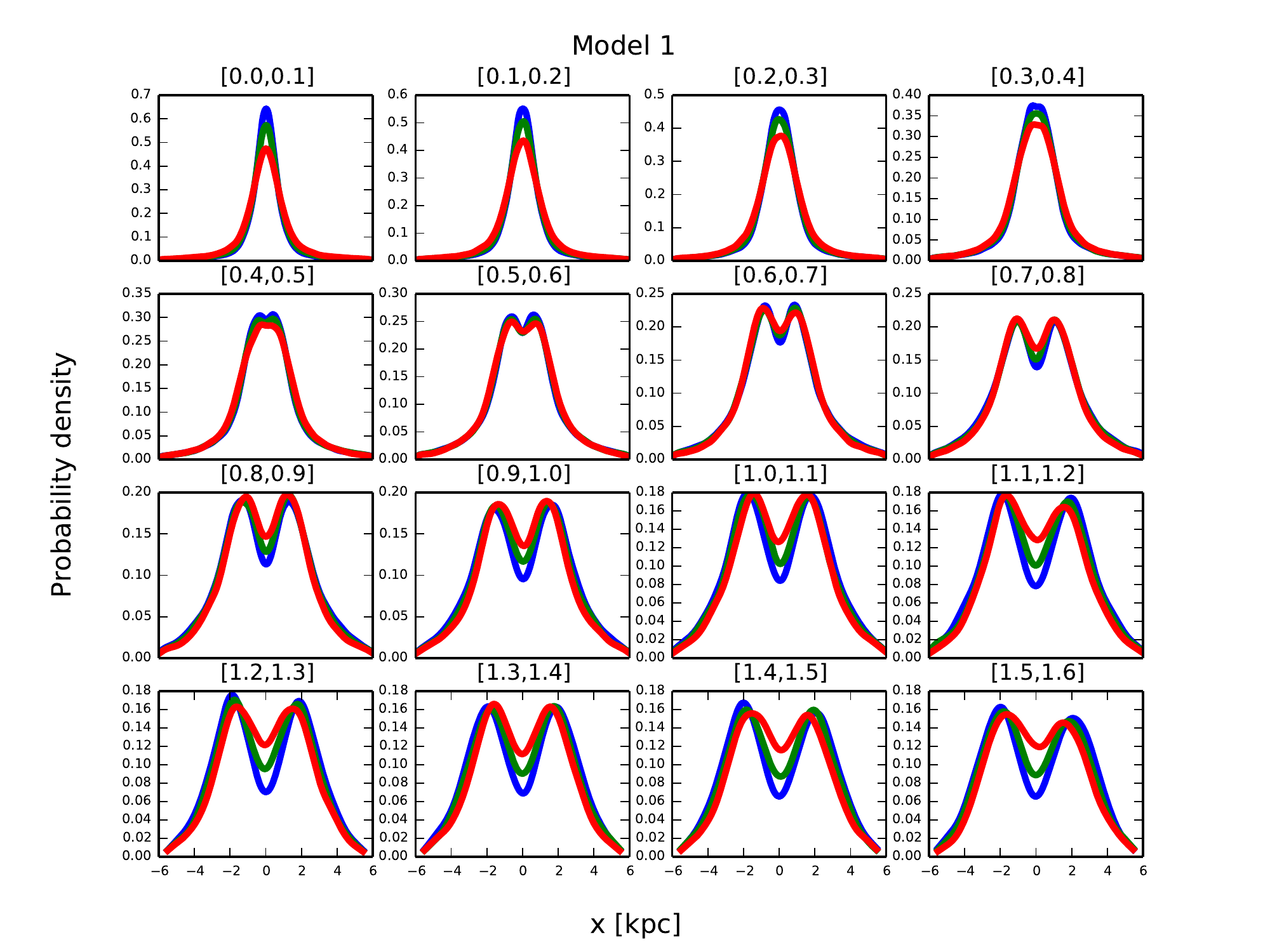}
\includegraphics[clip=true, trim = 5mm 0mm 5mm 0mm, width=0.5\linewidth]{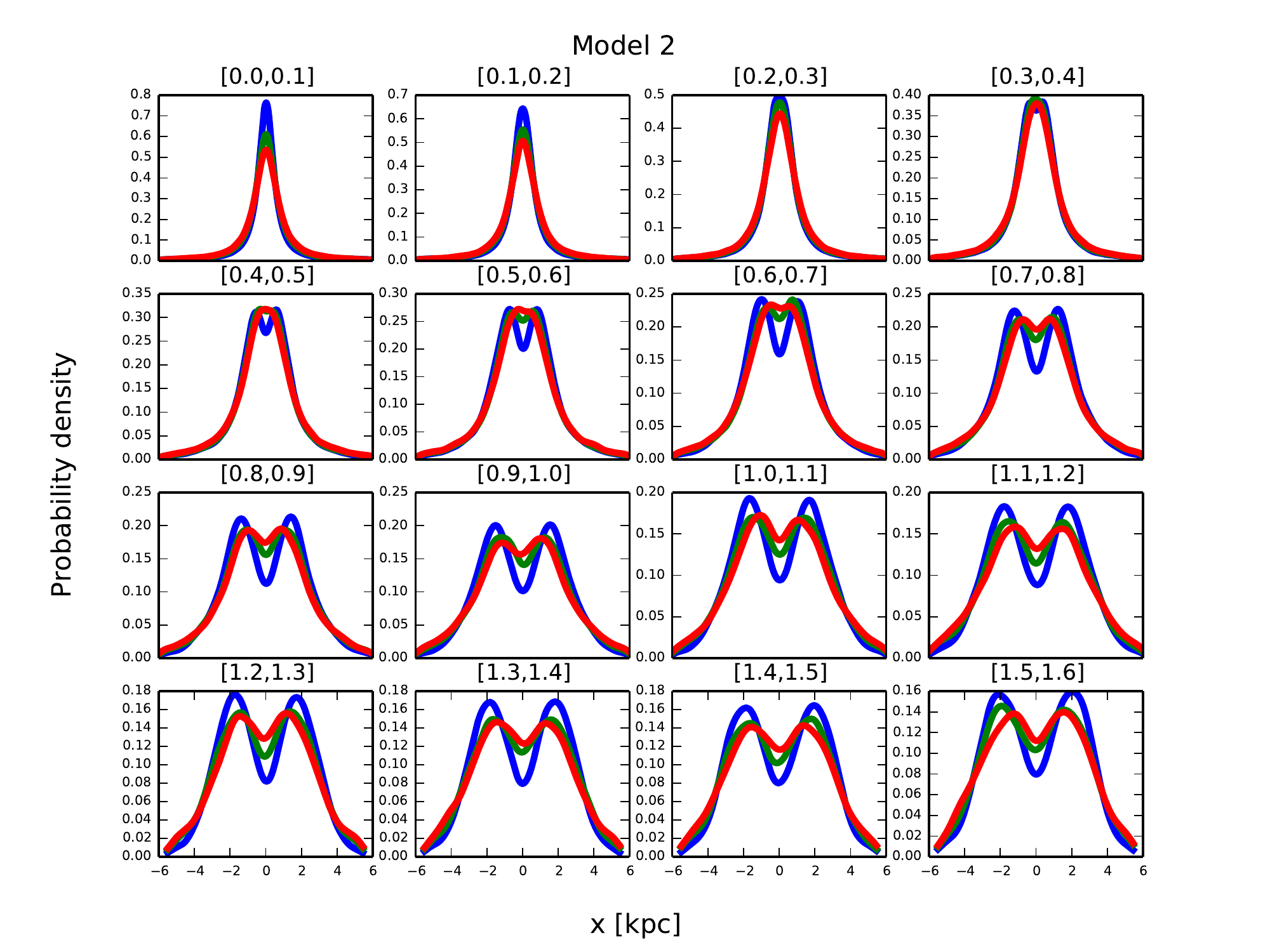}
\caption{Distribution of stars along the bar major axis at different heights above the plane, as indicated on the top of each panel.  The distribution of cold, intermediate and hot stars in the B/P structure are indicated, respectively, with blue, green and red histograms. The distribution appears bimodal for all populations at the same height above the plane in Model 1 (\emph{left panel}) , while in Model 2 (\emph{right panel})  the bimodality appears at larger heights for the hot population than for the colder ones.  }\label{zcuts}
\end{figure*}

\subsection{Multiple populations with different in-plane or vertical kinematics: on the shape and strength of the B/P bulge}\label{strength}

Despite that  Figs.~\ref{ModI_xyzmaps} and  \ref{ModII_xyzmaps} show  similar trends with height from the plane,  we notice that the initial disc kinematics imprints some significant differences in the intrinsic morphology of the final B/P bulges.\\
A first way to quantify these differences is shown in Fig.~\ref{Bsix}, where we estimate the strength of the B/P structure by means of the sixth Fourier component ($B_6$) of edge-on density isophotes 
in the bulge region \citep[see][for further details\footnote{The code is publicly available, see  https://github.com/BogdanCiambur/ISOFIT}]{ciambur15, ciambur16}. What we observe is that, while in Model 2, the amplitude of the  $B_6$ coefficient depends on the initial kinematics of the population -- the hotter the kinematics, the lower  $B_6$, and this over the whole bulge extent -- in Model  1, the  $B_6$ amplitude appears the same for all populations, independently on their initial kinematics -- both as a function of the projected radial length of the peanut, $R_\Pi$, and of its projected height, $z_\Pi$ \citep[for a definition of $R_\Pi$ and $z_\Pi$, see Fig.~1 in ][]{ciambur16}. In the regions where all three populations coexist in the bulge  (about $z_\Pi \le$1~kpc), the trends of $B_6$ with  $R_\Pi$ and $z_\Pi$ indeed are indistinguishable for Model 1. \\
A second way to appreciate the different B/P shapes in the cold and hot populations of the two models is illustrated in Fig.~\ref{normiso}, where we show  isodensity contour maps of the B/P bulge in Models~1 and 2. For both models and for all populations, we have normalized the isodensity contours to their central density value. The normalized density levels shown by the contours have been chosen in order to probe similar regions of the B/P bulge for hot and cold populations. The difference between the two models appears very clearly: while in Model~1 the isodensity shapes  inside the B/P lobes, i.e. $x\le 2$~kpc, are the same at all heights, for both cold and hot populations, in Model~2 the shape of the isodensity contours depends on the initial kinematics of the population, the hotter it is initially, the rounder and less peanuty the contour is, at all $z$.\footnote{We refer the reader to Appendix~\ref{initparam-app}, for a more extensive discussion on the populations dominating the innermost bulge regions of  Model~2.}
Fig.~\ref{normiso} thus shows that in Model~1 it is the thickness of the B/P structure that moderately changes with the initial kinematics of the population, but not its intrinsic shape.
The trends observed for Model~1 in Fig.~\ref{ModI_xyzmaps} -- namely the decreasing contribution of the cold disc population with height above the plane, when compared to the hot population -- and in Fig.~\ref{stellarpop} -- the X-shaped metallicity/age maps -- are rather due to final different vertical density gradients of the cold and hot populations, and thus to different thicknesses,  rather  than to an intrinsic difference in their isodensity shape. This result is interesting, because it shows that X-shaped metallicity maps are not only the result of the spatial superposition of populations with different metallicities and different intrinsic peanut/X-shapes, but can also be produced when in a B/P bulge multiple populations with different metallicities, but similar X-shapes, co-exist. In this latter case, the X-shape metallicity and age maps result simply because of a different vertical density decay of these populations.\\ 
Another consequence of the similarity in the isodensity contours found for Model~1 is that the peanut appears at the same height above the plane, at about 600~pc, for all populations, independently on their initial in-plane kinematics (see also Fig.~\ref{zcuts}). This  trend seems at odds with what is found in the MW bulge, where the strength and appearance of the B/P shape depends on the metallicity of bulge stars \citep{ness13spop, rojas14}. According to our experiment, this observational result favors a scenario where metal-rich and metal-poor populations in the MW bulge initially  had different initial vertical random motions, i.e. before being trapped by the B/P bar. If the metal-poor and metal-rich populations were different only in their in-plane random motions, but not in the vertical ones,  the bimodality in the stellar density distribution would show a weak or even null dependence on the height above the plane, as we find for Model~1. \\

But why is the B/P shape of hot and cold populations so similar in a disc whose populations are only differentiated by their in-plane motions, as in Model~1, and how general this result is? While it is premature to generalize, and a larger statistics of Model~1-like simulations is required before deriving final conclusions, we can give here some first explanations of the observed similarities in the B/P morphology of cold and hot populations in Model~1. In this model, all populations have initially the same scale lengths, and thus similar guiding radii, and they differ only in their initial radial velocity dispersions. In the epicyclic approximation, orbits in an axisymmetric potential, with different radial velocity dispersions $\sigma_R$ but same guiding radius $R_g$, have same azimuthal frequency $\Omega$, where $\Omega=V_c/R_g$,  $V_c$  being the circular velocity at $R=R_g$. Also when the epicyclic approximation is not valid,  because the departure from the orbit circularity is significant, still kinematically hot and cold populations, with similar guiding radii, have similar azimuthal frequencies (with differences typically less than 10\%), as we show in Fig.~\ref{freqhisto} in Appendix~\ref{freq-app}.  The distribution of $\Omega$ being similar for hot and cold populations, and because in Model~1 the distribution of vertical frequencies $\nu$ is the same by construction -- all populations have the same thickness -- this implies that the fraction of stars that satisfies the inequality $\nu \ge 2(\Omega-\Omega_p)$ \citep{merritt94}, that is that can respond to the bending/buckling instability which initiates the B/P formation, is similar in all populations,  for any given value of the bar pattern speed $\Omega_p$. Also, because of the initial similarity of $\Omega$ and $\nu$ at any given radius, the percentage of resonant material with the vertical Inner Lindblad resonance \citep{combes90} is similar for all populations. Finally, in Model~1 we observe that the bar formation rapidly raises the radial velocity dispersion of the cold disc to values similar to those of the hot population (see Fig.~\ref{sigmavstimeM1} in Appendix~\ref{discs-app}), thus erasing their initial kinematic differences. The similarity of $\Omega$ and $\nu$, and the convergence of the in-plane velocity dispersions to similar values, imply a similar B/P shapes in all populations, independently on their initial in-plane kinematics.

\section{Conclusions}\label{conclusions}

In this paper we have made use of dissipationless N-body simulations of boxy/peanut-shaped bulges formed from composite stellar discs, made of kinematically cold and hot stellar populations, to discuss  the main drivers of the trends observed in a B/P bulge. To this aim, we have analyzed two extreme models, the first  made of disc populations with the same initial vertical random motions, $\sigma_z$ but different in-plane random motions, $\sigma_R$, the second made of disc populations with different initial  thickness but equal in-plane random motions. We have chosen such extreme initial conditions to separately study the role of in-plane versus vertical kinematics  in determining the final properties of a B/P bulge. \\ 

While a larger sample of N-body simulations is needed to understand how representative our findings are of the two classes of models studied,  we can derive some first conclusions.\\
At first order, disc populations with different initial $\sigma_z$ and same $\sigma_R$ are subject  to a similar differential mapping into a boxy/peanut-shaped bulge, as that experienced by stellar populations with equal initial $\sigma_z$ and different $\sigma_R$. In both cases:
\begin{itemize}
\item when viewed face-on, the colder initially the disc population, the stronger the bar; when viewed edge-on, the hotter initially the disc population, the thicker the B/P structure. 
\item at a given height above the plane, the relative contribution of stars in the B/P-shaped bulge depends on their initial kinematics, and the relative weight of stars initially in the cold disc population decreases with height (and viceversa, that of the hot disc population increases with height). 
\item because of this differential distribution of initially cold and hot disc populations in the B/P-shaped bulge, vertical metallicity(/age) gradients are naturally formed along the bulge minor axis, and X-shape metallicity/age distributions as well,  if a velocity dispersion-metallicity(/age) relation exists in the disc, initially. 
\end{itemize}
As a consequence, the vertical, as well as the radial velocity dispersions, are \emph{both} responsible for generating similar trends in a boxy/peanut-bulge, like that of the Milky Way, and we cannot conclude that  in-plane random motions play a more dominant role than vertical random motions in determining these trends, as recently suggested.

However, some important differences exist in the characteristics of the B/P bulges, and their surrounding discs, as generated with these two models.
 \begin{itemize}
\item We find that the model where disc populations have initially only different in-plane random motions, but similar thickness, results into a boxy/peanut bulge where all populations have a similar peanut shape, independently on their initial kinematics, or metallicity. As a consequence, the peanut appears at the same height above the plane, for all populations.  This is at odds with the trends observed in the Milky Way bulge. 
\item Finally, a model where the disc stellar populations are initially differentiated only in their in-plane kinematics, generates discs with no relation between metallicity/age and vertical velocity dispersion and where all populations converge to the same final radial velocity dispersion. This model is thus not applicable to the Milky Way.
\end{itemize}
On the basis of these two models, we conclude that  a metal-poor, kinematically (radial and \emph{vertical}) hot component, that is a thick disc, is necessary in the Milky Way before bar formation.

\section*{Acknowledgments}
The authors are grateful to the referee for their comments, which helped improving the clarity of the manuscript and the presentation of the results. This work has been supported by the ANR (Agence Nationale de la Recherche) through the MOD4Gaia project (ANR-15-CE31-0007, P.I.: P. Di Matteo). During this work, FF was supported by a postdoctoral grant from the Centre National d'Etudes Spatiales (CNES), SK was supported by a postoctoral grant by ANR. This work was granted access to the HPC resources of CINES under the allocation A0020410154 made by GENCI. PDM and MH  acknowledge the generous hospitality of Valerie de Lapparent and the `Galaxies' team at the Institut d'Astrophysique de Paris, where this work has been partially developed.

\bibliographystyle{aa}
\bibliography{biblio}

\begin{appendix}

\section{On the initial conditions and their impact on the final bulge morphology}\label{initparam-app}
\begin{figure*}
\begin{center}
\includegraphics[clip=true, trim = 0mm 0mm 0mm 0mm, width=0.9\linewidth]{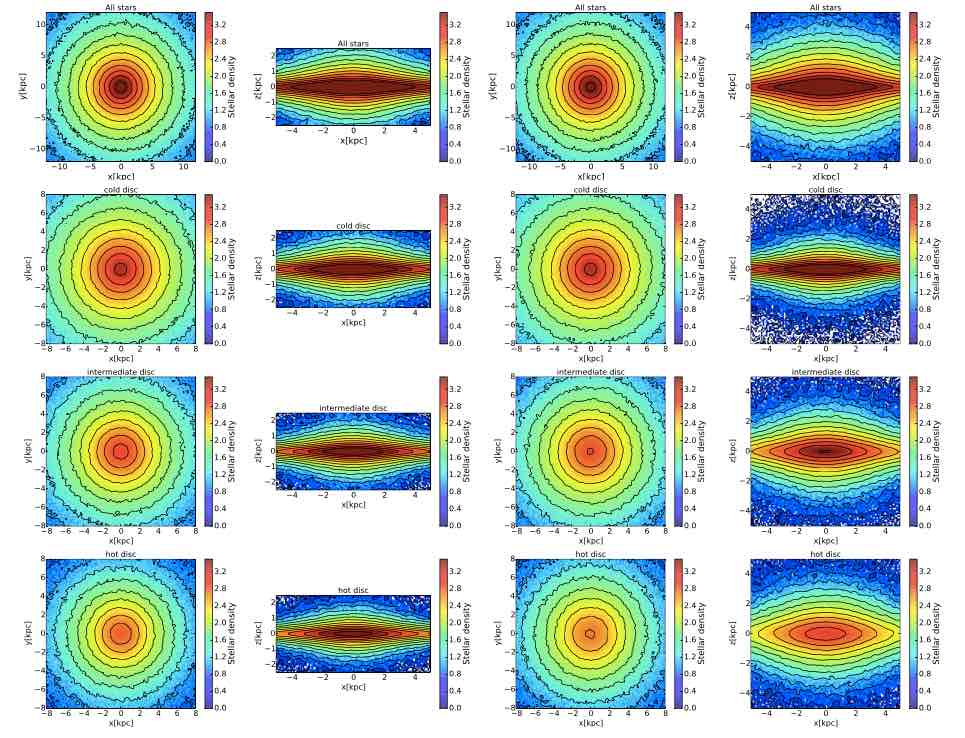}
\end{center}
\caption{\emph{Morphology of the bulge region at time $t=0$, for Model~1 and Model~2. First and third column, from top to bottom:} face-on absolute stellar densities of all stars (\emph{top panel}), and stars initially in the cold (\emph{second panel}), intermediate (\emph{third panel}) and hot (\emph{bottom panel}) discs for Model~1 ((\emph{col~1})) and Model~2 (\emph{col~3}) .  \emph{Second and fourth column, from top to bottom:} edge-on absolute stellar densities of all stars (\emph{top panel}), and stars initially in the cold (\emph{second panel}), intermediate (\emph{third panel}) and hot (\emph{bottom panel}) discs for Model~1 ((\emph{col~2})) and Model~2 (\emph{col~ 4}). In the edge-on maps, only stars with $|x|\le 5.5$~kpc and $|y|\le 3$~kpc have been selected.} 
\label{Mods_xyzmaps_0}
\end{figure*}

\begin{figure*}
\begin{center}
\includegraphics[clip=true, trim = 0mm 0mm 0mm 0mm, width=0.9\linewidth]{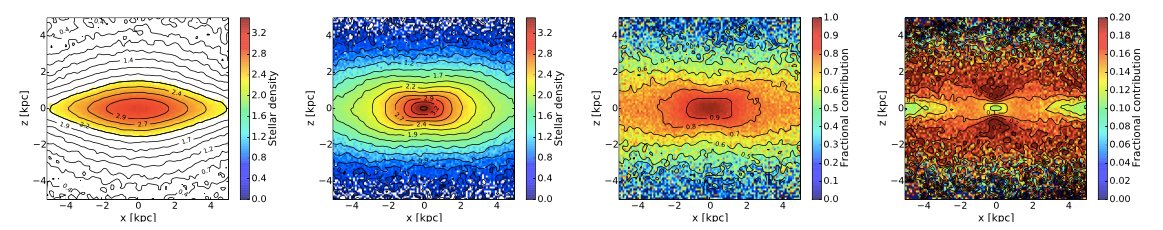}
\end{center}
\caption{\emph{From left to right, First panel:} Edge-on stellar density map of particles in the hot disc of Model~2 at $t=0$. The density contours trace the whole density distribution of the hot disc in the central regions, the colored map corresponds to the bulge-like part of this distribution (see text for details). \emph{Second panel:} Edge-on stellar density map, at the final time of the simulation, of hot disc particles initially inside the bulge-like region. \emph{Third panel}: Relative contribution of hot disc particles initially inside the bulge-like region to the final B/P bulge, as defined by hot disc particles only.  \emph{Fourth panel}: Relative contribution of hot disc particles initially inside the bulge-like region to the final B/P bulge, as defined by all disc particles, i.. cold and intermediate discs also included.}   
\label{init_bulgeselection}
\end{figure*}

In Fig.~\ref{Mods_xyzmaps_0}, the face-on and edge-on stellar density maps of stars in the inner regions are shown, for Models~1 and ~2.  
To generate the initial positions of the particles in the two models,  for each disc component, we have randomly extracted a number $N$ of particles, according to the corresponding Miyamoto-Nagai density distribution of characteristic radius $a$ and characteristic height $h$  (see Table~\ref{galparamtable} for the number of particles and characteristic spatial scales adopted for the discs in the two models). For this random realization, a cut in the in-plane, $R$, and vertical, $z$, distances has been adopted, so that only particles with positions inside $(R_{max}, z_{max})=(10\times a_{hot~disc}, 10\times h_{hot~disc})$ are retained. This choice corresponds to a density cut  at approximately $10^{-3}$ times the central stellar density, for all discs in the two models (see Fig.~1 in \citet{miyamoto75}). The adoption of these initial cuts explains why, in Model~1, the initial vertical extension of all discs is smaller than that of the corresponding discs of Model~2, since in Model~1 all discs have the same characteristic height.\footnote{Note that the limiting density induced by  this  cut is low enough to guarantee that no significant artificial effect in the further evolution of the system is observed. As an example of this assertion, the cold discs in Model~1 and ~2 have initially both the same mass, and same characteristic radii, but differ only in the vertical cut adopted for the generation of the initial conditions, $z_{max}$ being respectively equal to 2.5~kpc and 9~kpc,  for Model ~1 and ~2.  Despite this different initial vertical extension, the final B/P bulge morphology for the two cold populations is remarkably similar in the two models  at the final time of the simulations (cfr Figs.~\ref{ModI_xyzmaps} and \ref{ModII_xyzmaps} ).  }\\
As pointed out by \citet{miyamoto75} (see, again, their Fig.~1), and as it is also visible in the edge-on maps shown in our Fig.~\ref{Mods_xyzmaps_0}, for high $h/a$ ratios (b/a ratios, in  \citet{miyamoto75} nomenclature)  a central bulge-like part appears in the initial edge-on density maps. In the following, we estimate the contribution of this central bulge-like structure to the final B/P morphology of the bulge for the hot disc component of Model~2. Because of its highest  $h/a$ ratio, this is indeed the component with the most pronounced central bulge-like part in our models. To estimate the contribution of the central bulge-like component to the final B/P morphology, we have selected all  particles in the hot disc of Model~2, which have initially $|x|\le 5.5$~kpc and $|y|\le 3$~kpc -- corresponding to the final B/P bulge region --   and which, in particular, lie inside the bulge-like part of the hot disc, that we define as the region delimited by a density level equal to 0.1~times the central  density of the hot disc  (see Fig.~\ref{init_bulgeselection}, left panel).  At time $t=0$, the fraction of  particles inside this bulge-like region constitutes on average about 80\% of the whole hot disc inside the final B/P region, but only  15\% of the whole stellar distribution. The edge-on distribution of the bulge-like particles at the final time is shown in the second panel of Fig.~\ref{init_bulgeselection}. It appears that, despite their initial bulge-like morphology, these particles do also respond to the B/P instability, and show a (weak) peanut-like morphology at the end of the simulation.  It is also interesting to note that these particles tend to preferentially populate the innermost regions of the final B/P bulge described by the hot disc population. This is similar to the finding of \citet{dimatteo14, gomez16}, who showed that stars born in the innermost regions of a disc tend to populate the innermost parts of a B/P bulge, while stars born at larger distances tend to be preferentially redistributed in the outskirt of the B/P structure, preserving their Jacoby energy, as already shown by \citet{martinez13}.   Overall, the bulge-like component represents a significant fraction of the hot disc in the innermost B/P region, especially for $|z|\le 1$~kpc (see Fig.~\ref{init_bulgeselection}, third panel). However it does represent only a marginal fraction of the whole B/P bulge populations, which includes also particles formerly in the cold and intermediate discs (see Fig.~\ref{init_bulgeselection}, fourth panel). Thus, while this central bulge-like component certainly dominates the morphology of the kinematically hot part of the B/P bulge made of hot disc stars only, it only marginally affects the final, whole B/P morphology and its spatial extension.

\section{On the convergence of the iterative method used to generate the initial conditions}\label{rodionov} 

\begin{figure*}
\includegraphics[clip=true, trim = 5mm 6mm 15mm 5mm,, width=0.3\linewidth]{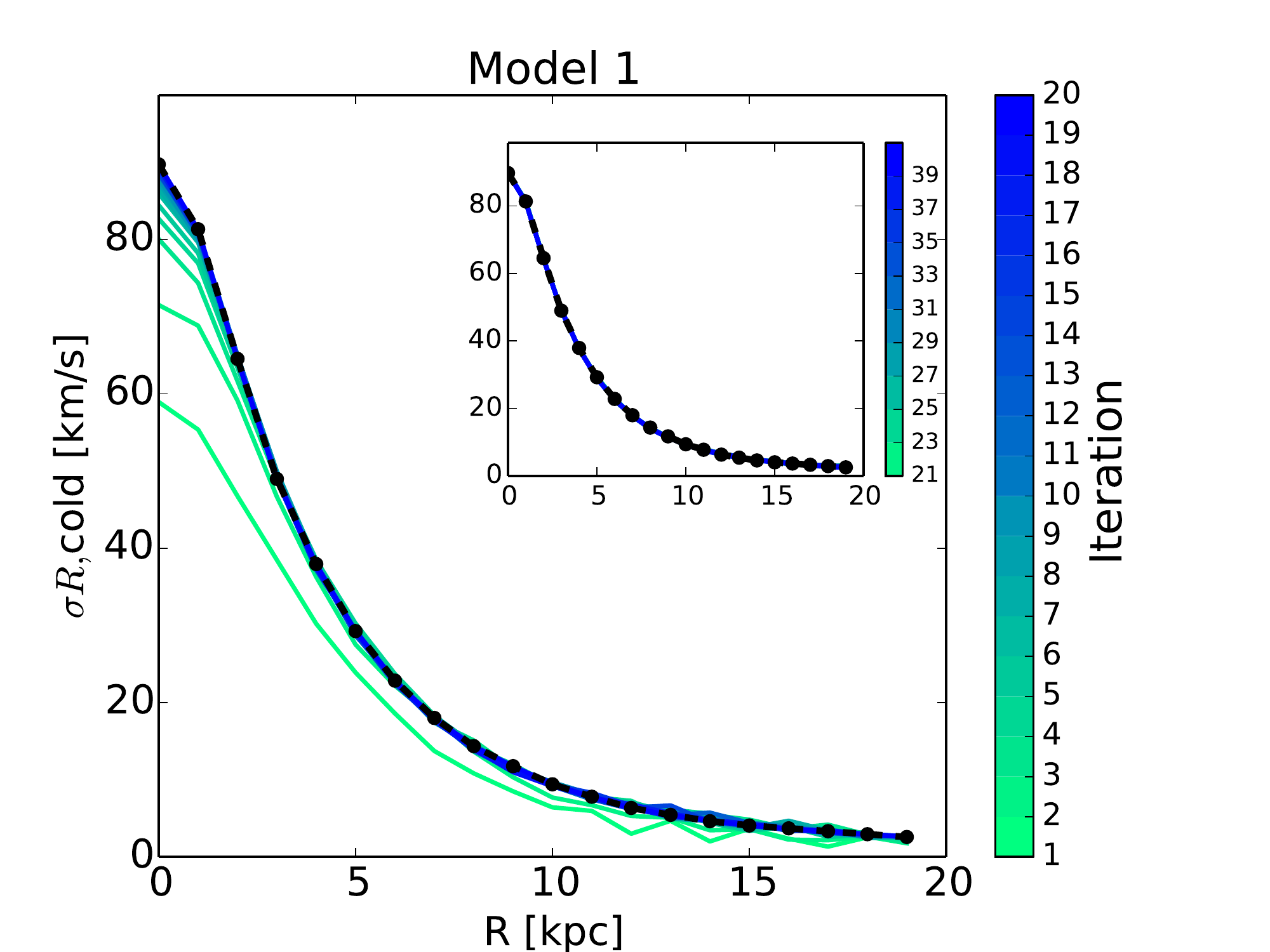}
\includegraphics[clip=true, trim = 5mm 6mm 15mm 5mm,, width=0.3\linewidth]{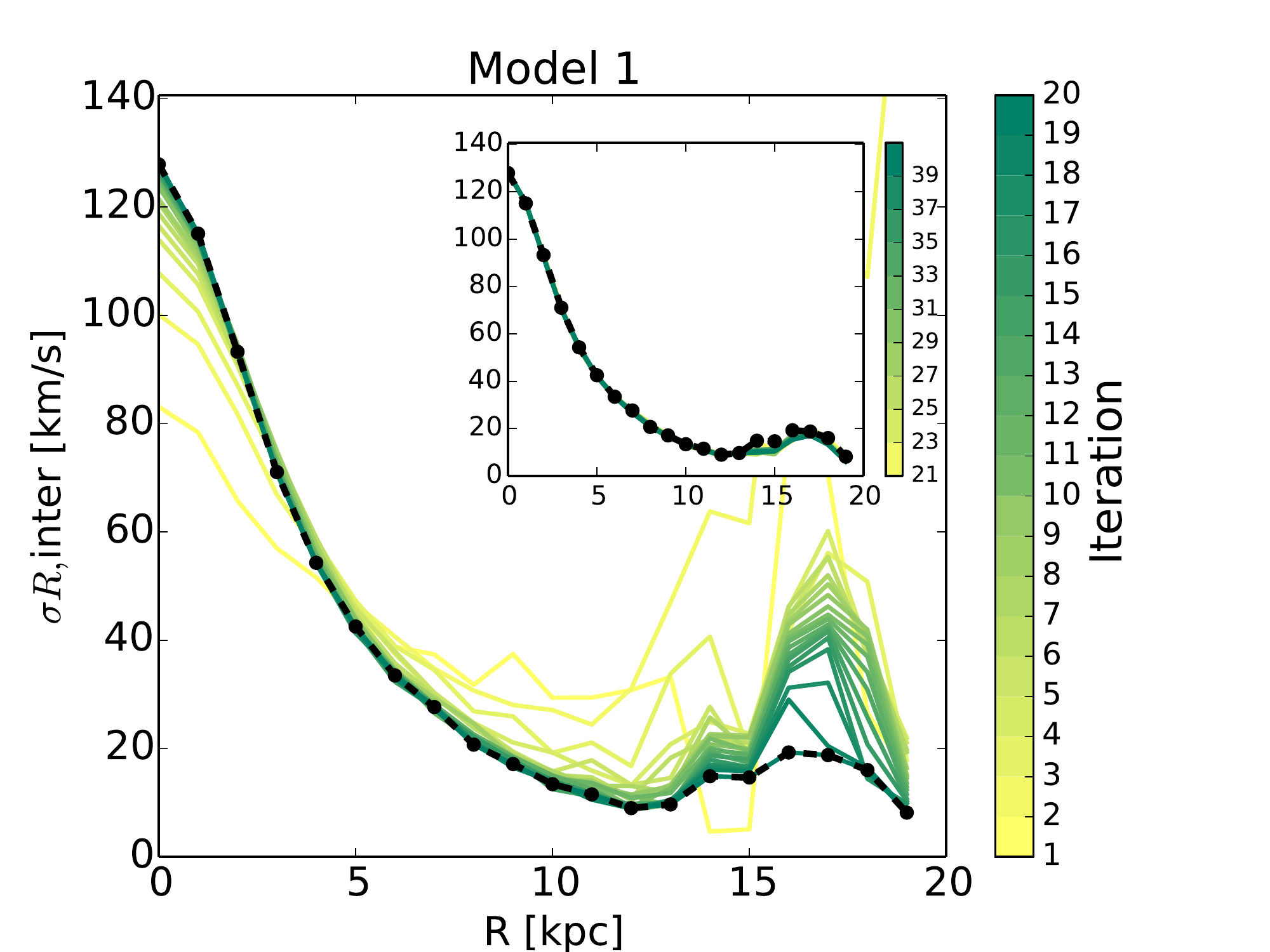}
\includegraphics[clip=true, trim = 5mm 6mm 15mm 5mm,, width=0.3\linewidth]{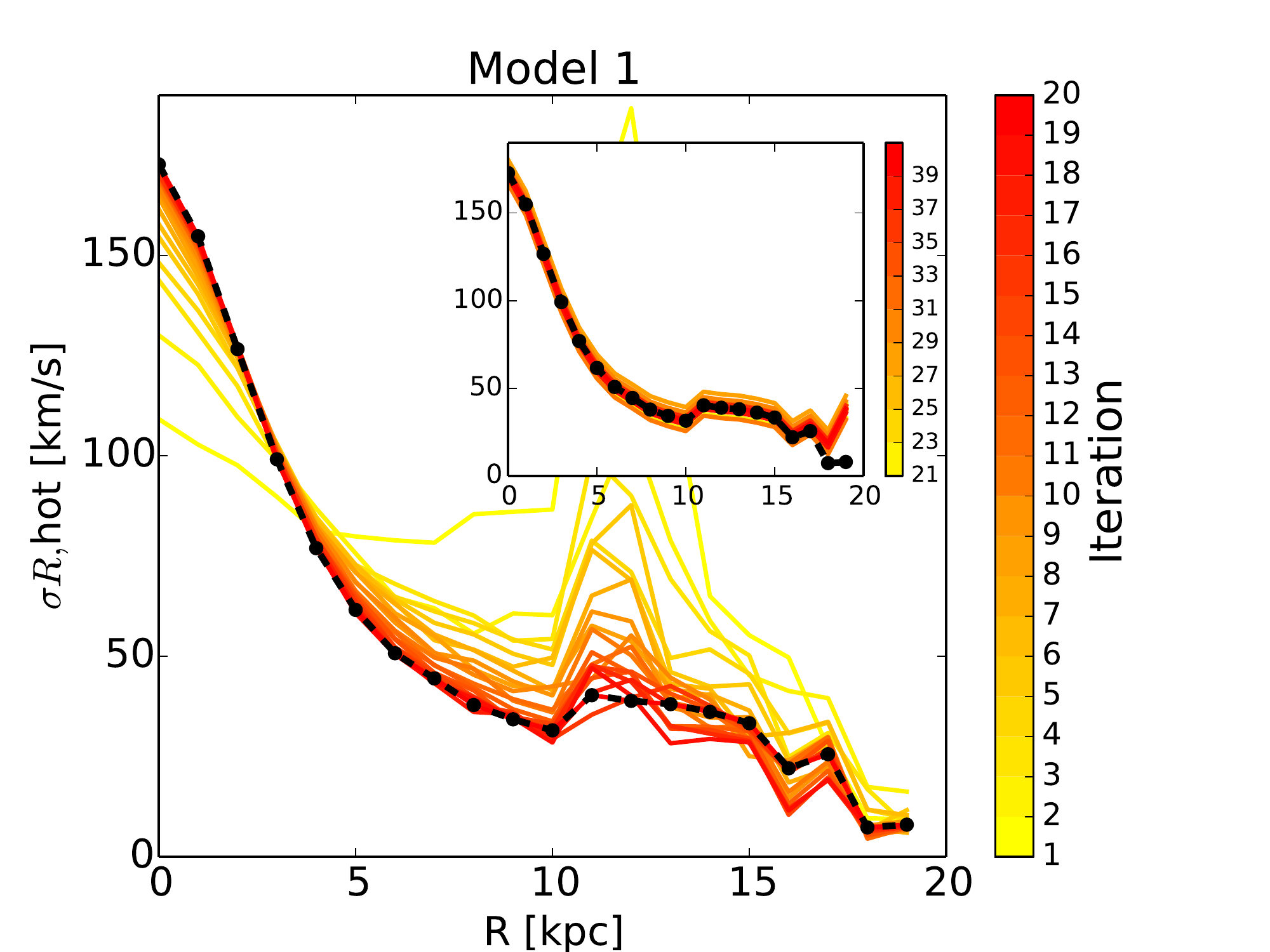}

\includegraphics[clip=true, trim = 5mm 6mm 15mm 15mm,, width=0.3\linewidth]{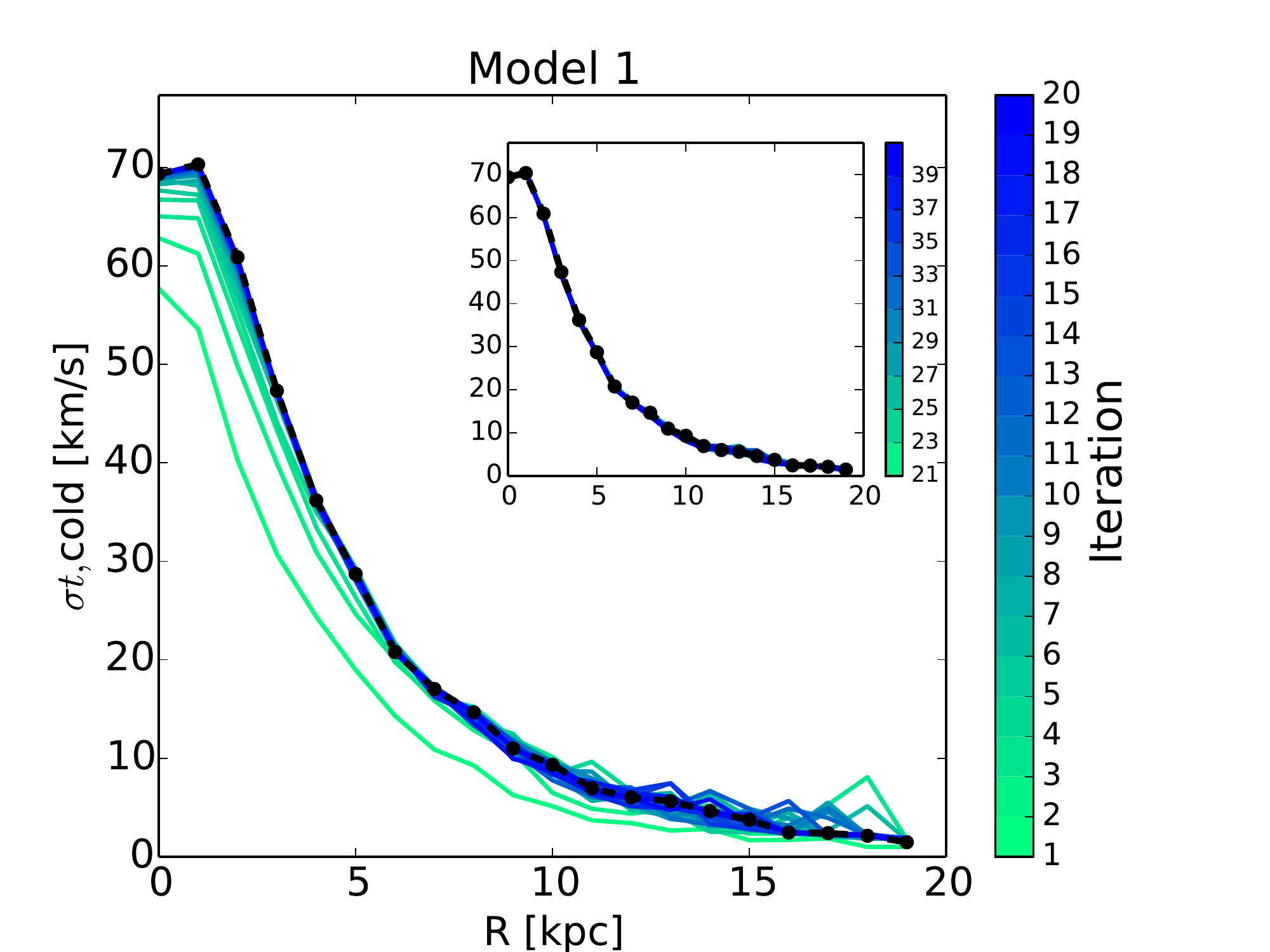}
\includegraphics[clip=true, trim = 5mm 6mm 15mm 15mm,, width=0.3\linewidth]{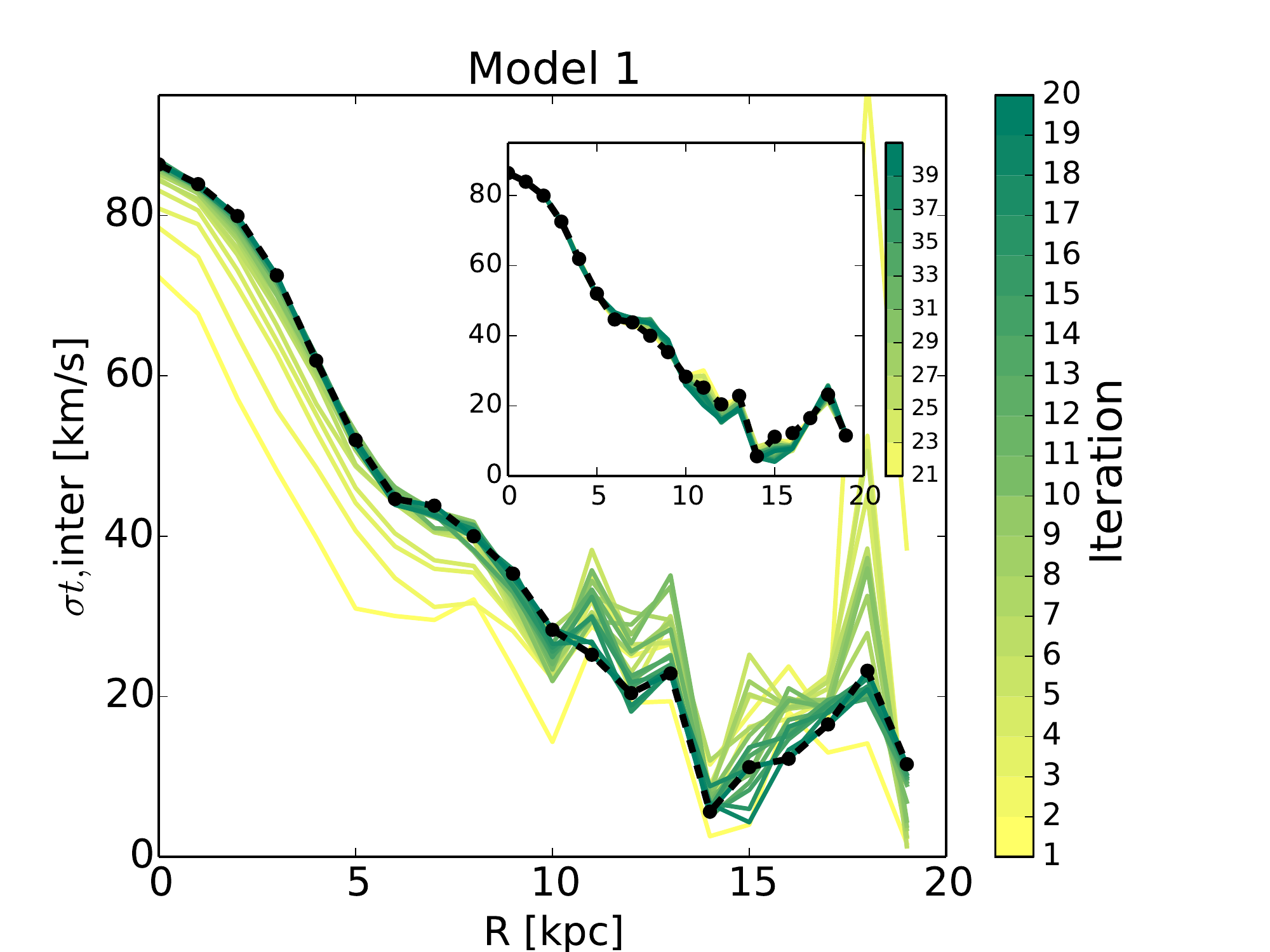}
\includegraphics[clip=true, trim = 5mm 6mm 15mm 15mm,, width=0.3\linewidth]{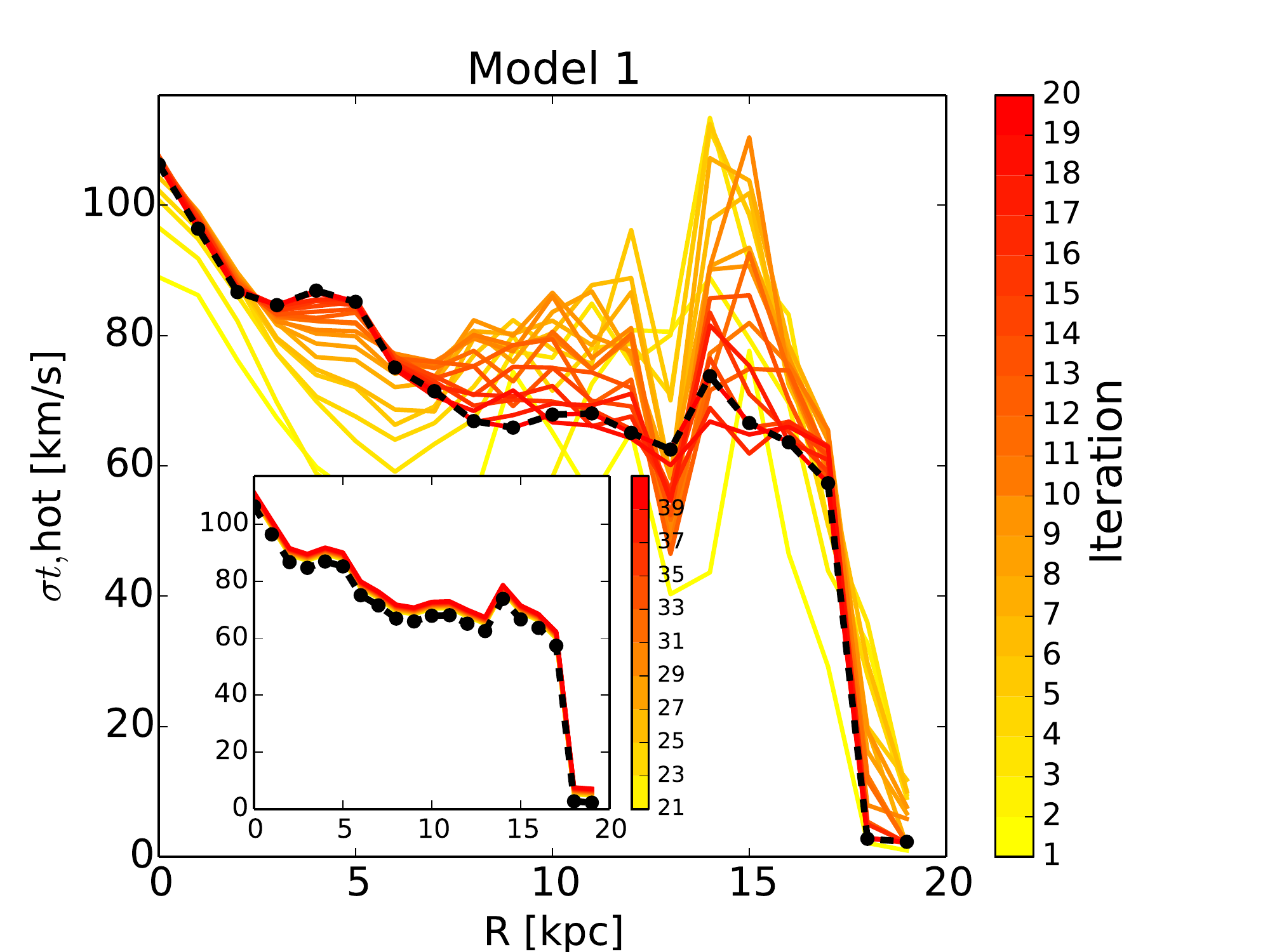}

\includegraphics[clip=true, trim = 5mm 0mm 15mm 15mm,, width=0.3\linewidth]{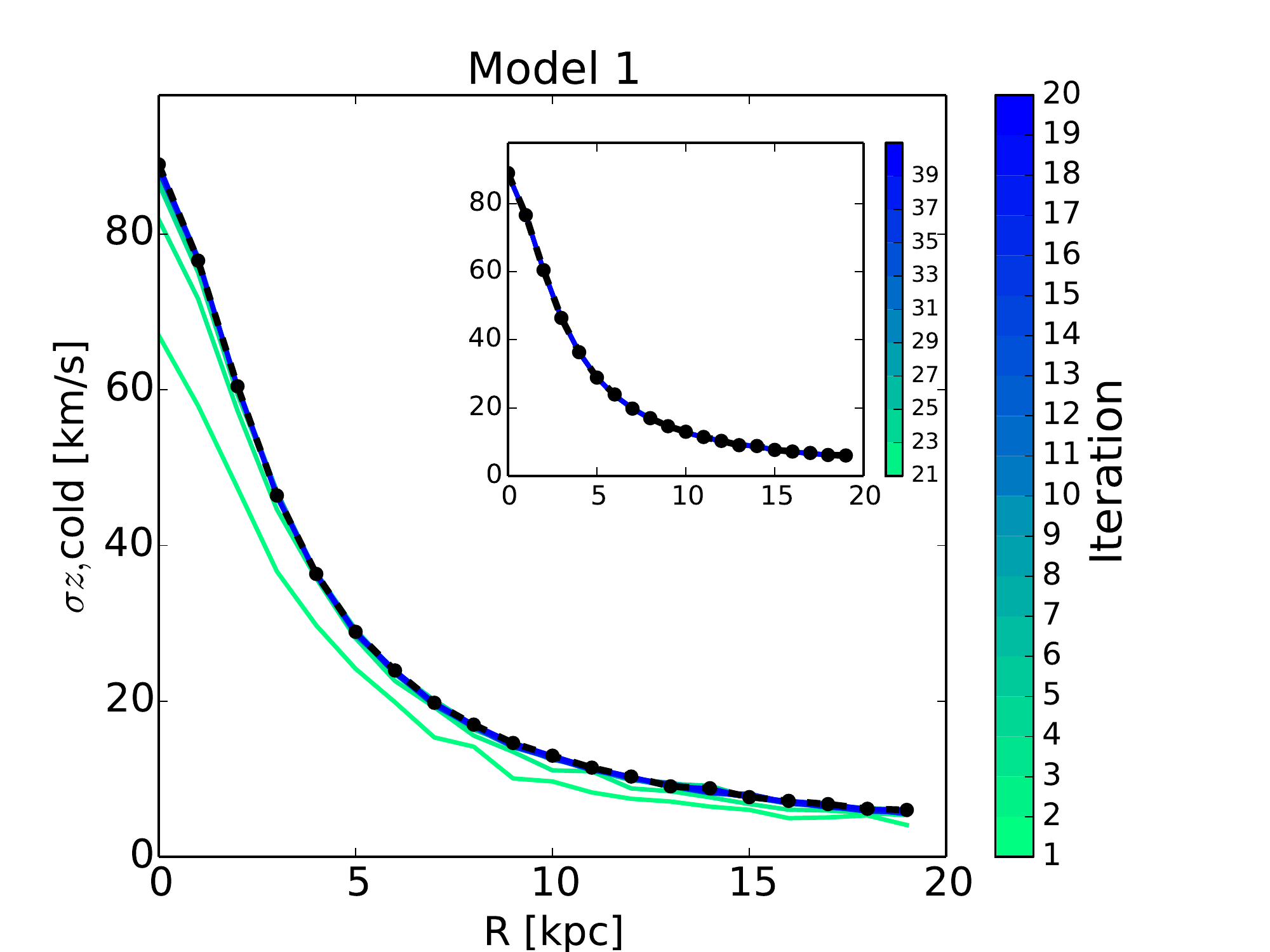}
\includegraphics[clip=true, trim = 5mm 0mm 15mm 15mm,, width=0.3\linewidth]{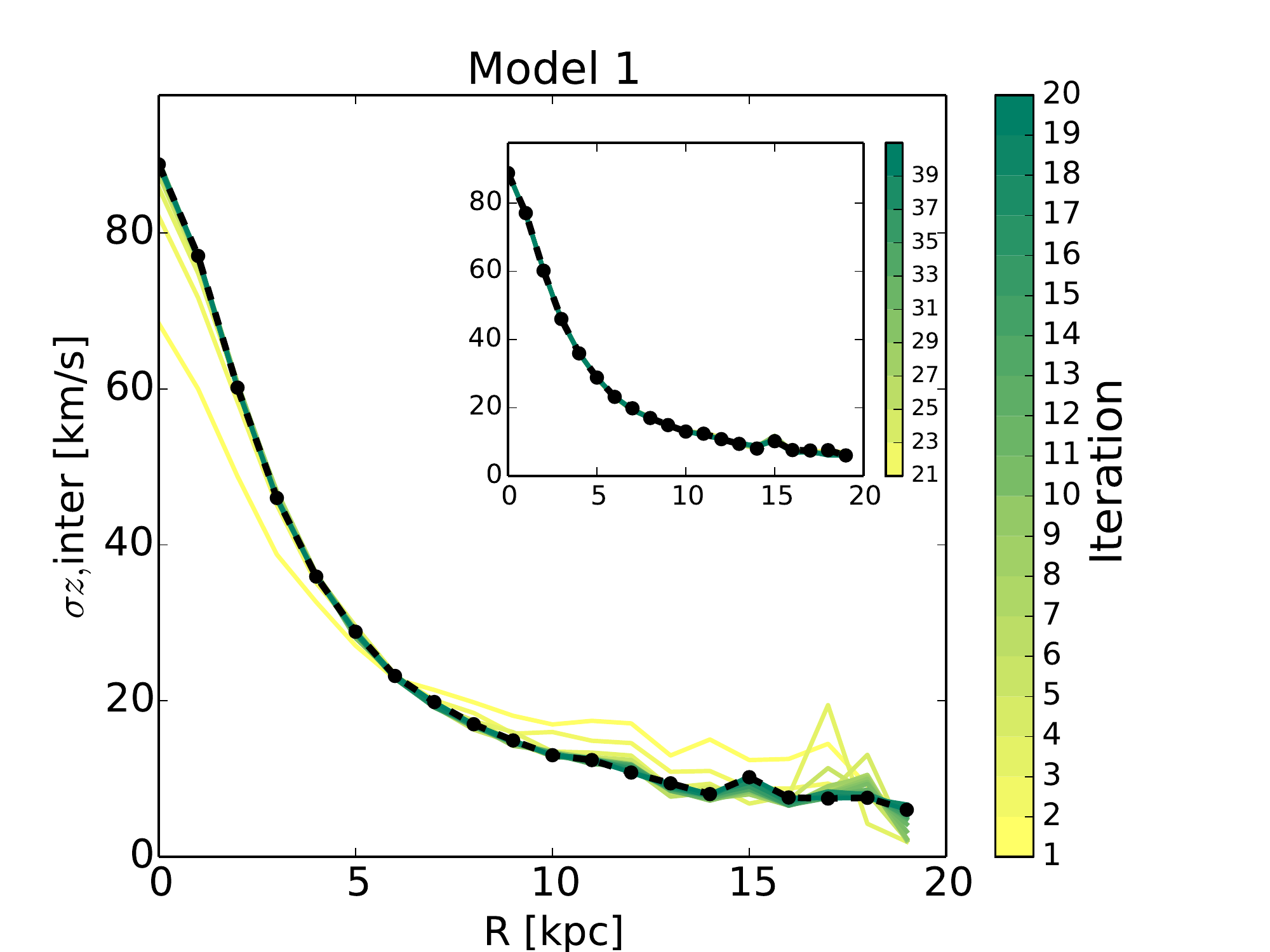}
\includegraphics[clip=true, trim = 5mm 0mm 15mm 15mm,, width=0.3\linewidth]{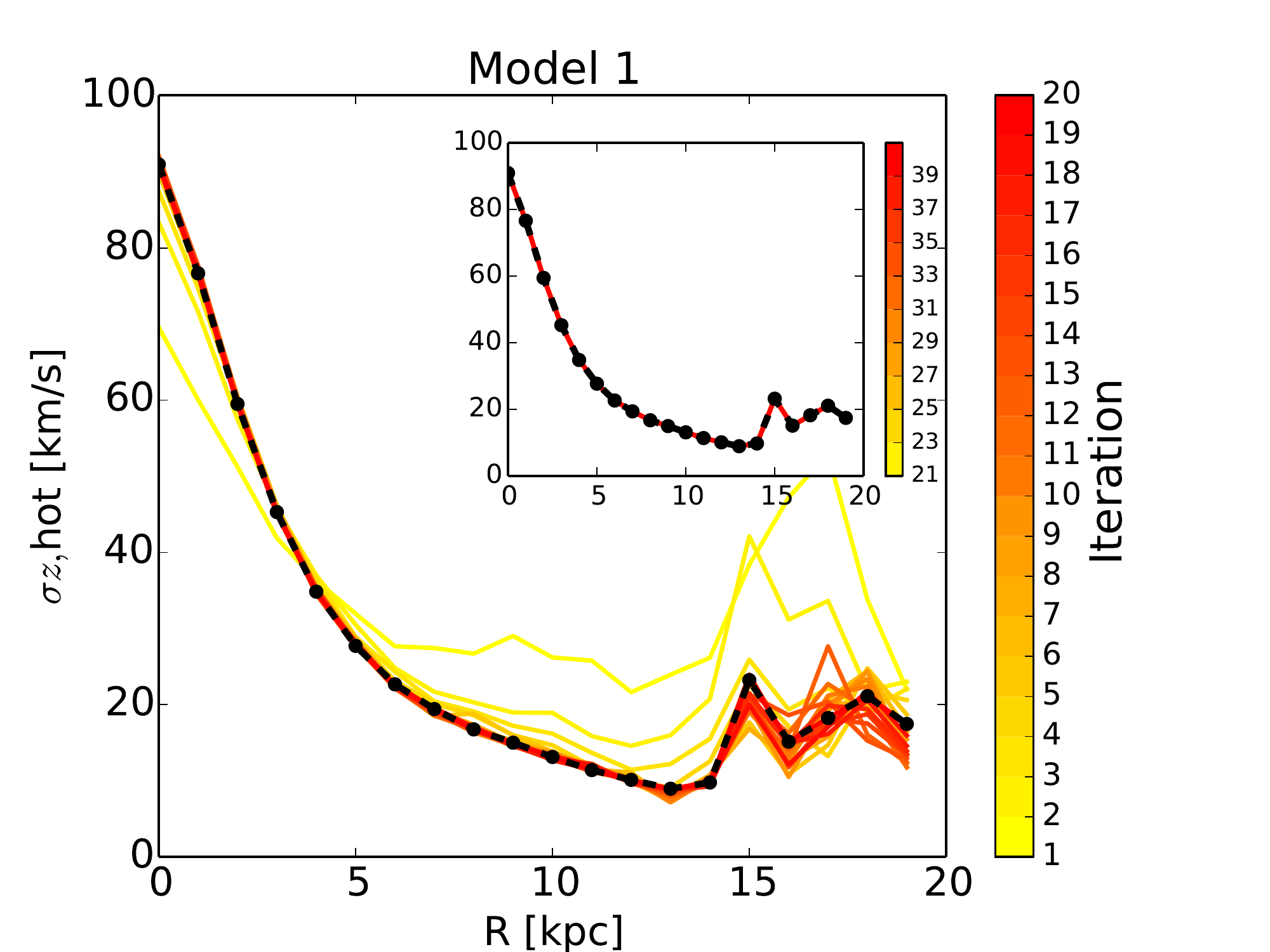}

\caption{\emph{From top to bottom:} Evolution of the radial, azimuthal and vertical velocity dispersion profiles, as a function of radius R, during the iterative procedure adopted to generate the initial conditions of Model~1. For each plot, and each inset in the plot, different colors represent different steps of the iteration, as indicated by the the color bar. In all plots, and insets, the black dotted cuve indicates the velocity dispersion after 20 iterations, when we consider the system converged to an equilibrium solution. Cold, intermediate and hot discs are shown, respectively, in the left, middle, and right columns, as indicated. }  \label{convergenceI}
\end{figure*}

\begin{figure*}
\includegraphics[clip=true, trim = 5mm 6mm 15mm 5mm,, width=0.3\linewidth]{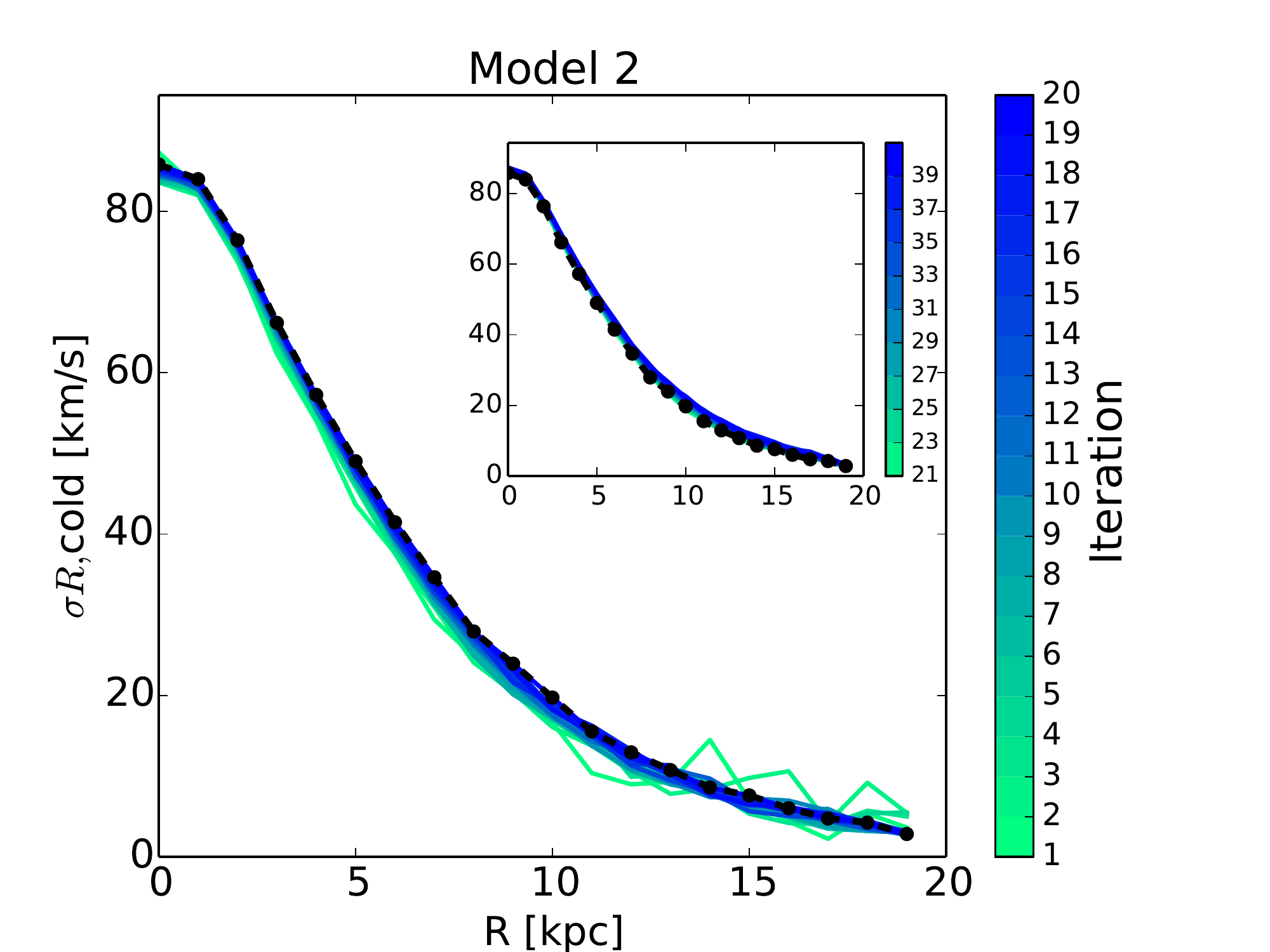}
\includegraphics[clip=true, trim = 5mm 6mm 15mm 5mm,, width=0.3\linewidth]{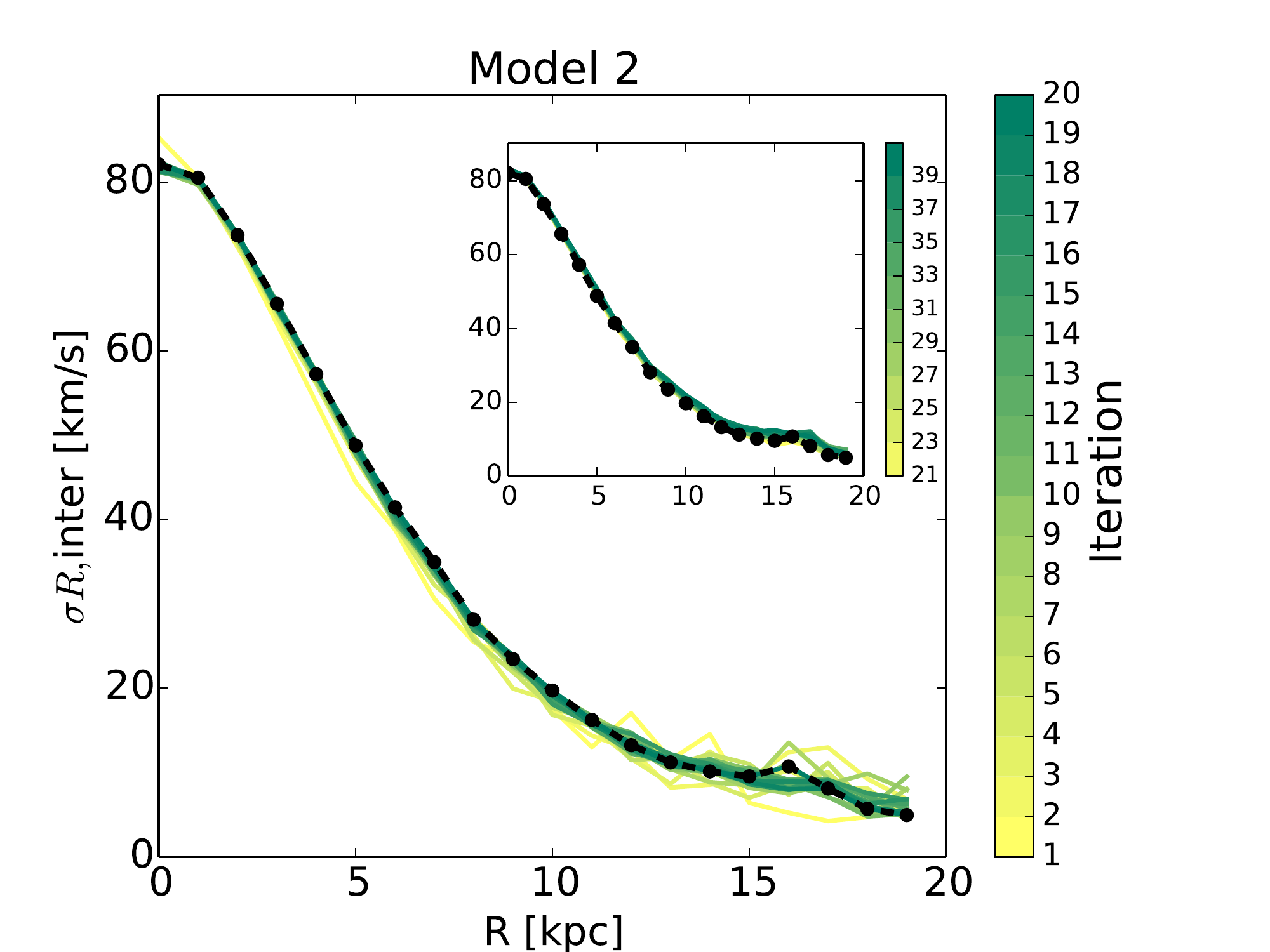}
\includegraphics[clip=true, trim = 5mm 6mm 15mm 5mm,, width=0.3\linewidth]{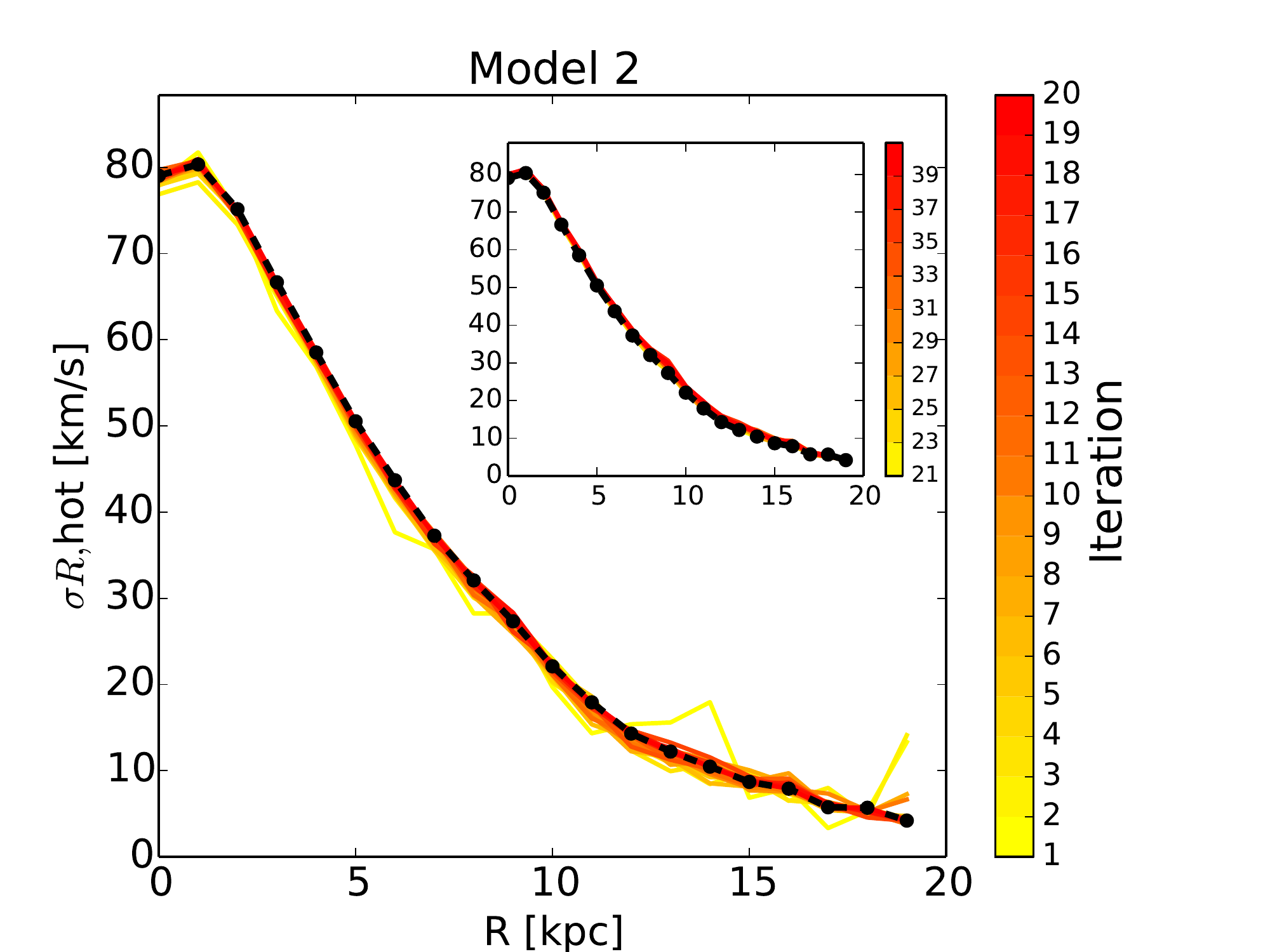}

\includegraphics[clip=true, trim = 5mm 6mm 15mm 15mm,, width=0.3\linewidth]{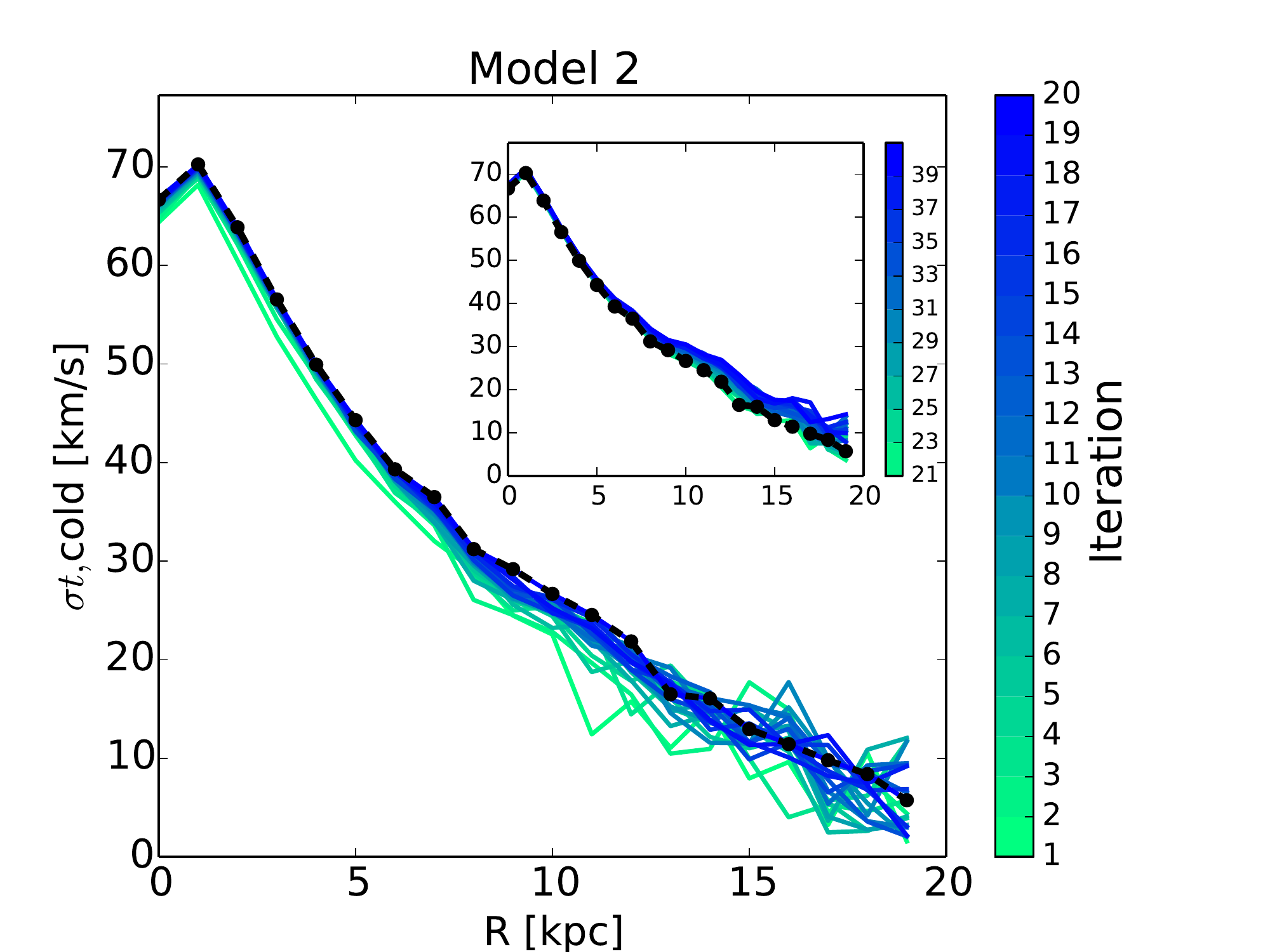}
\includegraphics[clip=true, trim = 5mm 6mm 15mm 15mm,, width=0.3\linewidth]{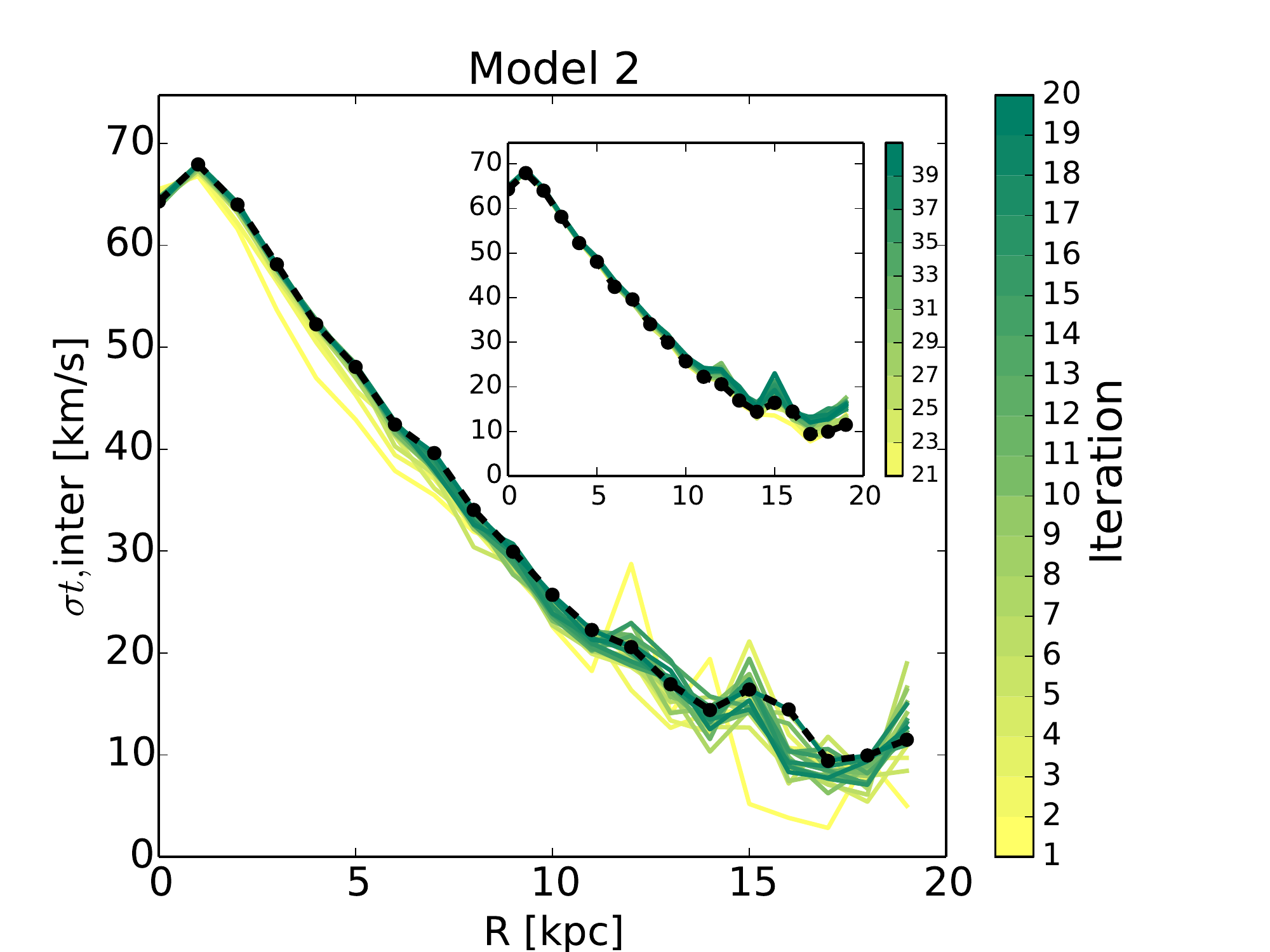}
\includegraphics[clip=true, trim = 5mm 6mm 15mm 15mm,, width=0.3\linewidth]{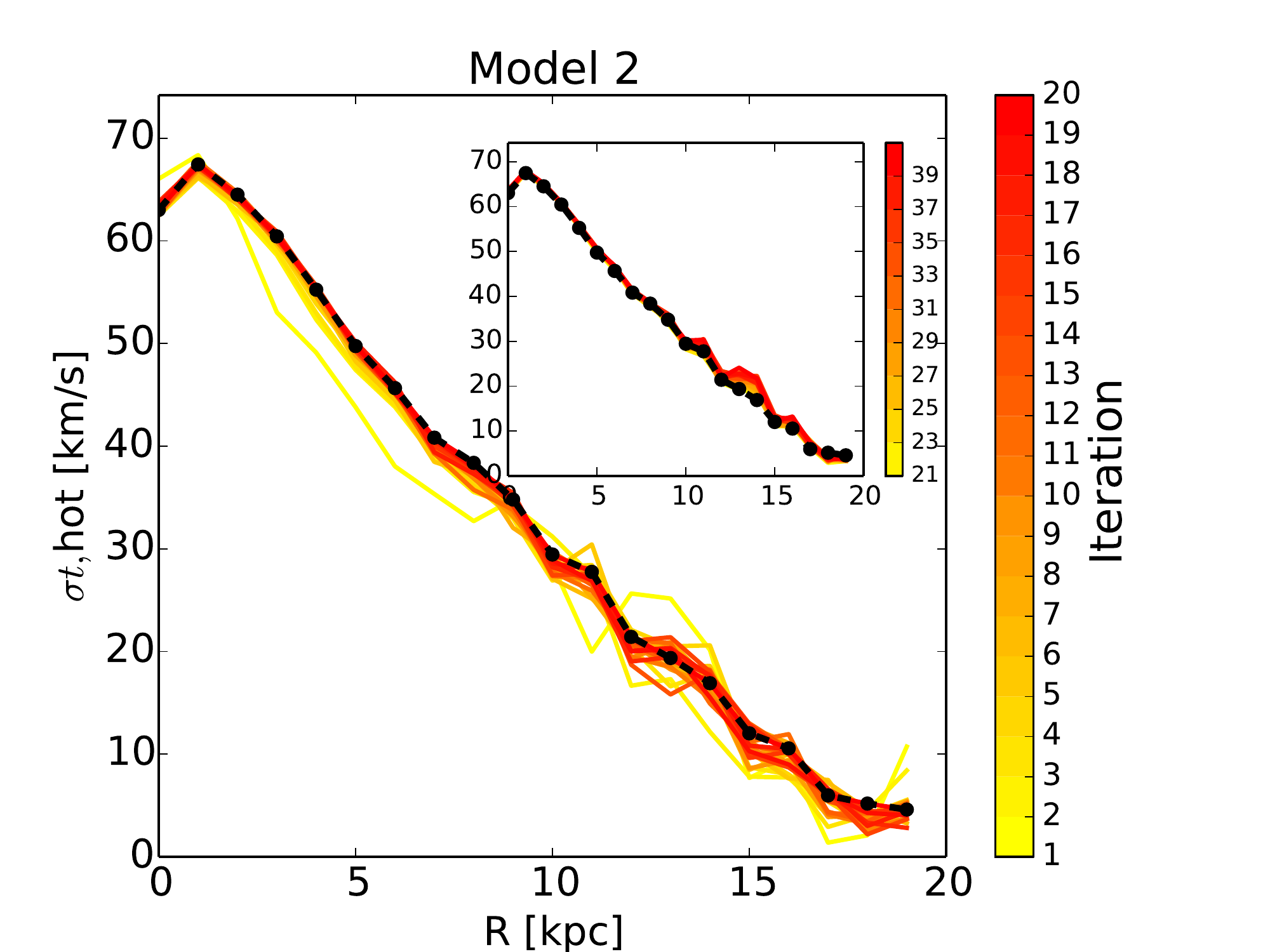}

\includegraphics[clip=true, trim = 5mm 0mm 15mm 15mm,, width=0.3\linewidth]{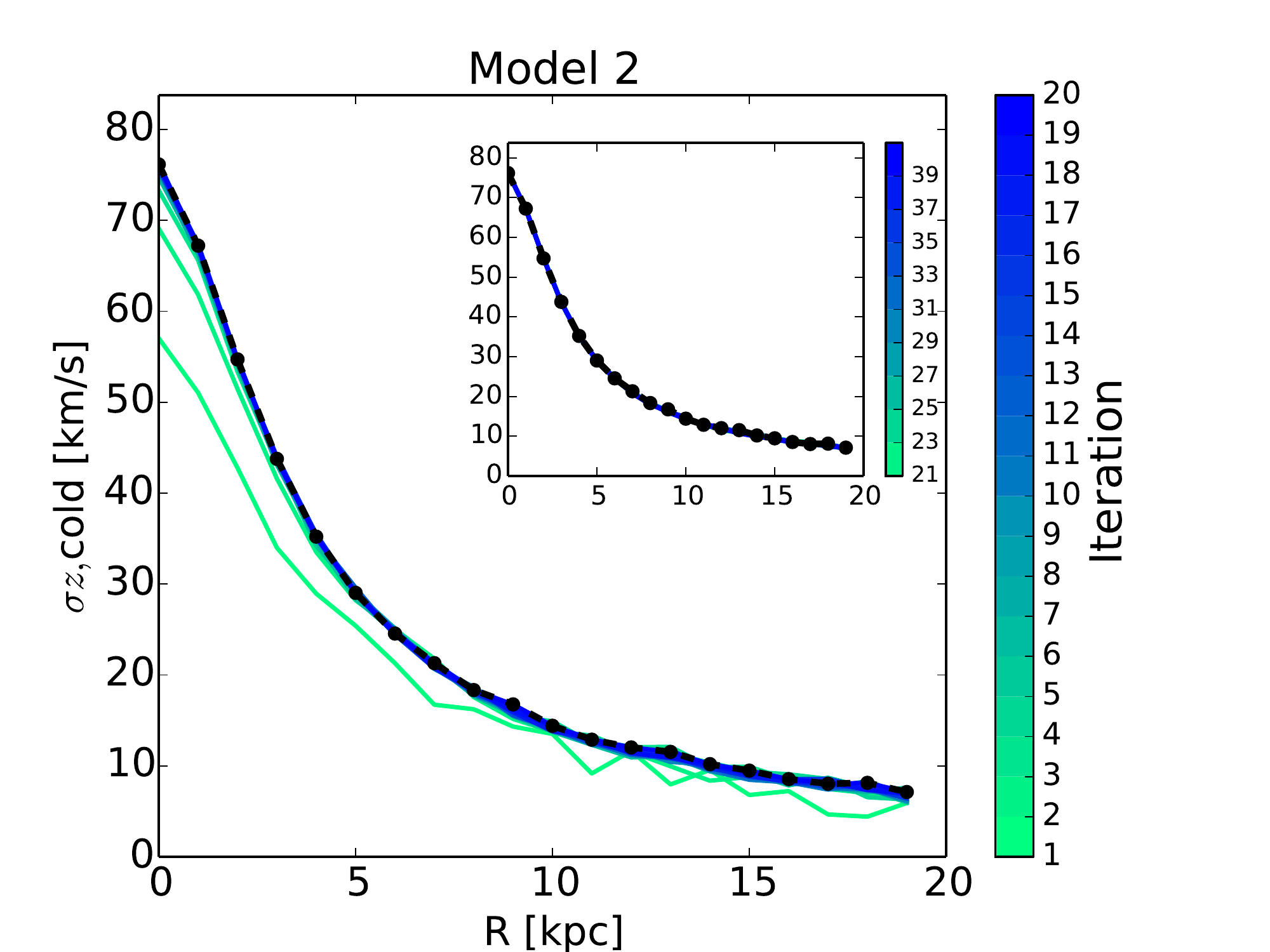}
\includegraphics[clip=true, trim = 5mm 0mm 15mm 15mm,, width=0.3\linewidth]{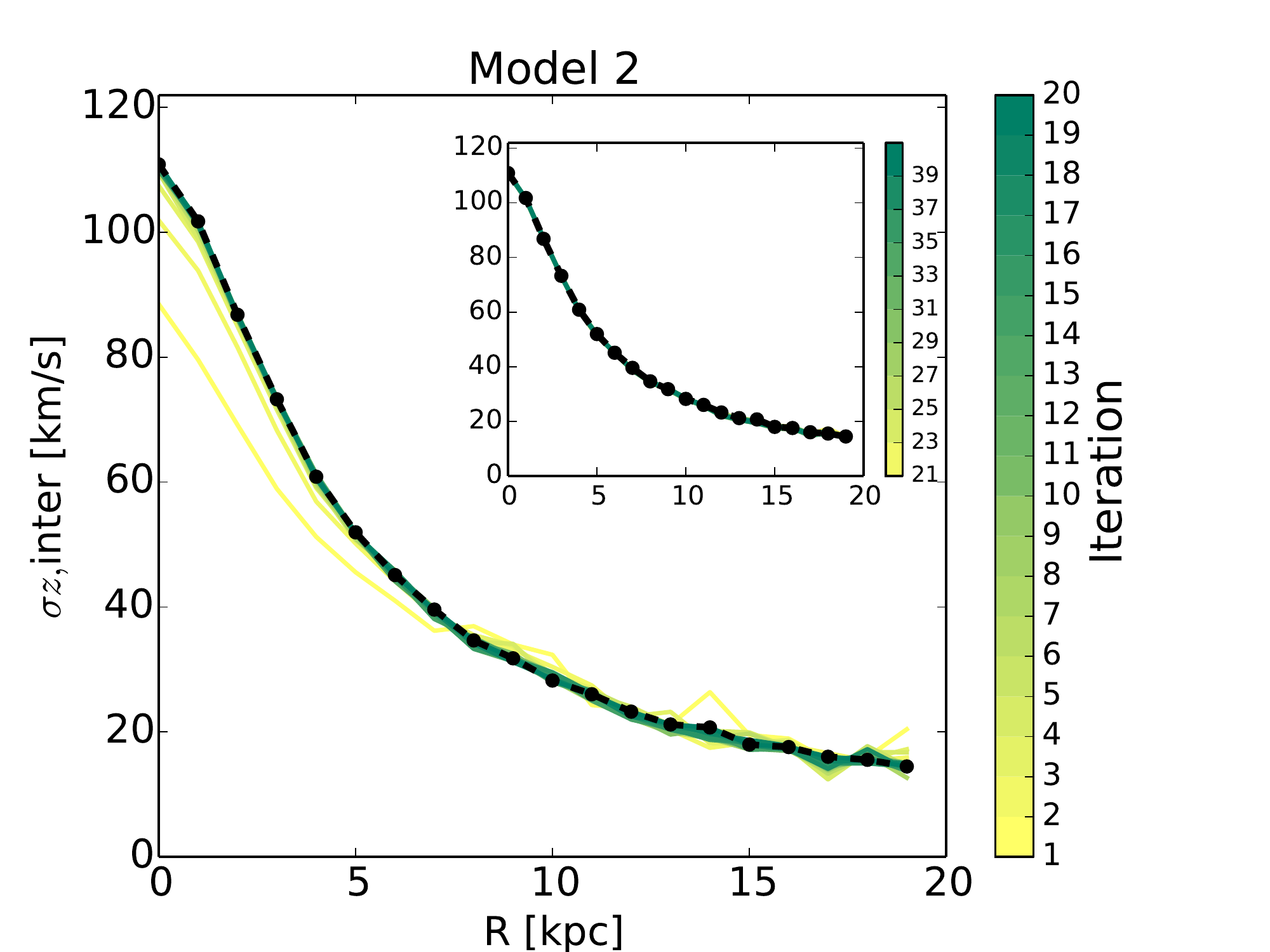}
\includegraphics[clip=true, trim = 5mm 0mm 15mm 15mm,, width=0.3\linewidth]{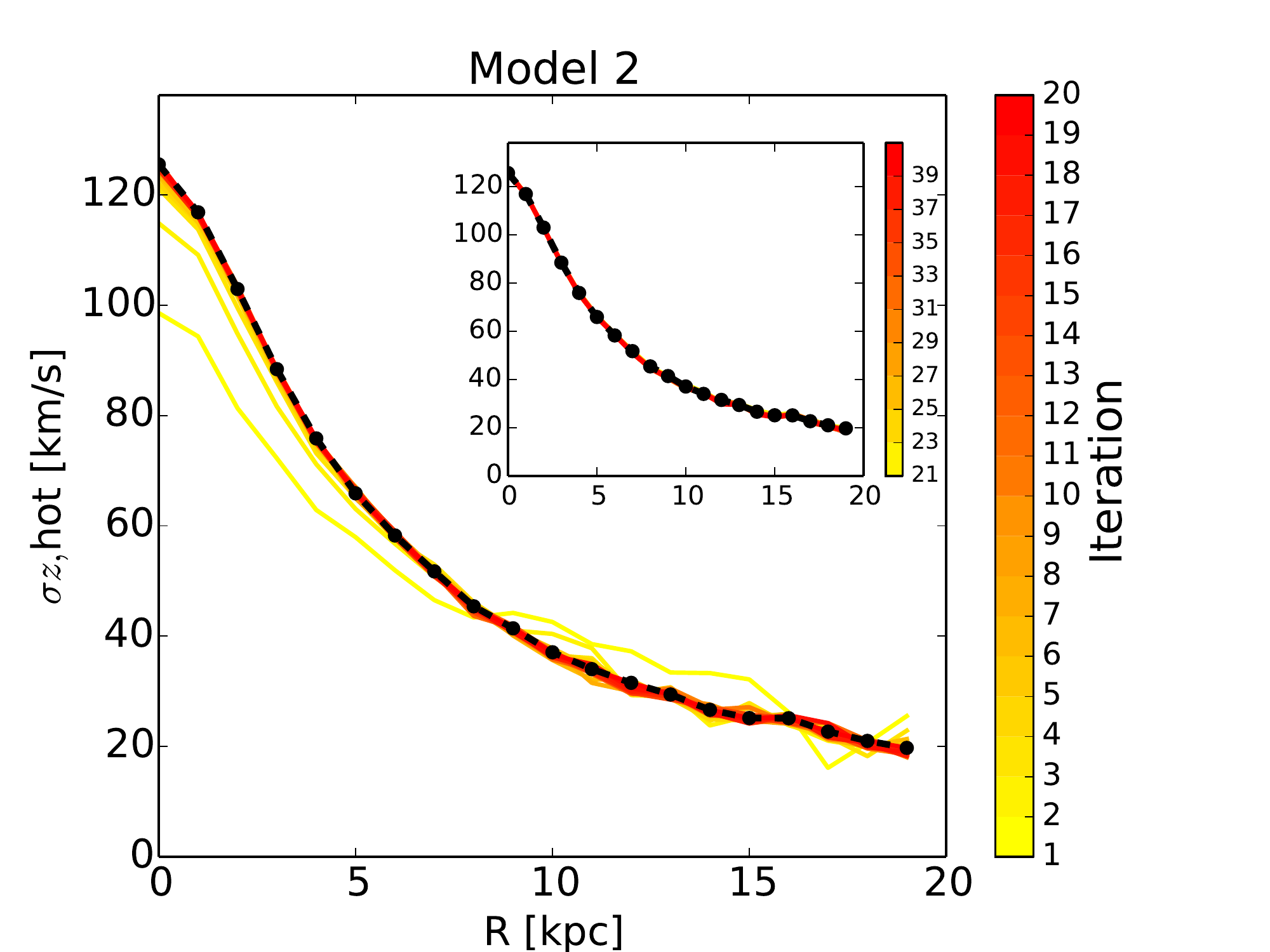}
\label{convergenceII}
\caption{Same as Fig.~\ref{convergenceI}, but for Model~2}  \label{convergenceII}
\end{figure*}

To generate initial conditions  at equilibrium for the two models, we employed the iterative method described in \citet{rodionov09}. \\
All discs of Model~1 and 2  have Myamoto-Nagai density profiles, with characteristic scale lengths and heights given in Table~\ref{galparamtable}. The dark matter halos follow a Plummer distribution, with characteristic parameters also given in Table~\ref{galparamtable}.  We did not impose any restriction on the kinematics of the system -- as done in some of the examples reported in  \citet{rodionov09} -- that is  none of the galactic components is required to converge to any specific velocity dispersion profile. To initialize the procedure, we have assigned the initial azimuthal velocities of each disc component in such a way to have an initial constant -- that is, independent on radial distance from the centre -- rotational lag\footnote{The rotational lag is defined as $v_{lag}=v_{circ}-<v_t>$, with $v_{circ}$ the rotational velocity of the modeled galaxy, and $<v_t>$ the mean azimuthal velocity of the disc component.} equal  to 0, 30 and 60 km/s, respectively, for the cold, intermediate and hot disc of Model~1. This guarantees the different in-plane kinematics of the three discs in Model~1 at the end of the iterative procedure. For Model~2, no initial rotational lag has been imposed to any of the disc components, and their different vertical kinematics is the consequence of their different characteristic heights. 
A number of 20 iterations  has been sufficient to obtain initial conditions at equilibrium in the galactic potentials. As an example of this convergence, in Figs.~\ref{convergenceI} and  \ref{convergenceII} we show the radial, azimutal and vertical velocity dispersion profiles of the cold, intermediate and hot discs in Model~1 and 2, respectively, at each step of the iteration, from 1 to 20.  In both figures, the insets show the evolution of these profiles for  20 supplementary iterations, from $step=21$ to $step=40$, to demonstrate that already after 20 iterations all our models have converged, and that no further significant change in the velocity dispersion profiles is observed.

\section{On the initial distribution of azimuthal and vertical frequencies}\label{freq-app}

\begin{figure*}
\begin{flushleft}

\includegraphics[clip=true, trim = 0mm 0mm 0mm 0mm, width=0.24\linewidth]{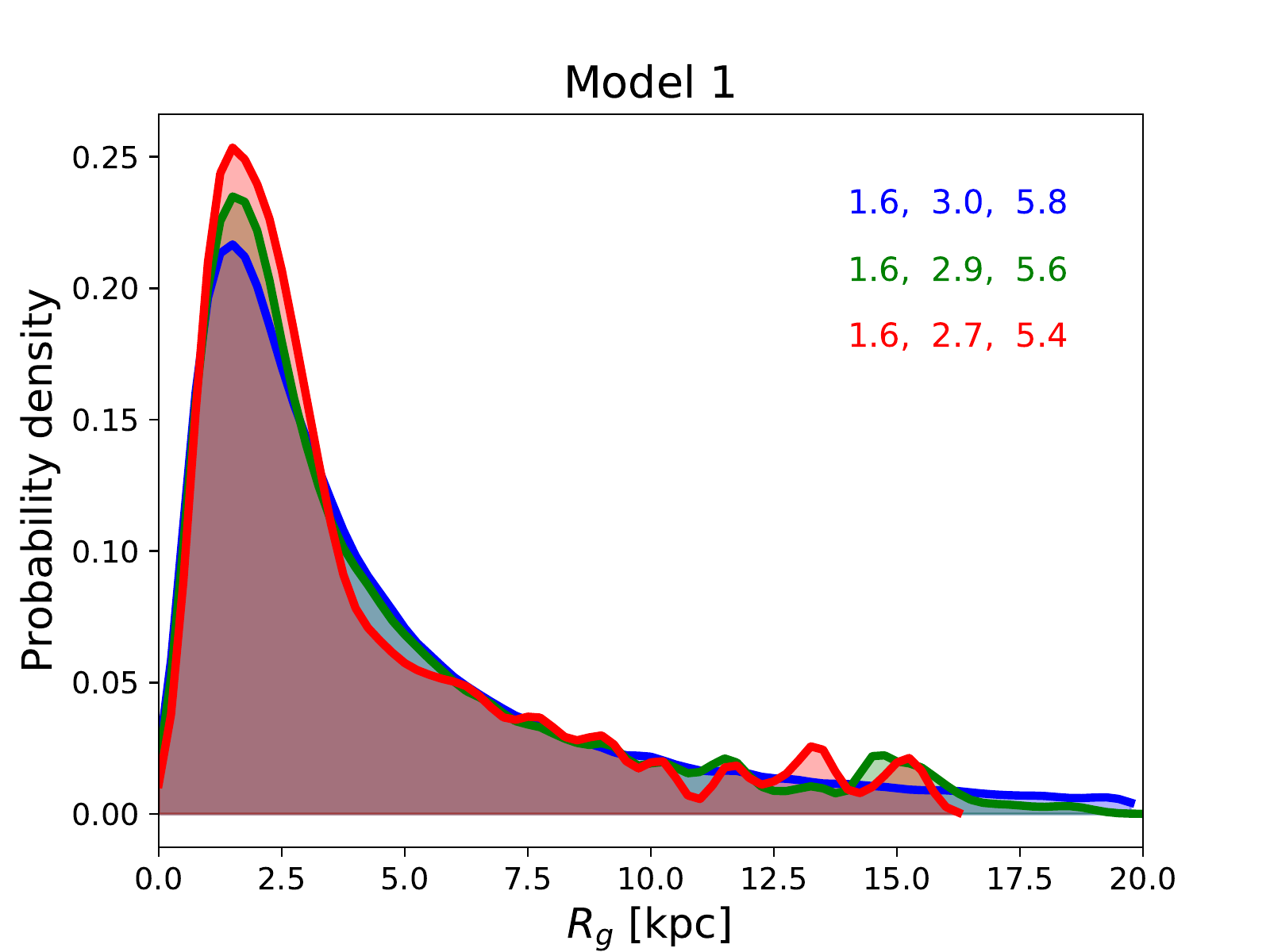}
\includegraphics[clip=true, trim = 0mm 0mm 0mm 0mm, width=0.24\linewidth]{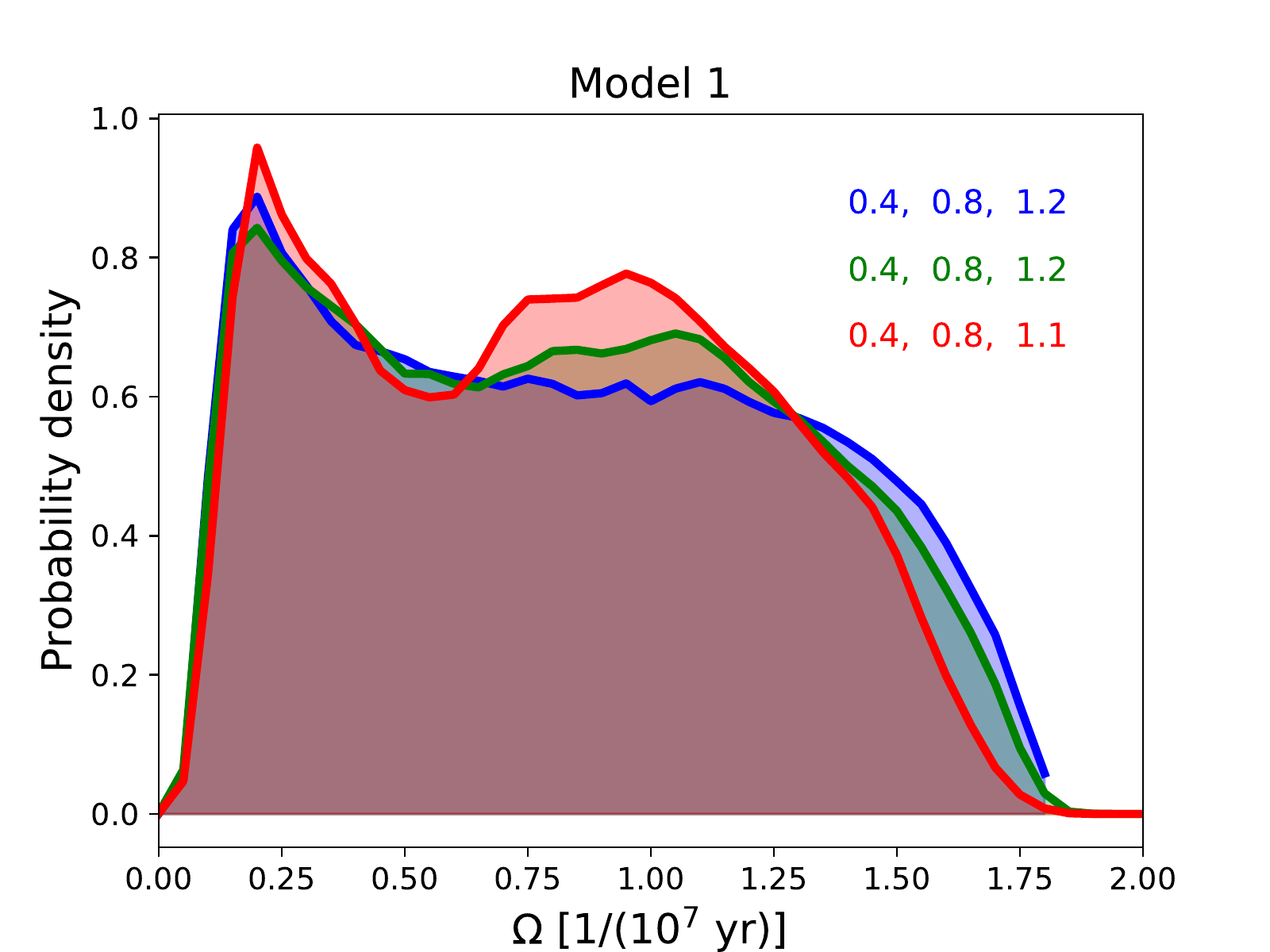}
\includegraphics[clip=true, trim = 0mm 0mm 0mm 0mm, width=0.24\linewidth]{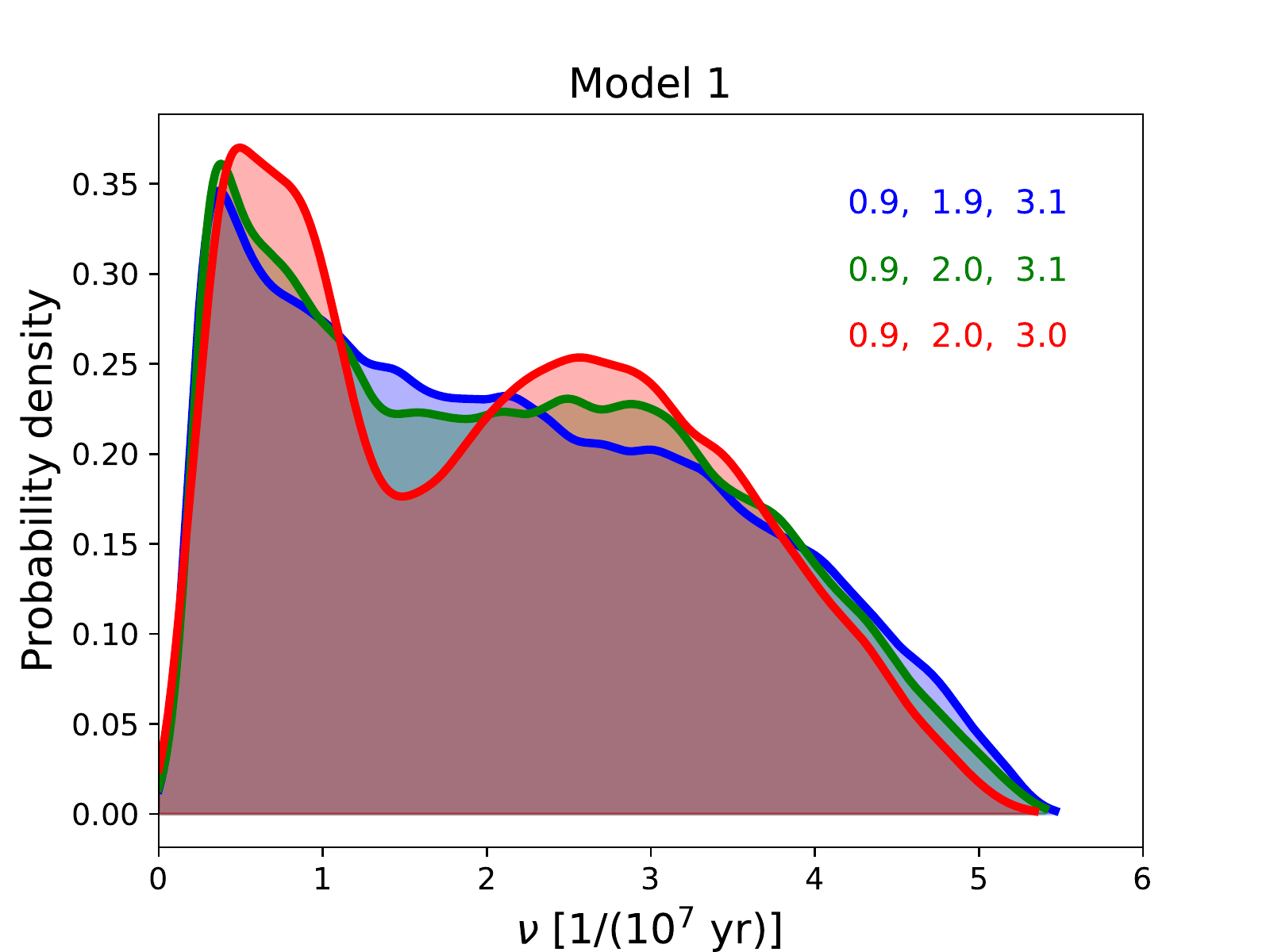}
\includegraphics[clip=true, trim = 0mm 0mm 0mm 0mm, width=0.24\linewidth]{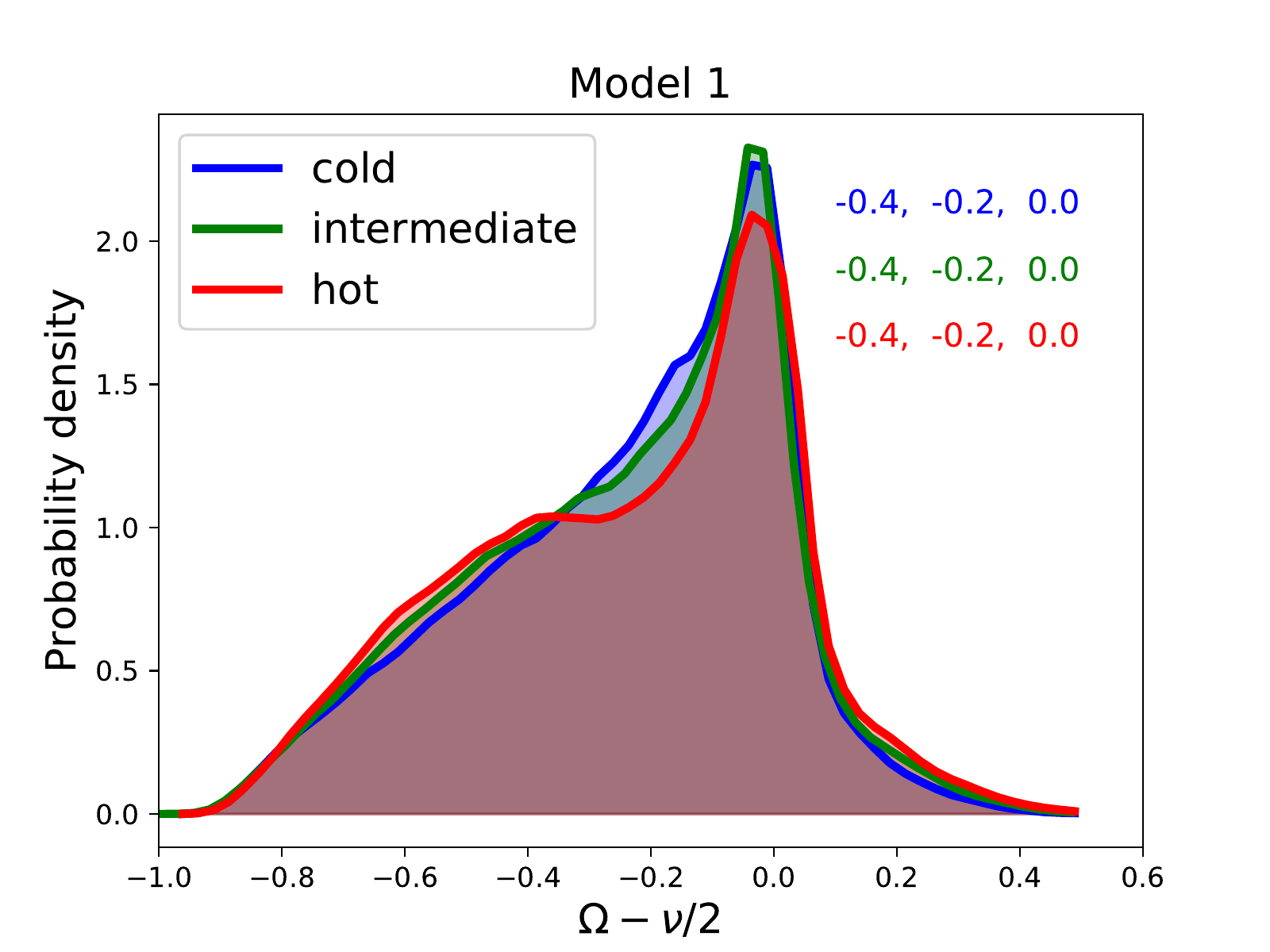}

\includegraphics[clip=true, trim = 0mm 0mm 0mm 0mm, width=0.24\linewidth]{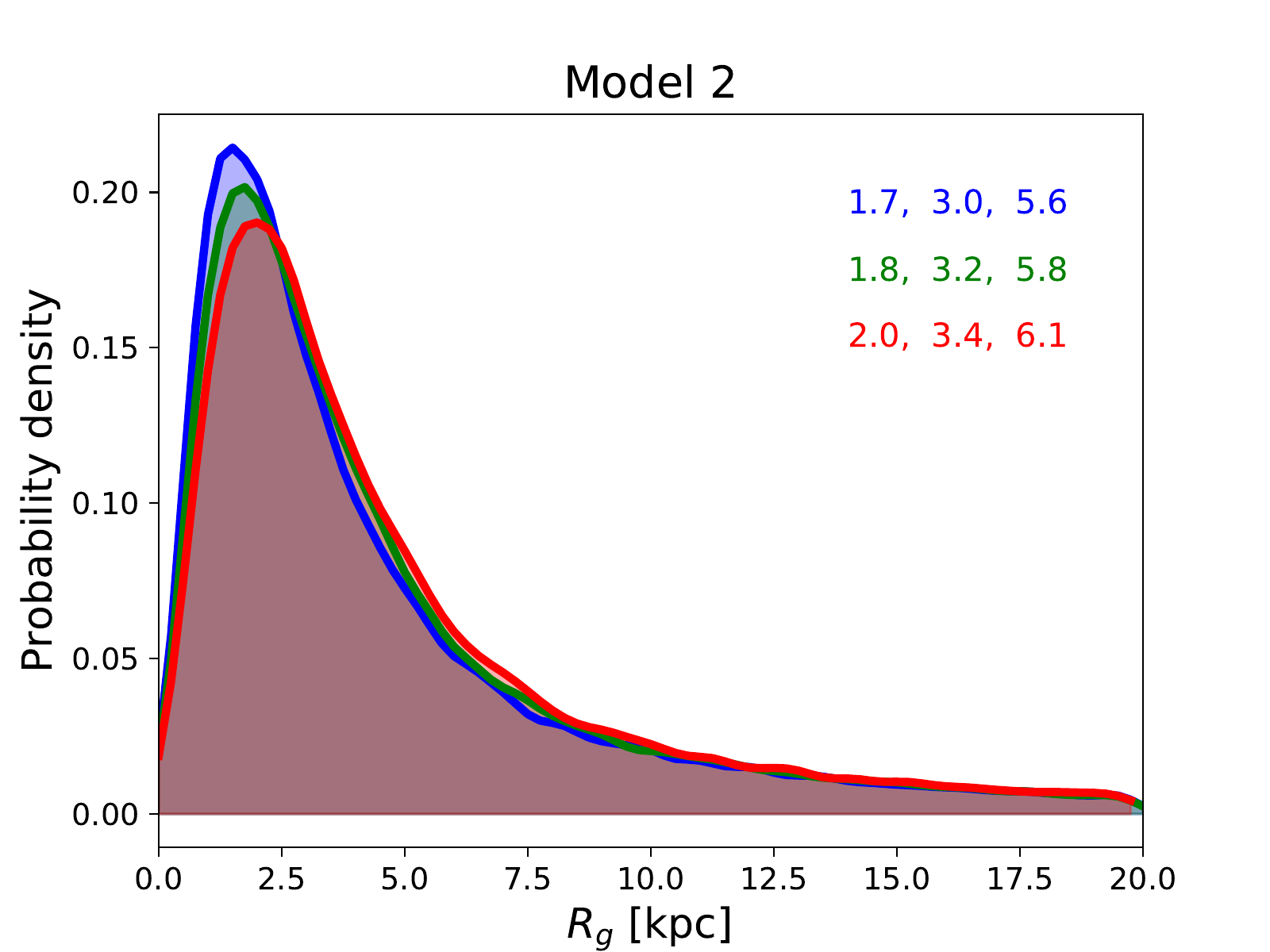}
\includegraphics[clip=true, trim = 0mm 0mm 0mm 0mm, width=0.24\linewidth]{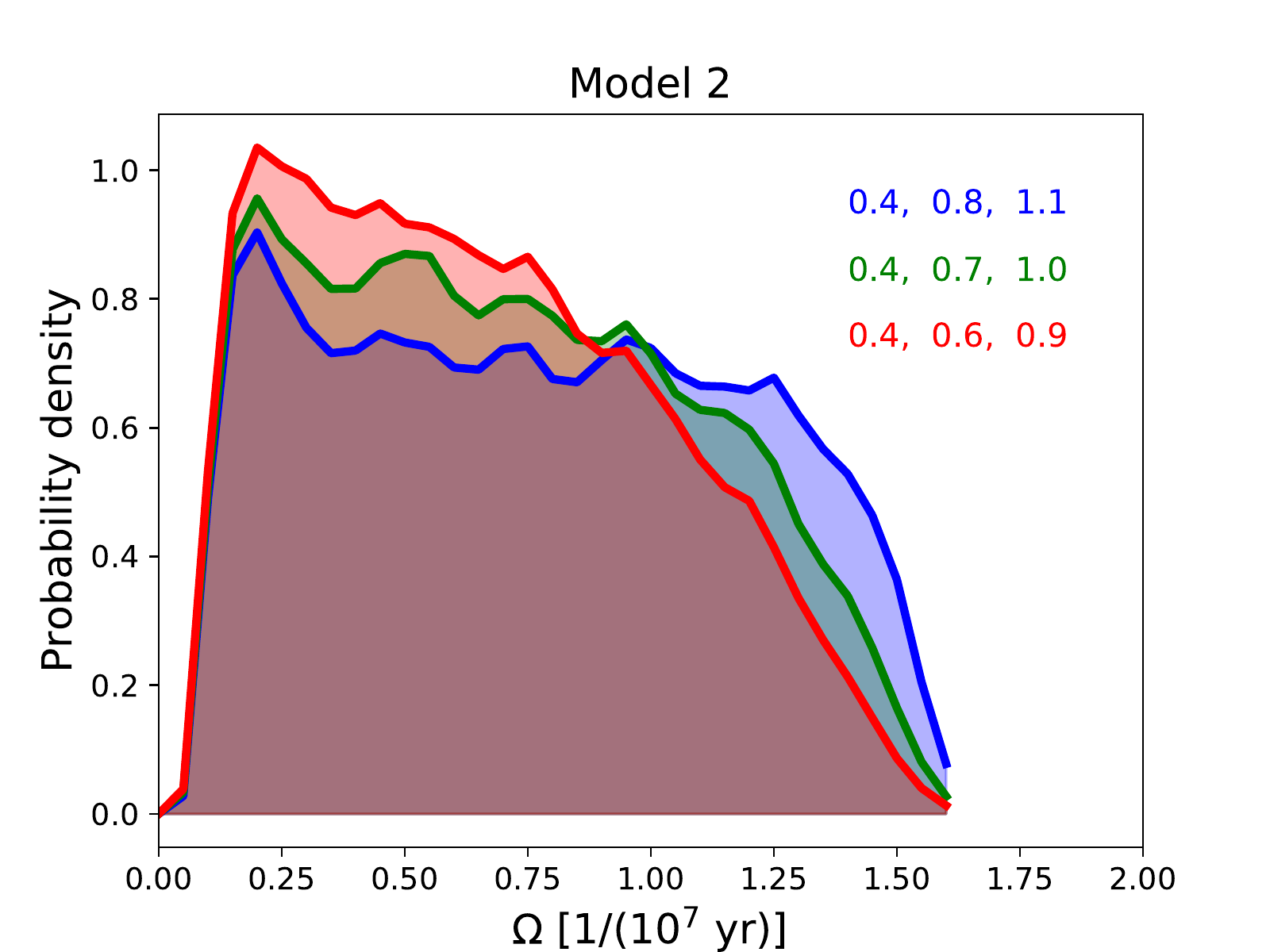}
\includegraphics[clip=true, trim = 0mm 0mm 0mm 0mm, width=0.24\linewidth]{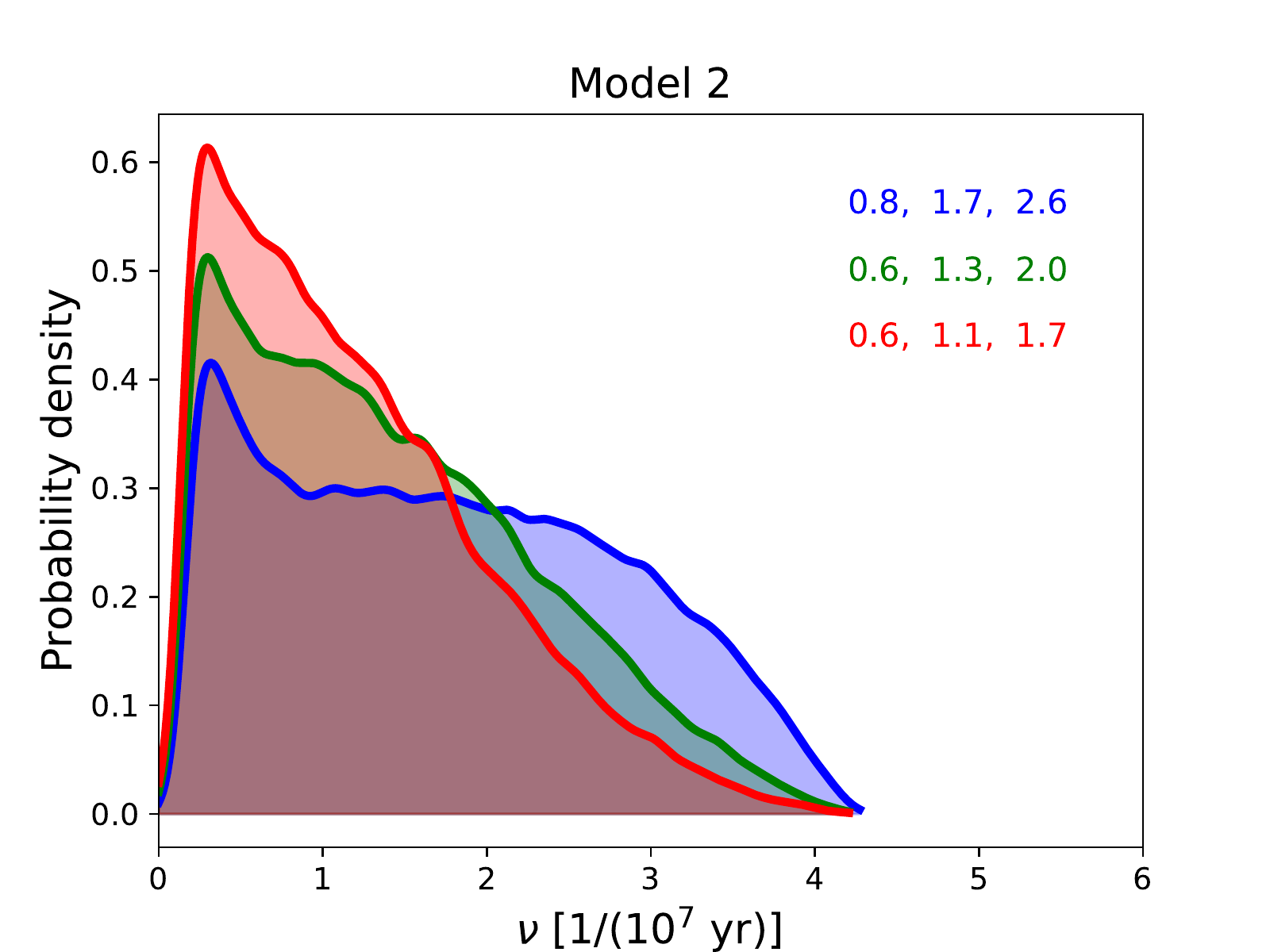}
\includegraphics[clip=true, trim = 0mm 0mm 0mm 0mm, width=0.24\linewidth]{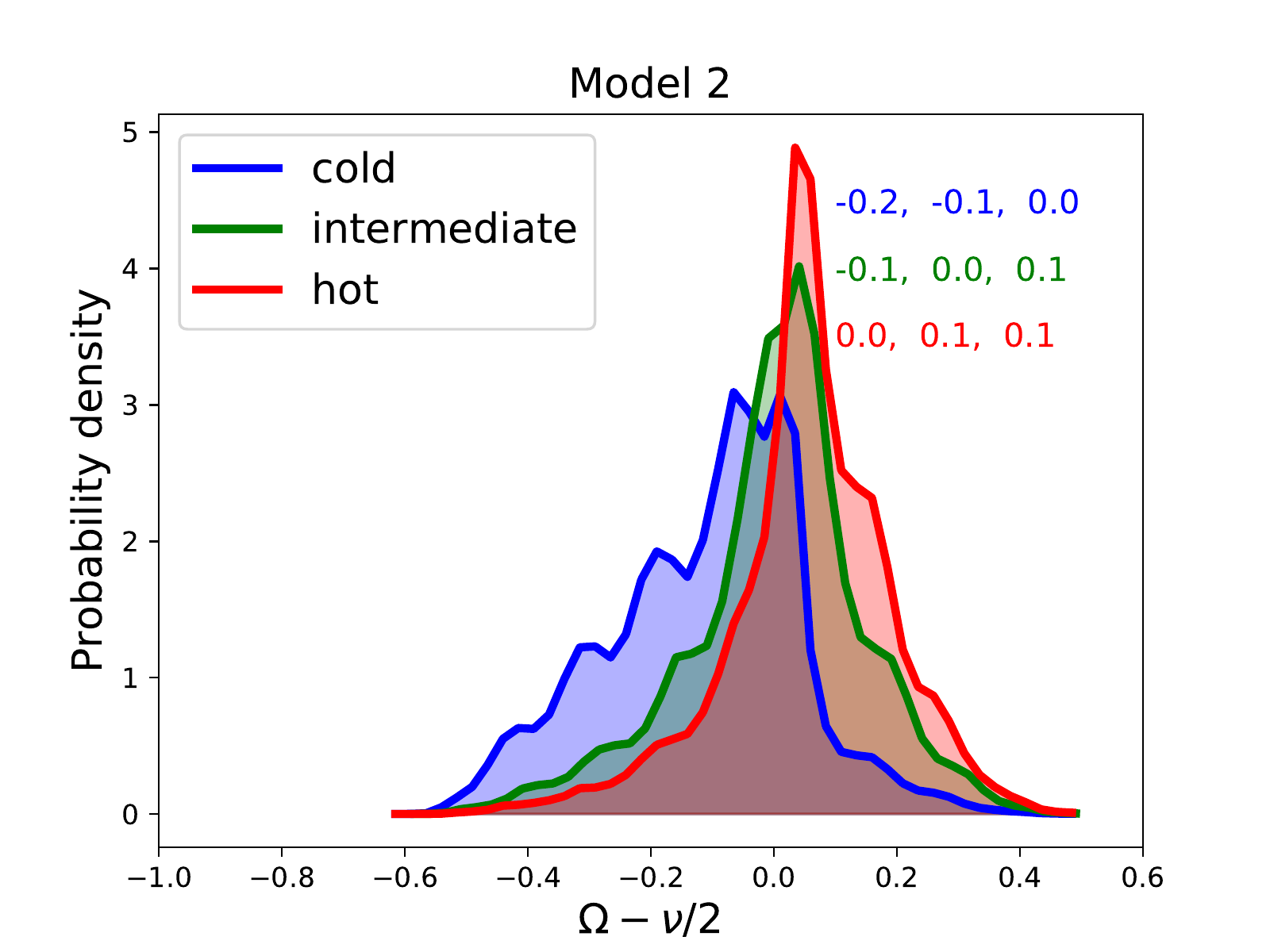}

\end{flushleft}
\caption{\emph{From left to right:} Distribution of guiding radii, azimuthal frequencies $\Omega$, vertical frequencies $\nu$ and of the differences $\Omega-\nu/2$, for Model 1 (\emph{top panels}), and Model  2 (\emph{bottom panels}), at t=0. In each plot, the blue, green and red curves correspond, respectively, to the cold, intermediate and hot disc components, as indicated. In each plot, we also give the  25$th$, 50$th$ and 75$th$ percentile of the distributions, with colors corresponding to each of these components.}
\label{freqhisto}
\end{figure*}

In Fig.~\ref{freqhisto}, we show the distributions of guiding radii, azimuthal, $\Omega$, and vertical frequencies, $\nu$, for the two simulations, at t=0, when the disc is still axisymmetric. While in Model~2, there is a significant difference in the distribution of the vertical frequencies for the three populations, in Model~1 all distributions appear remarkably similar. As a consequence, in Model~1 also the distribution of  $\Omega-\nu/2$ is similar for the three discs, implying that the fraction of stars that can respond  to the buckling instability, that is the fraction of stars which satisfies the inequality $\Omega_p \ge \Omega-\nu/2$, is similar in the three discs. In the previous inequality  $\Omega_p$ indicates the bar pattern speed.   

\section{Evolution of the in-plane and vertical velocity dispersions with time}\label{discs-app}

\begin{figure*}
\begin{center}
\includegraphics[clip=true, trim = 0mm 2mm 10mm 5mm, width=0.4\linewidth]{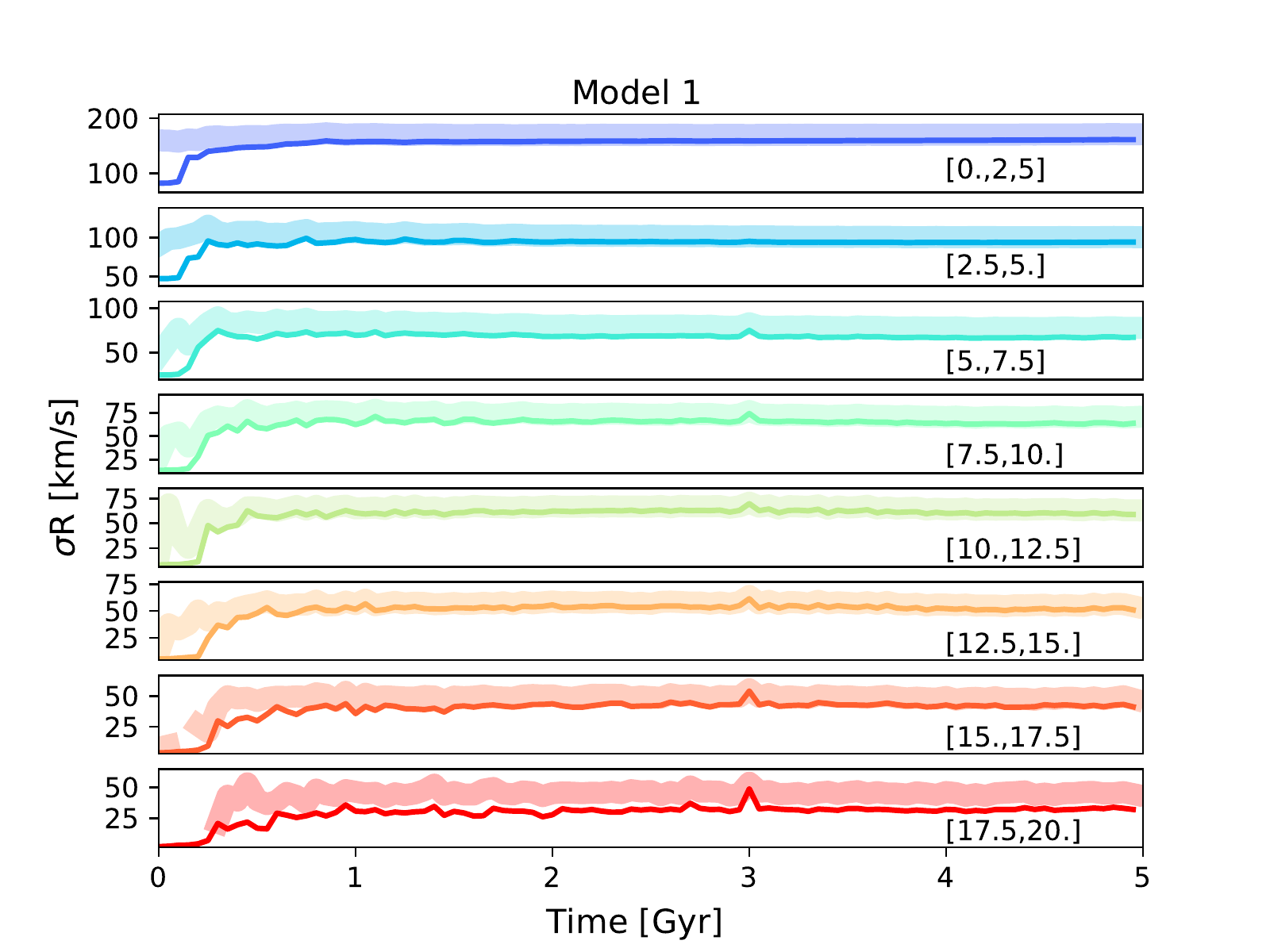}
\includegraphics[clip=true, trim = 0mm 2mm 10mm 5mm, width=0.4\linewidth]{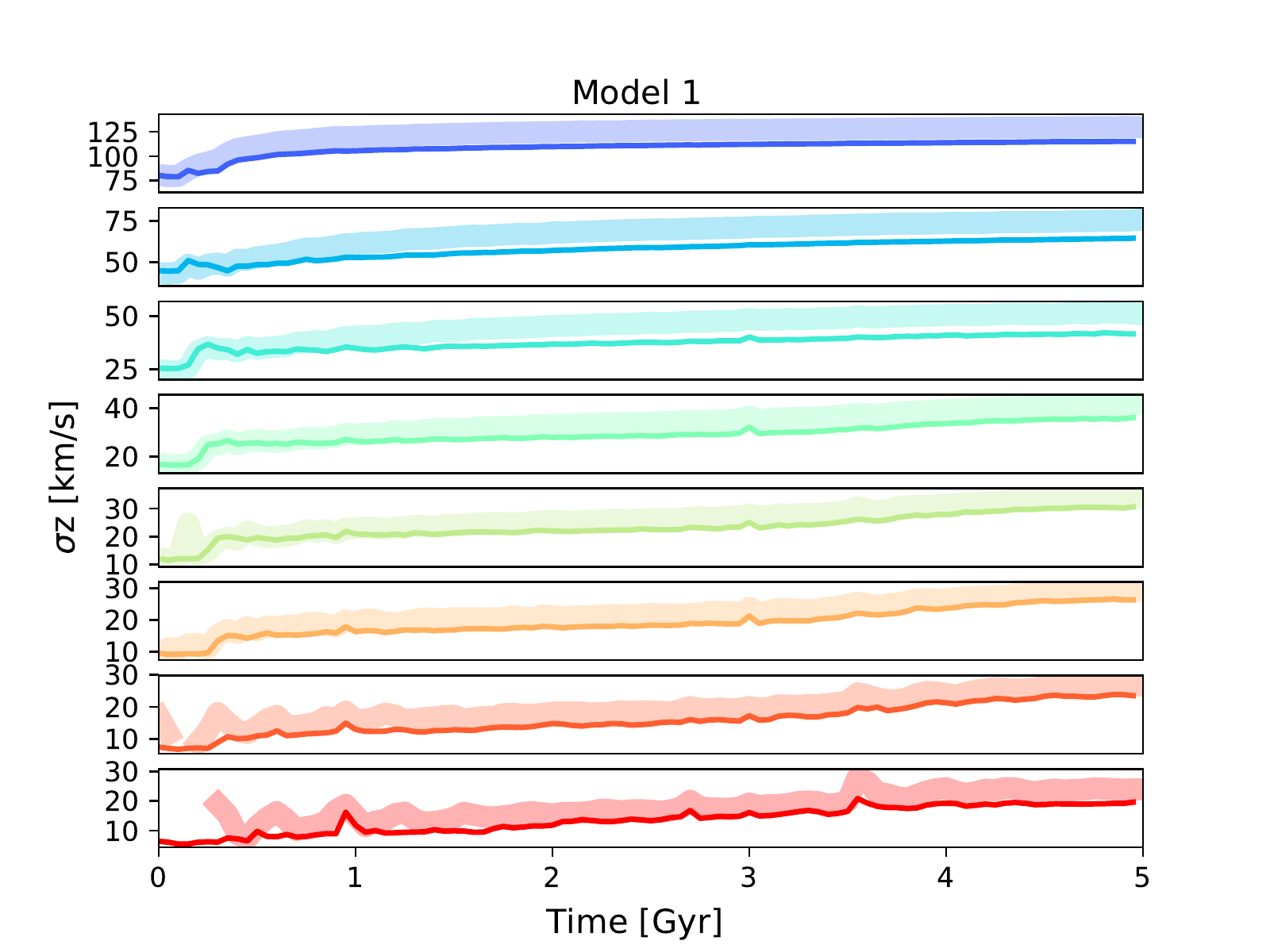}

\includegraphics[clip=true, trim = 0mm 2mm 10mm 5mm, width=0.4\linewidth]{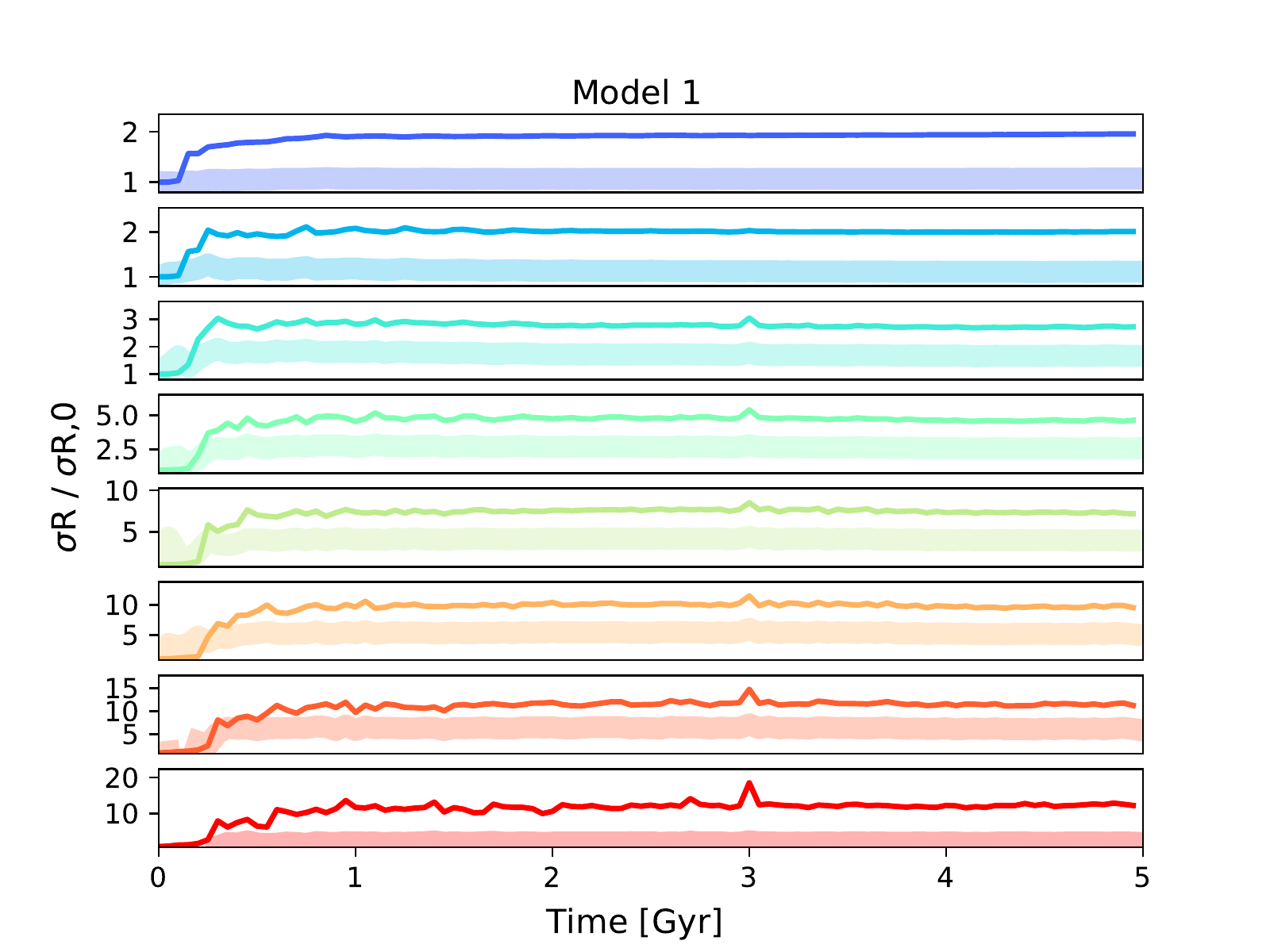}
\includegraphics[clip=true, trim = 0mm 2mm 10mm 5mm, width=0.4\linewidth]{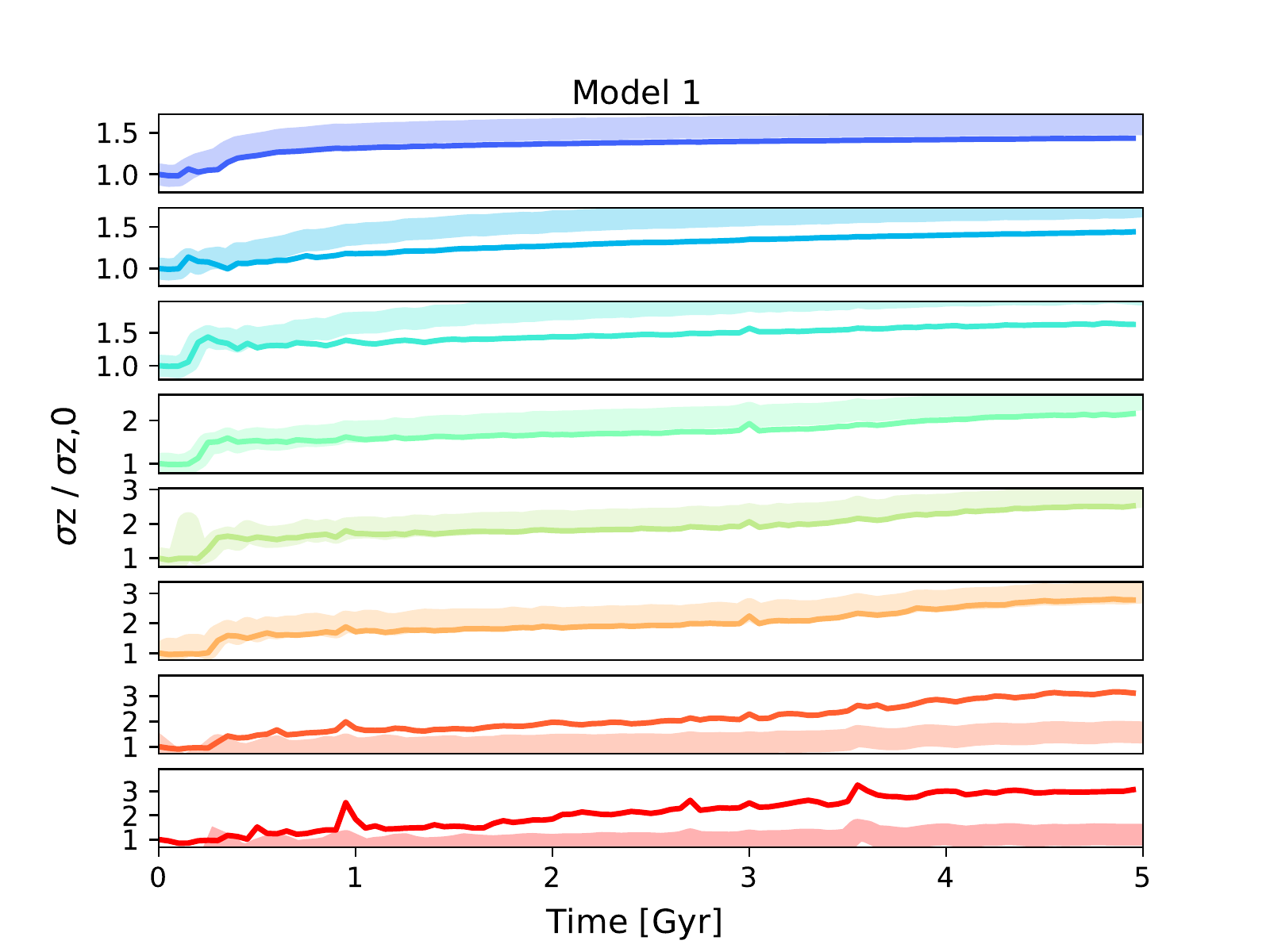}

\includegraphics[clip=true, trim = 0mm 2mm 10mm 5mm, width=0.4\linewidth]{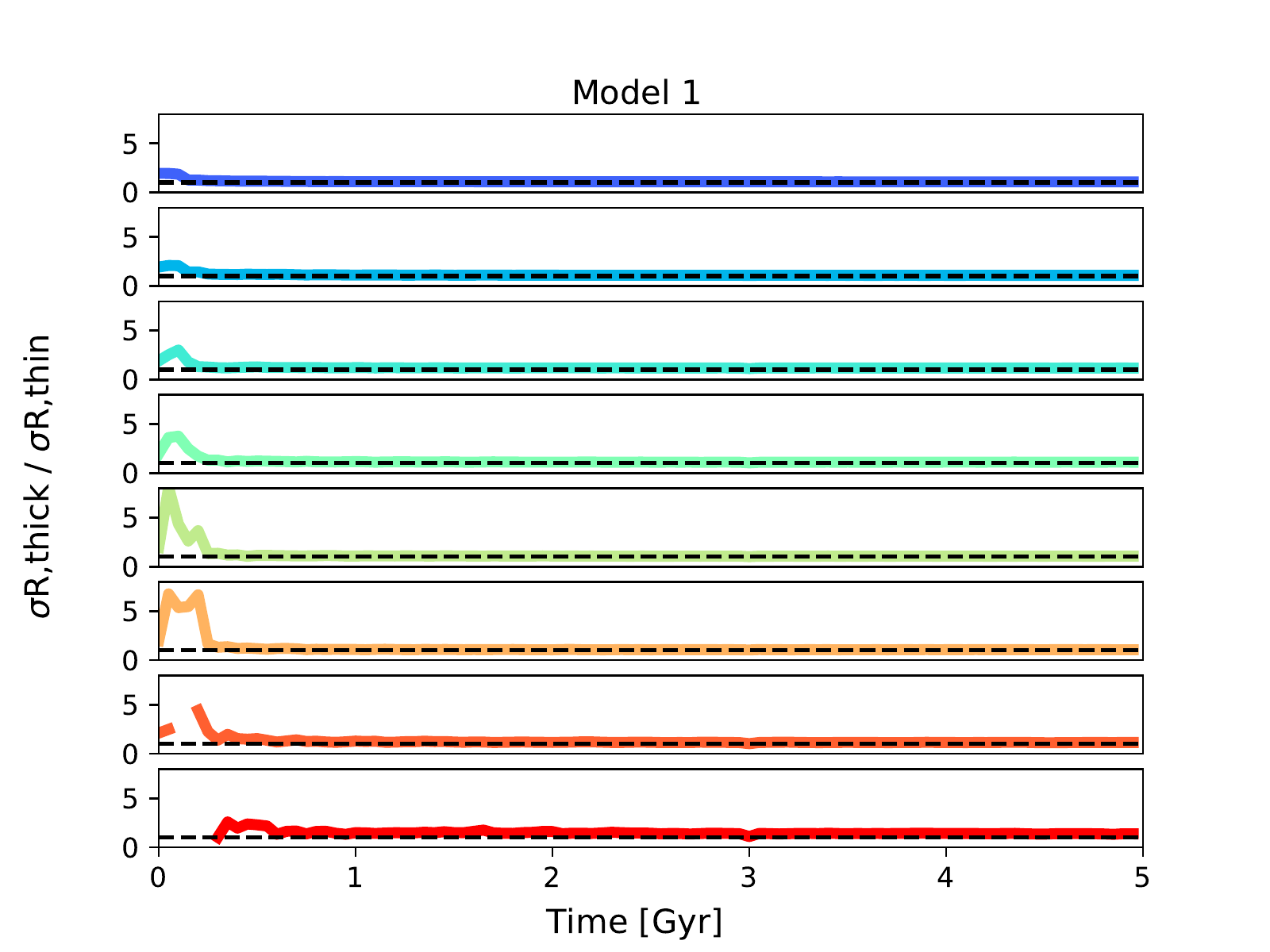}
\includegraphics[clip=true, trim = 0mm 2mm 10mm 5mm, width=0.4\linewidth]{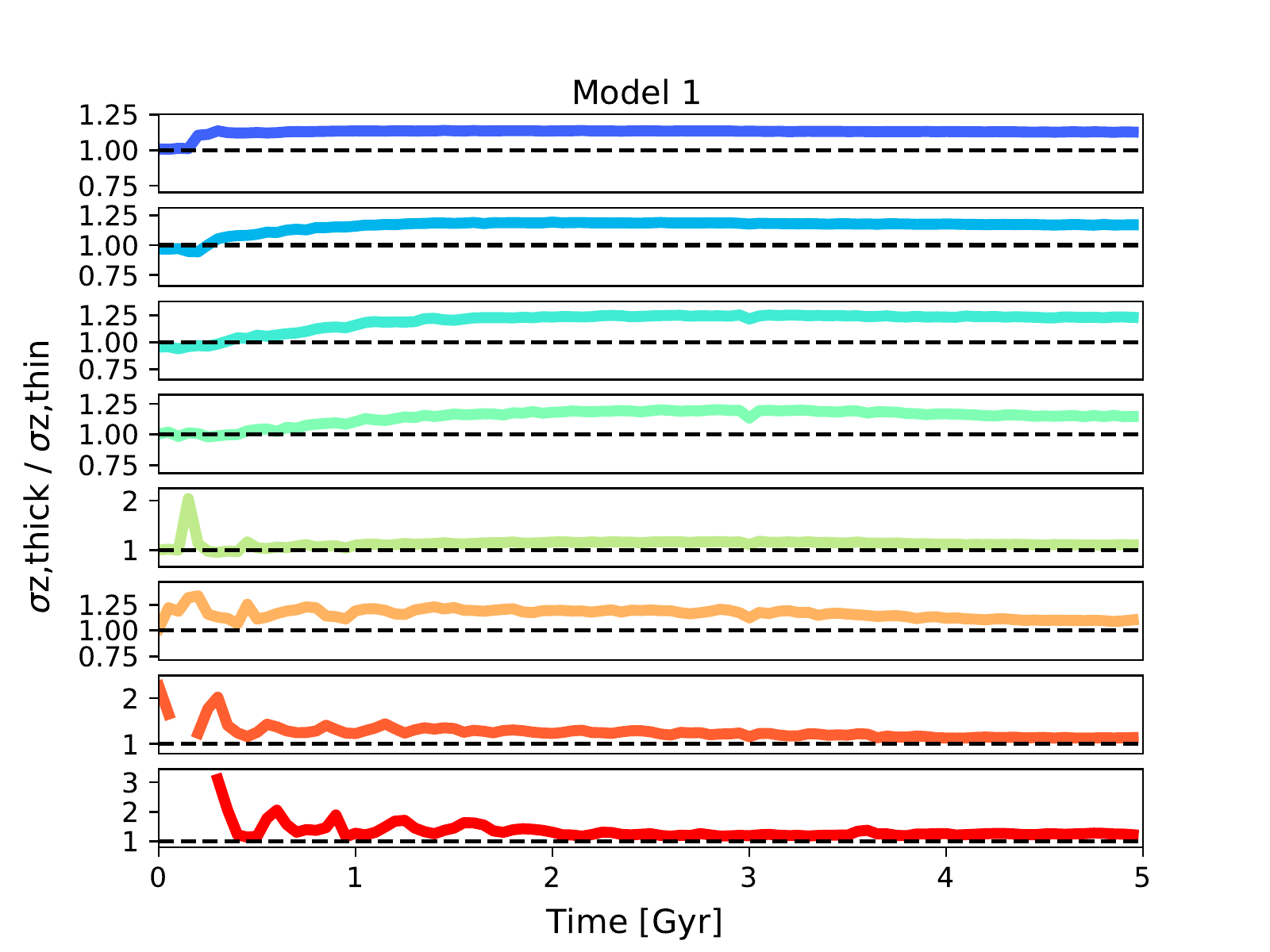}

\end{center}
\caption{ \emph{Top and middle rows:} Radial (\emph{left panels}) and vertical  (\emph{right panels}) velocity dispersions as a function of time for stars initially in the cold (thin curves) and hot (thick curves) discs of Model 1. Velocity dispersions are shown for 8 different radial annulli in the modeled disc, as indicated in the top-left panel, which are also represented by different colors. In the middle panels, velocity dispersions  are normalized to their corresponding value at the initial time of the simulation. \emph{Bottom row:} Ratio of the radial (\emph{left panel}) and vertical (\emph{right panel}) velocity dispersions of stars in the initially hot and cold discs, for the 8 different radial annulli, as a function of time.  The dashed line indicates a ratio equal to unity.  }
\label{sigmavstimeM1}
\end{figure*}

\begin{figure*}
\begin{center}
\includegraphics[clip=true, trim = 0mm 2mm 10mm 5mm, width=0.4\linewidth]{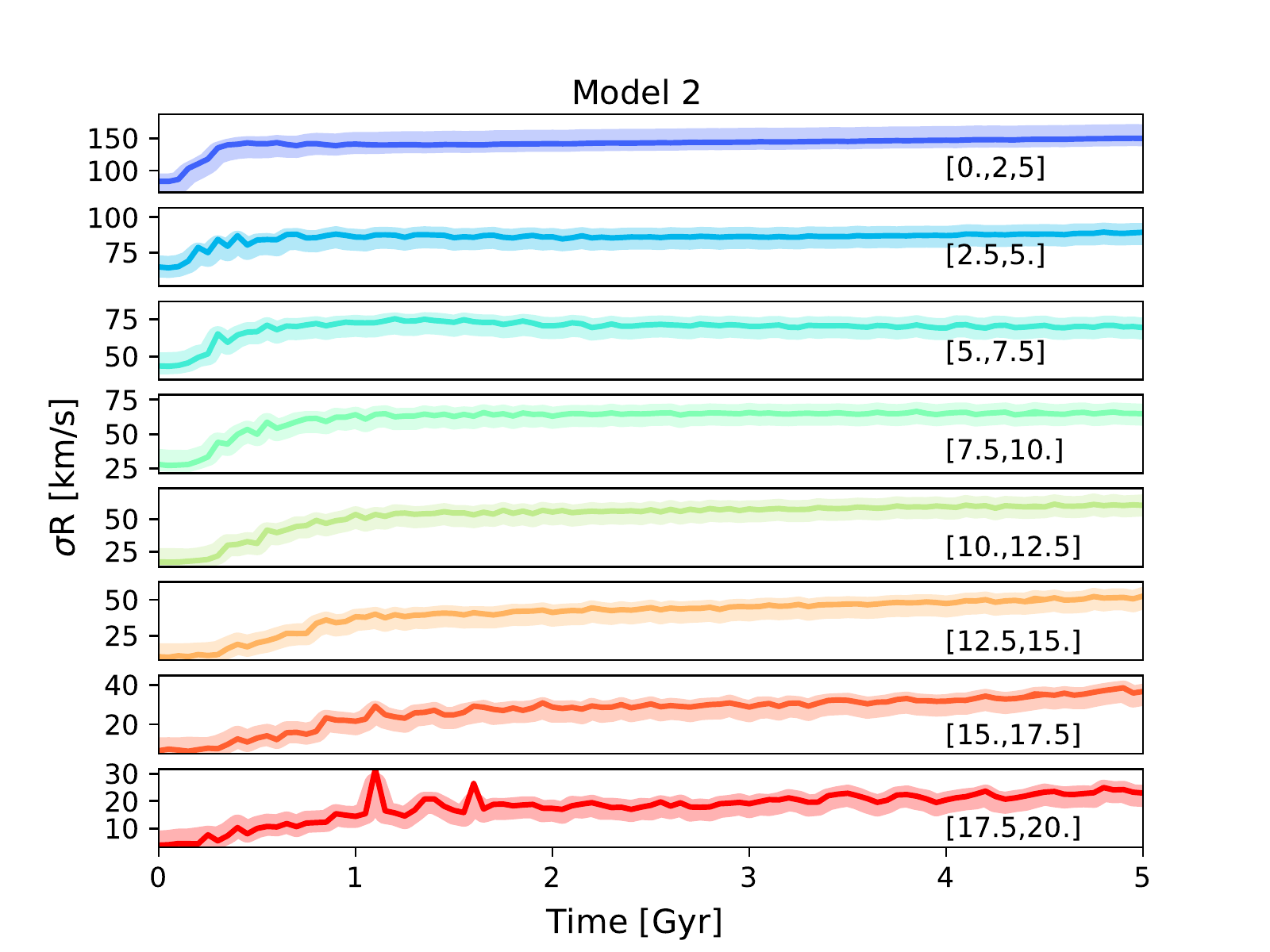}
\includegraphics[clip=true, trim = 0mm 2mm 10mm 5mm, width=0.4\linewidth]{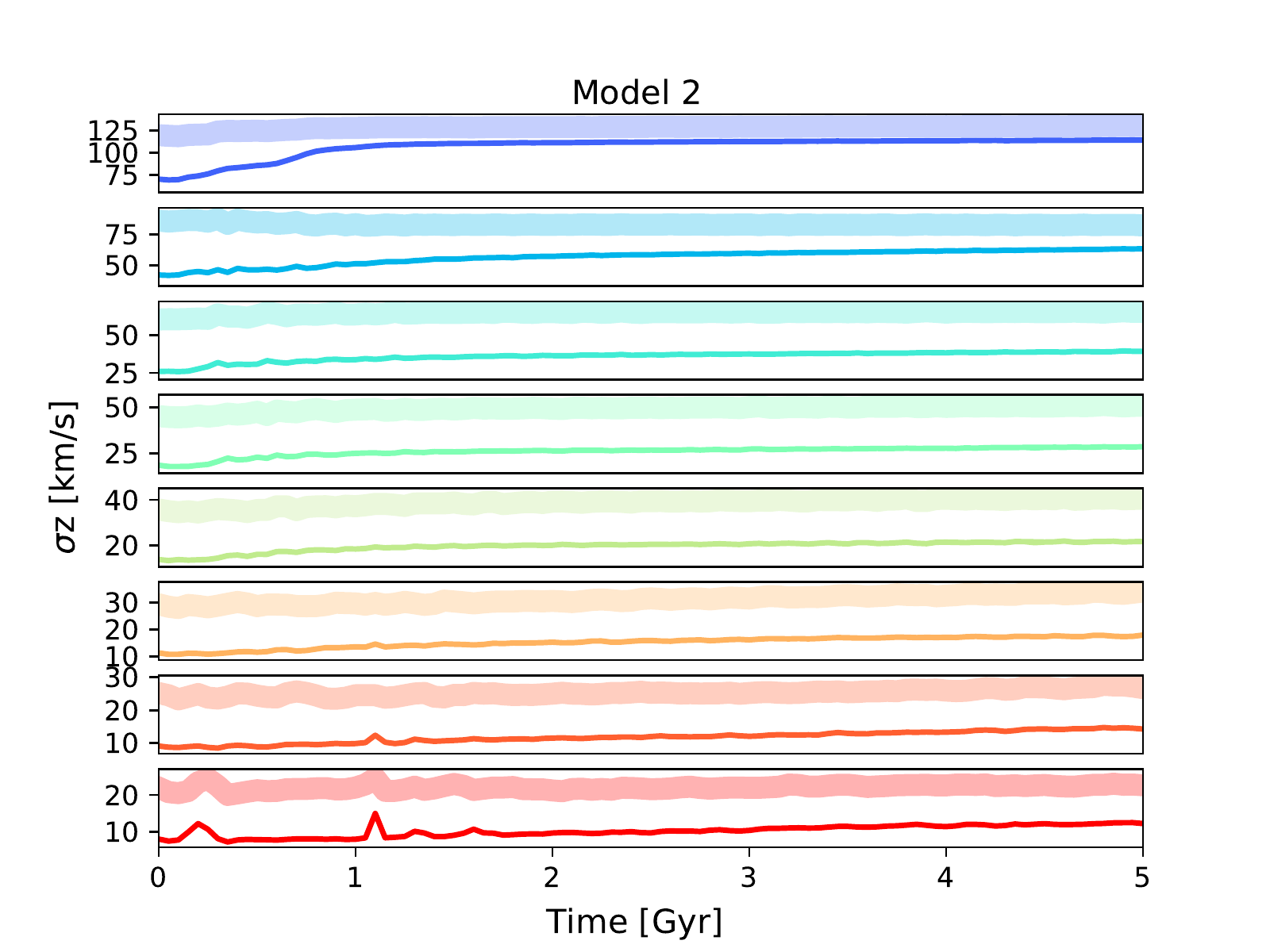}

\includegraphics[clip=true, trim = 0mm 2mm 10mm 5mm, width=0.4\linewidth]{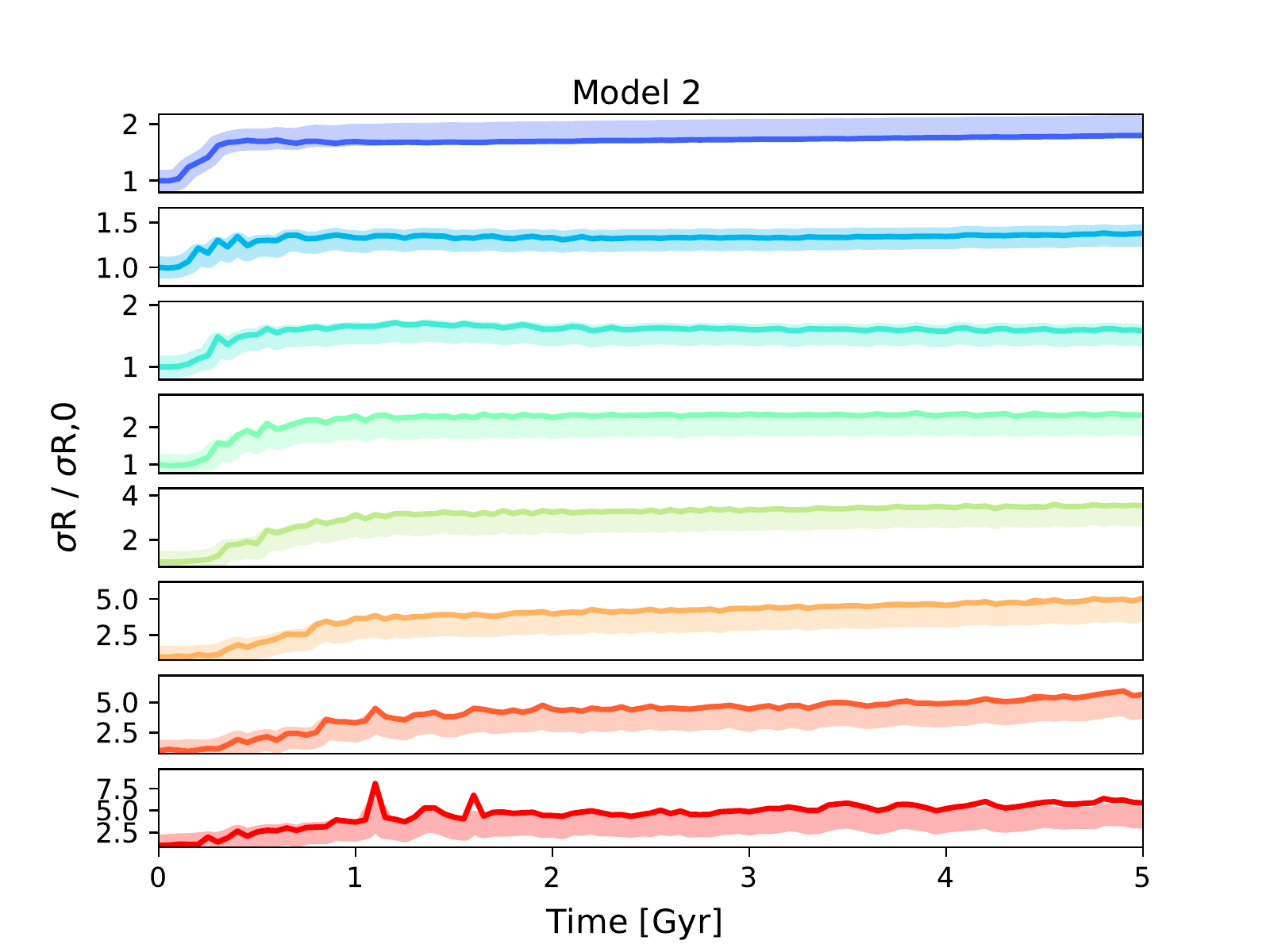}
\includegraphics[clip=true, trim = 0mm 2mm 10mm 5mm, width=0.4\linewidth]{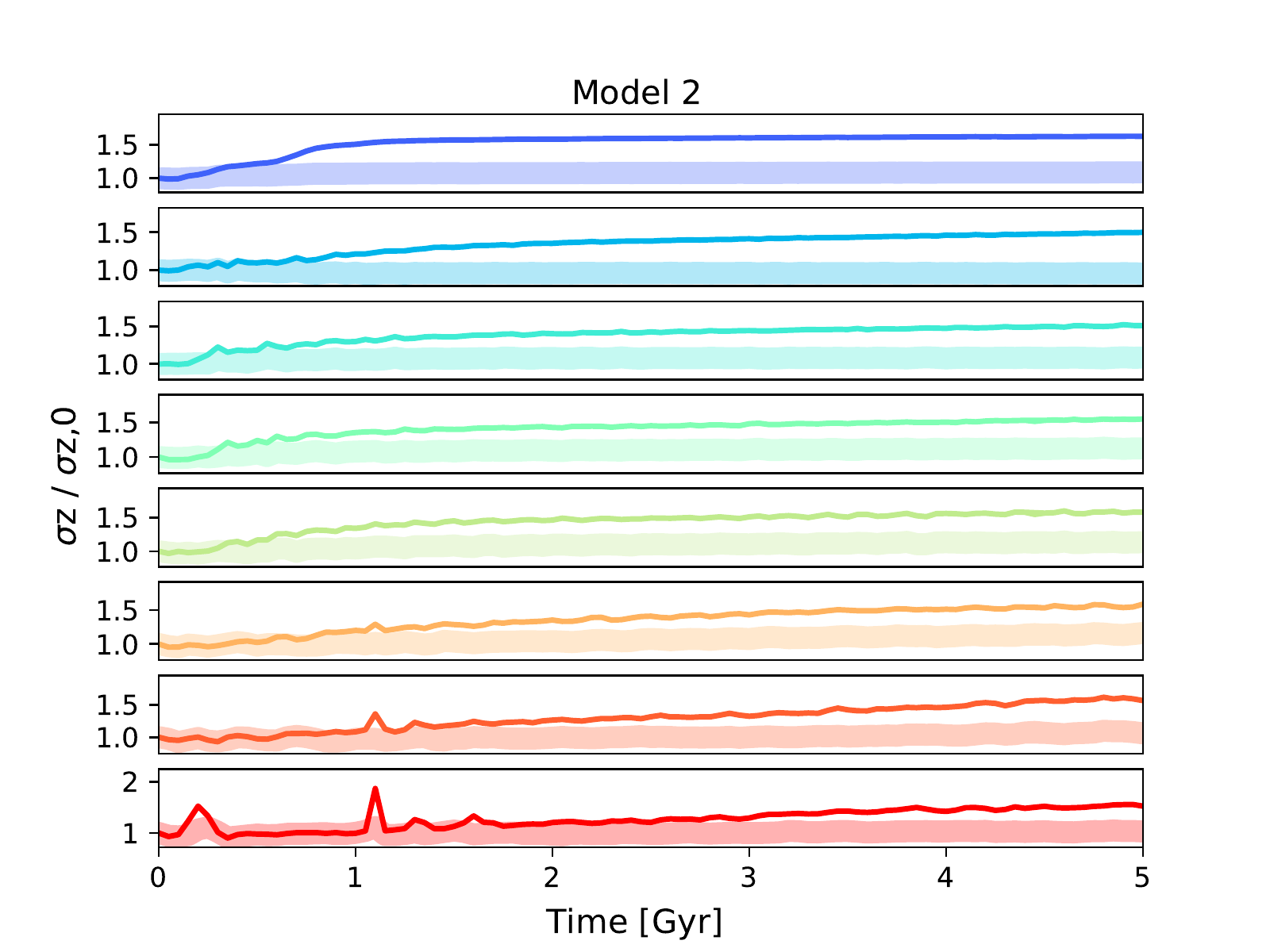}

\includegraphics[clip=true, trim = 0mm 2mm 10mm 5mm, width=0.4\linewidth]{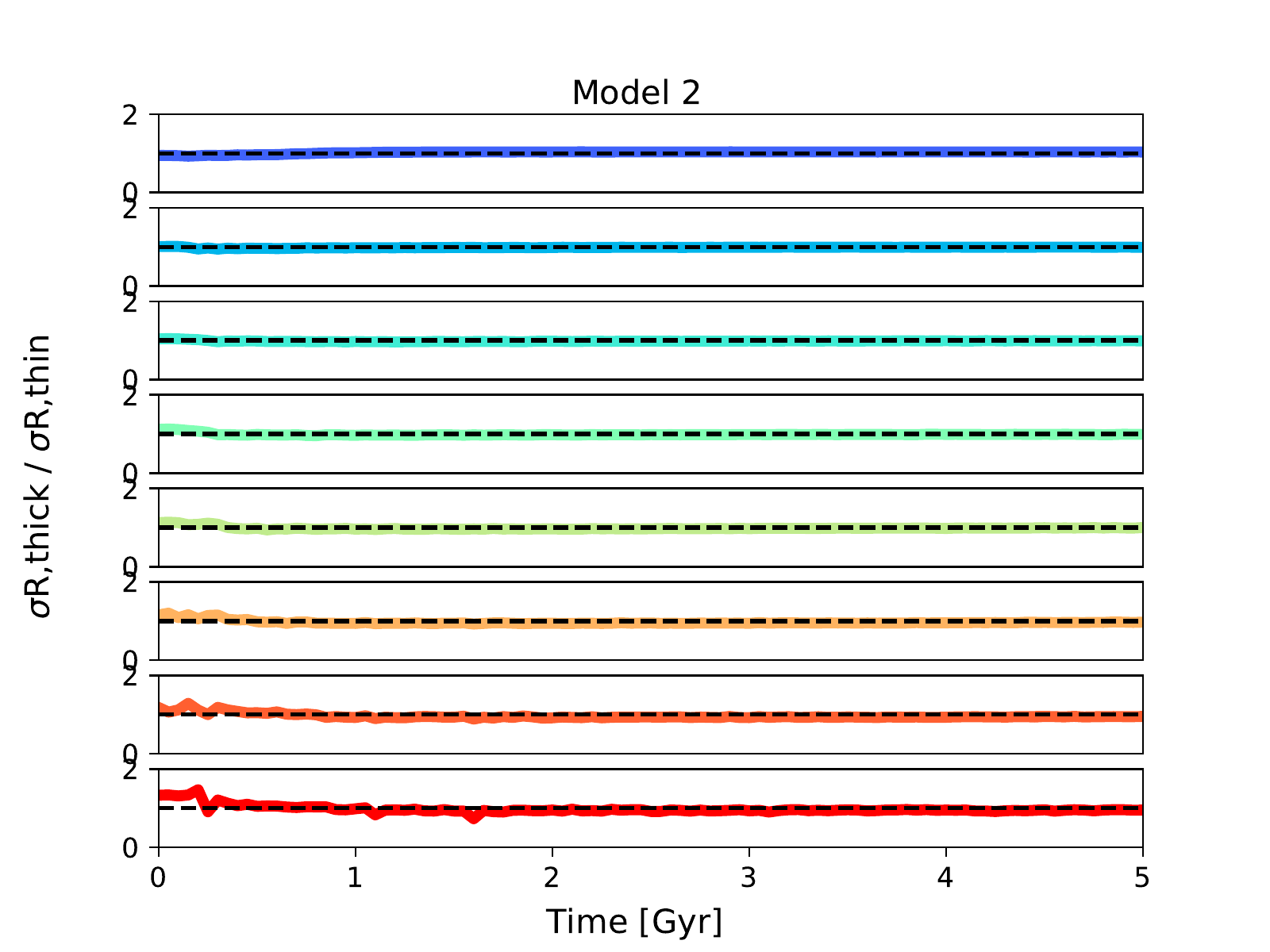}
\includegraphics[clip=true, trim = 0mm 2mm 10mm 5mm, width=0.4\linewidth]{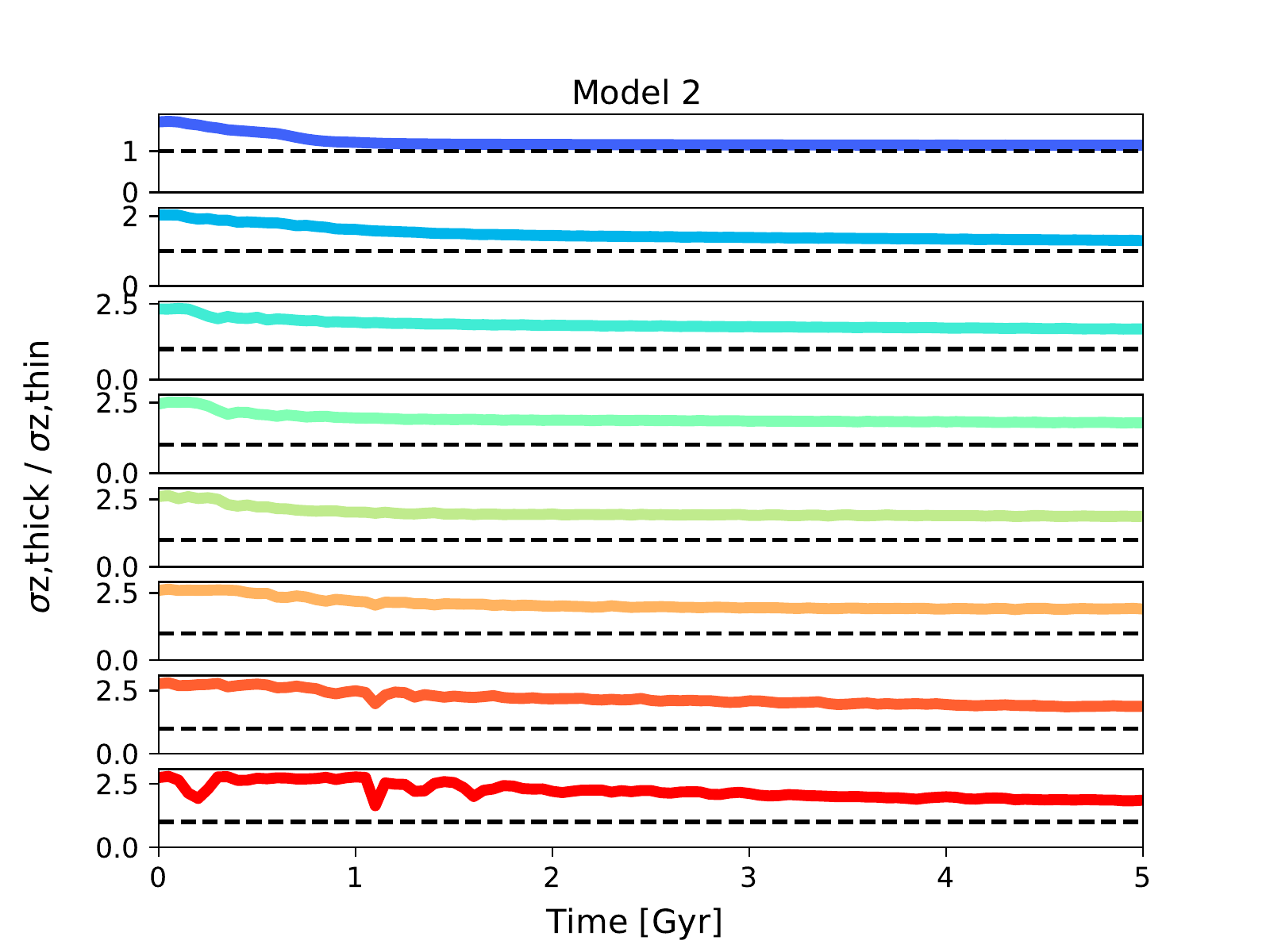}

\end{center}
\caption{Same as Fig.~\ref{sigmavstimeM1}, but for Model 2.}
\label{sigmavstimeM2}
\end{figure*}

In this Section, we show the temporal evolution of the radial and vertical velocity dispersion of the initially kinematically cold and hot discs, for  Models 1 and 2, in Figs.~\ref{sigmavstimeM1} and ~\ref{sigmavstimeM2}, respectively.
In both plots, these velocity dispersions are shown for  8 different radial annuli, covering the whole extent of the simulated galaxy, from its inner regions to its outskirts. 
For clarity, in these plots we do not show the intermediate disc, but we checked that its temporal  behavior is always bracketed between that of the initially hot and cold populations. 
We refer the reader to Sect.~\ref{bp} for a discussion about the implications of these plots on the vertical metallicity gradients  outside the B/P bulges of Model~1 and 2, and to the comprehensive works of \citet{aumer16, aumer17, aumer17b} for the evolution of vertical and in-plane random motions in composite thin and thick stellar discs. 

\end{appendix}

\end{document}